%% file: CP_N.tex
\documentclass[final,1p,12pt]{elsarticle}
\usepackage[T1]{fontenc}
\usepackage[utf8]{inputenc}
\usepackage[ngerman,english]{babel}
\usepackage{amsmath}
\usepackage{amsfonts}
\usepackage{amssymb}
\usepackage{amsthm}
\usepackage{enumitem}
\usepackage{dcolumn}
\usepackage{bm}
\usepackage{bbm}
\usepackage{cancel}
\usepackage{tikz}
\usetikzlibrary{calc,shapes,arrows,patterns,snakes,positioning,automata}
\usepackage{tabularx}
\usepackage{booktabs}
\usepackage{tensor}
\usepackage{scalerel}
\usepackage[small]{caption}
\usepackage{xcolor}
\usepackage{mathtools}
\usepackage{float}
\usepackage{xfrac}
\usepackage[pdfborder={0 0 0}]{hyperref}
\definecolor{darkblue}{rgb}{0,0,.5}
\definecolor{darkgreen}{rgb}{0,0.5,.5}
\definecolor{darkyellow}{rgb}{0.5,0.5,0}
\definecolor{fhl}{rgb}{1,0,0}
\hypersetup{colorlinks=true, breaklinks=true, linkcolor=darkblue, menucolor=darkblue, urlcolor=darkblue, linktocpage=true}
\usepackage{times}
\usepackage{geometry}
\usepackage{textpos}
\DeclareMathSizes{12}{10}{8}{6} 
\usepackage[titletoc,title]{appendix}

\makeatletter 
\newsavebox\myboxA 
\newsavebox\myboxB 
\newlength\mylenA 
 
\newcommand*\xoverline[2][0.75]{%
    \sbox{\myboxA}{$\m@th#2$}%
    \setbox\myboxB\null
    \ht\myboxB=\ht\myboxA%
    \dp\myboxB=\dp\myboxA%
    \wd\myboxB=#1\wd\myboxA
    \sbox\myboxB{$\m@th\overline{\copy\myboxB}$}
    \setlength\mylenA{\the\wd\myboxA}
    \addtolength\mylenA{-\the\wd\myboxB}%
    \ifdim\wd\myboxB<\wd\myboxA%
       \rlap{\hskip 0.5\mylenA\usebox\myboxB}{\usebox\myboxA}%
    \else 
        \hskip -0.5\mylenA\rlap{\usebox\myboxA}{\hskip 0.5\mylenA\usebox\myboxB}%
    \fi}
\makeatother
\numberwithin{equation}{section}

\let\originalleft\left
\let\originalright\right
\renewcommand{\left}{\mathopen{}\mathclose\bgroup\originalleft}
\renewcommand{\right}{\aftergroup\egroup\originalright}
\newcommand{\e}{\operatorname{e}}
\newcommand{\SU}[1]{\operatorname{SU}\left(#1\right)}
\newcommand{\On}[1]{\operatorname{O}\left(#1\right)}
\newcommand{\Un}[1]{\operatorname{U}\left(#1\right)}
\newcommand{\CP}[1]{\operatorname{CP}\left(#1\right)}

\newcommand{\CPn}[1]{\operatorname{\mathbb{C}P}^{#1}}

\newcommand{\of}[1]{\left(#1\right)}

\newcommand{\roint}[1]{\left[#1\right)}
\newcommand{\bof}[1]{\biggl(\bigg.#1\bigg.\biggr)}
\newcommand{\sof}[1]{\bigl(\big.#1\big.\bigr)}
\newcommand{\ssof}[1]{(#1)}
\newcommand{\fof}[1]{\left[#1\right]}

\newcommand{\sfof}[1]{\bigl[\big.#1\big.\bigr]}
\newcommand{\cof}[1]{\left\{#1\right\}}
\newcommand{\bcof}[1]{\biggl\{\bigg.#1\bigg.\biggr\}}

\newcommand{\avof}[1]{\left\langle #1\right\rangle}
\newcommand{\savof}[1]{\big\langle #1\big\rangle}
\newcommand{\ssavof}[1]{\small\langle #1\small\rangle}

\newcommand{\ii}{\mathrm{i}}
\newcommand{\idd}[2]{\mathrm{d}^{#2}#1}
\newcommand{\dd}{\mathrm{d}}
\newcommand{\DD}[1]{\mathcal{D}\bigl[#1\bigr]}
\newcommand{\cD}{\mathrm{D}}
\newcommand{\totd}[3]{\frac{\dd^{#3} #1}{\dd #2^{#3}}}

\newcommand{\partd}[2]{\frac{\partial #1}{\partial #2}}
\newcommand{\partdm}[3]{\frac{\partial^{#3} #1}{\partial #2^{#3}}}
\newcommand{\spartd}[3]{\frac{\partial^{2} #1}{\partial #2 \partial #3}}

\newcommand{\order}[1]{\mathcal{O}\big(#1\big)}

\newcommand{\obs}{\mathcal{O}}

\newcommand{\abs}[1]{\left| #1\right|}
\newcommand{\ssabs}[1]{| #1|}
\newcommand{\sabs}[1]{\big| #1\big|}

\newcommand{\op}[1]{\operatorname{#1}}

\renewcommand*\[{\begin{equation}}
\renewcommand*\]{\end{equation}}
\let\obar\bar
\renewcommand*\bar[1]{\ThisStyle{\xoverline{\SavedStyle #1}}}
\let\ohat\hat
\renewcommand*\hat[1]{\widehat{#1}}
\let\oldstackrel\stackrel
\renewcommand*\stackrel[2]{{\scriptstyle\oldstackrel{#1}{#2}}}

\definecolor{emphcol}{RGB}{0,0,0}
\let\oldemph\emph
\renewcommand*\emph[1]{\oldemph{\textcolor{emphcol}{#1}}}
\newcommand{\ucases}[1]{\begin{cases}#1\end{cases}}

\newlength{\hatchspread}
\newlength{\hatchthickness}
\newlength{\hatchshift}
\newcommand{\hatchcolor}{}
\tikzset{hatchspread/.code={\setlength{\hatchspread}{#1}},
         hatchthickness/.code={\setlength{\hatchthickness}{#1}},
         hatchshift/.code={\setlength{\hatchshift}{#1}},
         hatchcolor/.code={\renewcommand{\hatchcolor}{#1}}}
\tikzset{hatchspread=3pt,
         hatchthickness=0.4pt,
         hatchshift=0pt,
         hatchcolor=black}
\pgfdeclarepatternformonly[\hatchspread,\hatchthickness,\hatchshift,\hatchcolor]
   {custom north west lines}
   {\pgfqpoint{\dimexpr-2\hatchthickness}{\dimexpr-2\hatchthickness}}
   {\pgfqpoint{\dimexpr\hatchspread+2\hatchthickness}{\dimexpr\hatchspread+2\hatchthickness}}
   {\pgfqpoint{\dimexpr\hatchspread}{\dimexpr\hatchspread}}
   {
    \pgfsetlinewidth{\hatchthickness}
    \pgfpathmoveto{\pgfqpoint{\dimexpr0pt}{\dimexpr\hatchspread+\hatchshift}}
    \pgfpathlineto{\pgfqpoint{\dimexpr\hatchspread+0.15pt}{\dimexpr\hatchshift-0.15pt}}
    \ifdim \hatchshift > 0pt
      \pgfpathmoveto{\pgfqpoint{\dimexpr-0.15pt}{\dimexpr\hatchshift+0.15pt}}
      \pgfpathlineto{\pgfqpoint{\dimexpr\hatchshift+0.15pt}{\dimexpr-0.15pt}}
    \fi
    \pgfsetstrokecolor{\hatchcolor}
    \pgfusepath{stroke}
   }

\pgfdeclarepatternformonly[\hatchspread,\hatchthickness,\hatchshift,\hatchcolor]
   {custom north east lines}
   {\pgfqpoint{\dimexpr-2\hatchthickness}{\dimexpr-2\hatchthickness}}
   {\pgfqpoint{\dimexpr\hatchspread+2\hatchthickness}{\dimexpr\hatchspread+2\hatchthickness}}
   {\pgfqpoint{\dimexpr\hatchspread}{\dimexpr\hatchspread}}
   {
    \pgfsetlinewidth{\hatchthickness}
    \pgfpathmoveto{\pgfqpoint{0pt}{\dimexpr\hatchshift}}
    \pgfpathlineto{\pgfqpoint{\dimexpr\hatchspread+0.15pt}{\dimexpr\hatchspread+0.15pt+\hatchshift}}
    \ifdim \hatchshift > 0pt
      \pgfpathmoveto{\pgfqpoint{\dimexpr\hatchspread-\hatchshift-0.15pt}{\dimexpr-0.15pt}}
      \pgfpathlineto{\pgfqpoint{\dimexpr\hatchspread+0.15pt}{\dimexpr\hatchshift+0.15pt}}
    \fi
    \pgfsetstrokecolor{\hatchcolor}
    \pgfusepath{stroke}
   }

\pgfdeclarepatternformonly[\hatchspread,\hatchthickness,\hatchshift,\hatchcolor]
   {custom vertical lines}
   {\pgfqpoint{\dimexpr-2\hatchthickness}{\dimexpr-2\hatchthickness}}
   {\pgfqpoint{\dimexpr\hatchspread+2\hatchthickness}{\dimexpr\hatchspread+2\hatchthickness}}
   {\pgfqpoint{\dimexpr\hatchspread}{\dimexpr\hatchspread}}
   {
    \pgfsetlinewidth{\hatchthickness}
    \pgfpathmoveto{\pgfqpoint{\dimexpr\hatchshift}{0pt}}
    \pgfpathlineto{\pgfqpoint{\dimexpr\hatchshift}{\dimexpr\hatchspread+0.15pt}}
    \pgfsetstrokecolor{\hatchcolor}
    \pgfusepath{stroke}
   }

\pgfdeclarepatternformonly[\hatchspread,\hatchthickness,\hatchshift,\hatchcolor]
   {custom horizontal lines}
   {\pgfqpoint{\dimexpr-2\hatchthickness}{\dimexpr-2\hatchthickness}}
   {\pgfqpoint{\dimexpr\hatchspread+2\hatchthickness}{\dimexpr\hatchspread+2\hatchthickness}}
   {\pgfqpoint{\dimexpr\hatchspread}{\dimexpr\hatchspread}}
   {
    \pgfsetlinewidth{\hatchthickness}
    \pgfpathmoveto{\pgfqpoint{0pt}{\dimexpr\hatchshift}}
    \pgfpathlineto{\pgfqpoint{\dimexpr\hatchspread+0.15pt}{\dimexpr\hatchshift}}
    \pgfsetstrokecolor{\hatchcolor}
    \pgfusepath{stroke}
   }
 
\tikzset{cross/.style={cross out,draw,minimum size=2*(#1-\pgflinewidth),inner sep=0pt, outer sep=0pt}}
         
\DeclareRobustCommand\uocirc{\tikz{\node[draw,thick,circle,inner sep=1.75,color=white] at (0,0) {};\node[draw,thick,circle,inner sep=1.,color=red] at (0,0) {};}\,}
\DeclareRobustCommand\ufcirc{\tikz{\node[draw,thick,circle,inner sep=1.75,color=white] at (0,0) {};\node[draw,thick,circle,inner sep=1.,fill,color=red] at (0,0) {};}\,}
\DeclareRobustCommand\ucross{\tikz{\node[draw,thick,circle,inner sep=1.75,color=white] at (0,0) {};\node[draw,thick,cross,inner sep=1.5,color=red] at (0,0) {};}\,}
\DeclareRobustCommand\urcross{\tikz{\node[draw,thick,circle,inner sep=1.75,color=white] at (0,0) {};\node[draw,thick,cross,rotate=45,inner sep=1.5,color=red] at (0,0) {};}\,}

\begin{document}\selectlanguage{english}
\begin{frontmatter}
\title{\begin{textblock*}{100pt}(395pt,-95pt)
\textnormal{\small \texttt{CERN-TH-2016-209}}
\end{textblock*}Worm Algorithm for the $\CPn{N-1}$ Model}

\author[1,3]{Tobias Rindlisbacher}
\ead{tobias.rindlisbacher@helsinki.fi}
\author[1,2]{Philippe de Forcrand}
\address[1]{ETH Z\"urich, Institute for Theoretical Physics, Wolfgang-Pauli-Str. 27, CH - 8093 Z\"urich, Switzerland}
\address[2]{CERN, Physics Department, Theory Division, CH-1211 Gen\`eve 23, Switzerland}
\address[3]{present address: University of Helsinki, Department of Physics, P.O. Box 64, FI-00014 University of Helsinki, Finland.}

\begin{abstract}
The $\CPn{N-1}$ model in 2D is an interesting toy model for 4D QCD as it possesses confinement, asymptotic freedom and a non-trivial vacuum structure. Due to the lower dimensionality and the absence of fermions, the computational cost for simulating 2D $\CPn{N-1}$ on the lattice is much lower than that for simulating 4D QCD. However, to our knowledge, no efficient algorithm for simulating the lattice $\CPn{N-1}$ model for $N>2$ has been tested so far, which also works at finite density. To this end we propose a new type of worm algorithm which is appropriate to simulate the lattice $\CPn{N-1}$ model in a dual, flux-variables based representation, in which the introduction of a chemical potential does not give rise to any complications. In addition to the usual worm moves where a defect is just moved from one lattice site to the next, our algorithm additionally allows for worm-type moves in the internal variable space of single links, which accelerates the Monte Carlo evolution. We use our algorithm to compare the two popular $\CPn{N-1}$ lattice actions and exhibit marked differences in their approach to the continuum limit.
\end{abstract}


\end{frontmatter}
\clearpage
\section{Introduction}\label{sec:introduction}
The classical $\CPn{N-1}$ model was introduced in 1978 in different contexts \cite{Eichenherr,Cremmer,Golo}. Shortly afterwards, also the two-dimensional quantum theory was discussed independently in \cite{Adda} and \cite{Witten1}, where it was shown that (among other interesting properties) the model possesses a non-trivial vacuum structure with stable instanton solutions and that it incorporates the phenomena of confinement and asymptotic freedom, which are also key features of four-dimensional Yang-Mills theories. The two-dimensional $\CPn{N-1}$ model is presumably the simplest model which possesses all of these properties and is therefore an ideal toy model to study their interrelations.\\
After the model was studied perturbatively \cite{Adda,Witten1} and by means of a $1/N$ expansion in the continuum \cite{Adda,Witten1,Vecchia2} and on the lattice \cite{Vecchia,Vecchia2} as well as by means of a strong coupling expansion \cite{Rabinovici}, a crosscheck of some predictions by direct Monte Carlo simulations was attempted in \cite{Campostrini}. Although a sophisticated over-heat bath algorithm was used in the latter work, it was found that the simulations suffer from an exponential critical slowing-down of topological modes. About the same time, a cluster algorithm for the $\CPn{N-1}$ model was proposed in \cite{Jansen} and tested for $N=4,5$, but in contrast to the Ising or to $\On{N}$ models, where cluster algorithms solve the critical slowing-down problem almost completely, it was found that for $\CPn{N-1}$, the cluster algorithm does not help in overcoming critical slowing-down, which would be necessary in order to perform further non-perturbative checks of the large $N$ and continuum predictions.\\
After the introduction of the worm algorithm \cite{Prokofev} in 2001 as an alternative to cluster algorithms in order to overcome critical slowing down, a reformulation of the lattice $\CPn{N-1}$ partition function in terms of $\order{N^2}$ dual, integer-valued flux-variables per link was proposed in \cite{Chandrasekharan}, which at first sight seemed to be suitable to be updated by a worm algorithm. However, it was later found in \cite{Vetter} that an ordinary worm algorithm gives rise to an ergodicity problem when applied naively to this dual flux-variable formulation, as soon as $N>2$. The reason for this problem has been identified only recently and one of the main purposes of this paper is to present and test our solution to it, which consists of an extension of the ordinary worm algorithm, so that the worm does not just move from site to site but also moves in internal space (in a particular way) at intermediate steps.\\
Meanwhile two alternative flux-variable representations \cite{Wolff,Gattringer1} for the $\CPn{N-1}$ partition function have been proposed: the version in \cite{Wolff} comes out with just a single flux variable per link and the system is updated with a so-called loop algorithm (more precisely \cite{Wolff} describes two formulations and algorithms: one for positive integer $N$ and one for real $N$). Unfortunately, just as the cluster algorithm, also this loop algorithm was found to perform no better than the over-heat bath algorithm. The other dual formulation \cite{Gattringer1}, which describes configurations in terms of $\order{2\,N}$ flux-variables per link, has not been implemented and tested so far, which is why we will also shortly discuss and test an algorithm to simulate this dual formulation and compare it to our new algorithm for the $\order{N^2}$ d.o.f. per link version.\\
The reason why these dual formulations are interesting is the following: they do not give rise to a sign problem when introducing one or more chemical potentials and allow therefore numerical studies of finite density effects \cite{Gattringer1}. But there is also another reason: as already mentioned, it has been found in \cite{Campostrini} that in standard simulations of the $\CPn{N-1}$ model, topological modes suffer from exponential critical slowing down, which is most likely caused by quickly decreasing tunneling rates between different topological sectors for increasing system size. Now, since in the above dual formulations, the original configuration variables are integrated out analytically in order to obtain the weights for the dual configurations (and this integration also covers all topological sectors), each dual configuration already contains contributions from all topological sectors, so that tunneling between them is no longer required and the main source of critical slowing-down should therefore be absent.\\ 

The paper is organized as follows: in the remainder of this section, we review in Sec.~\ref{ssec:model} the two standard continuum actions for the $\CPn{N-1}$ model\cite{Witten1,Vecchia2} and in Sec.~\ref{ssec:latticeformulation} their lattice analogues and the corresponding partition functions \cite{Campostrini}. In Sec.~\ref{sec:dualformulation} we discuss the flux-variable representation of the $\CPn{N-1}$ lattice partition functions as introduced in \cite{Chandrasekharan} and how chemical potentials can be incorporated in this representation. In Sec.~\ref{sec:simulationmethods} we give an explanation for the apparent ergodicity problem \cite{Vetter} that occurs when using a "naive" worm algorithm to simulate the $\CPn{N-1}$ model in the flux-variable representation from \cite{Chandrasekharan}, and introduce our "internal space sub-worm algorithm" which solves the problem. In Sec.~\ref{sec:results} we present the result of some tests of correctness of our algorithm and discuss its efficiency. Sec.~\ref{ssec:largenlimit} contains a discussion of the large $N$ vs. large $V$ behavior of one of the lattice formulations of the $\CPn{N-1}$ model, followed by a summary in Sec.~\ref{sec:summary} . In the Appendix, we discuss, following \cite{Wolff}, one possibility to incorporate a topological term into the flux-variable formalism, and demonstrate that a single flux-variable configuration indeed contains contributions from all topological sectors, by showing that the topological susceptibility in the strong coupling limit is extracted from a single flux-variable configuration only.

\subsection{The Model}\label{ssec:model}
The (Euclidean) $\CPn{N-1}$ model in $d$ dimensional continuous space can be defined as the $\Un{1}$ gauged non-linear $\SU{N}$ sigma model (i.e. a non-linear sigma model with local $\Un{1}$ and global $\SU{N}$ symmetry)\cite{Cremmer,Witten1,Vecchia2}:
\[
S_{A}\,=\,-\frac{1}{g}\int\idd{x}{d}\,\of{\cD_{\mu}z}^{\dagger}\cdot\of{\cD_{\mu}z}\ ,\label{eq:caction1}
\] 
where $z\in\mathbb{C}^{N}$ is an $N$-component complex scalar field subject to the constraint $z^{\dagger}\,z\,=\,1$, $\cD_{\mu}\,=\,\partial_{\mu}\,+\,\ii\,A_{\mu}$ is the covariant derivative with respect to an auxiliary $\Un{1}$ gauge field $A_{\mu}$ and $g$ is the corresponding coupling strength. Using the Euler-Lagrange equations for $A_{\mu}$, the classical solution is found to be
\[
A_{\mu}\,=\,\frac{\ii}{2}\sof{z^{\dagger}\cdot\ssof{\partial_{\mu} z}\,-\,\ssof{\partial_{\mu} z^{\dagger}}\cdot z}\ ,\label{eq:cgaugefield}
\]
which can be substituted back into \eqref{eq:caction1} to obtain
\[
S_{Q}\,=\,-\frac{1}{g}\int\idd{x}{d}\,\bof{\ssof{\partial_{\mu}z^{\dagger}}\cdot\ssof{\partial_{\mu}z}\,+\,\frac{1}{4}\sof{z^{\dagger}\cdot\ssof{\partial_{\mu} z}\,-\,\ssof{\partial_{\mu} z^{\dagger}}\cdot{z}}^2}\ ,\label{eq:caction2}
\]
which contains a term quartic in the fields $z$ \cite{Golo,Witten1,Vecchia2}.\\
There is no kinetic term for the $\Un{1}$ gauge field $A_{\mu}$, so classically, it is just a dummy field. In the quantum theory however, quantum fluctuations generate a kinetic term for $A_{\mu}$ \cite{Witten1} and turn it into a dynamic field.

\subsection{Lattice Formulation}\label{ssec:latticeformulation}
There are two commonly used lattice actions for the $\CPn{N-1}$ model \cite{Campostrini}, obtained by discretizing the actions $S_{A}$ from \eqref{eq:caction1} and $S_{Q}$ from \eqref{eq:caction2}, respectively. The discretization of \eqref{eq:caction1} yields
\[
S_{A}\,=\,-\beta\,\sum\limits_{x,\mu}\of{z^{\dagger}\of{x}U_{\mu}\of{x}z\of{x+\hat{\mu}}\,+\,z^{\dagger}\of{x}U^{\dagger}_{\mu}\of{x-\hat{\mu}}z\of{x-\hat{\mu}}\,-\,2}\ ,\label{eq:laction1}
\] 
where $x$ labels the different lattice sites, $\hat{\mu}$ is the vector that points from one lattice site to its nearest neighbor in $\mu$-direction and $U_{\mu}\of{x} \in \Un{1}$ is the parallel transporter with respect to the gauge field $A_{\mu}$ from site $x$ to site $x+\hat{\mu}$ along the corresponding link. The full partition function then reads
\begin{multline}
Z_{A}\,=\,\int\DD{z^{\dagger},z,U}\,\exp\bof{\beta\,\sum\limits_{x,\mu}\of{z^{\dagger}\of{x}U_{\mu}\of{x}z\of{x+\hat{\mu}}\,+\,z^{\dagger}\of{x}U^{\dagger}_{\mu}\of{x-\hat{\mu}}z\of{x-\hat{\mu}}}}\\
\,=\,\int\DD{z^{\dagger},z}\prod\limits_{x,\mu}\cof{I_{0}\sof{2\,\beta\,\sabs{z^{\dagger}\of{x}\cdot z\of{x+\hat{\mu}}}}\,\e^{-2\,\beta}}\ ,\label{eq:lpartf1}
\end{multline}
where $\DD{z^{\dagger},z,U}=\DD{z^{\dagger},z}\DD{U}$ with $\DD{z^{\dagger},z}=\prod\limits_{x}\delta\sof{\abs{z\of{x}}^2-1}\idd{\bar{z}\of{x}}{N}\,\idd{z\of{x}}{N}$ and $\DD{U}=\prod\limits_{x,\mu}\dd U_{\mu}\of{x}$, with $\dd U_{\mu}\of{x}$ being the $\Un{1}$ Haar measure for the link variable $U_{\mu}\of{x}=\e^{\ii\,\theta_{x,\mu}}$. After the second equality sign, we made use of the well known identity for modified Bessel functions of the first kind,
\[
\e^{a\,\ssof{t+\frac{1}{t}}}\,=\,\sum\limits_{k=-\infty}^{\infty} I_{k}\of{2\,a}\,t^{k}\ ,\label{eq:besselid1}
\]
as was already done in \cite{Vecchia}, from which, setting $t=\e^{\ii\,\theta}$, it follows that
\[
\int\limits_{0}^{2\,\pi}\frac{\dd{\theta}}{2\,\pi}\,\e^{2\,a\,\cos\of{\theta}}\,=\,I_{0}\of{2\,a}\ .
\]

By discretizing \eqref{eq:caction2}, we find the quartic action
\[
S_{Q}\,=\,-\beta\,\sum\limits_{x,\mu}\sabs{z^{\dagger}\of{x}\cdot z\of{x+\hat{\mu}}}^2\ ,\label{eq:laction2}
\]
and the corresponding partition function
\begin{multline}
Z_{Q}\,=\,\int\DD{z^{\dagger},z}\exp\bof{\beta\sum\limits_{x,\mu}\sabs{z^{\dagger}\of{x}\cdot z\of{x+\hat{\mu}}}^2}\\
\,=\,\int\DD{z^{\dagger},z}\prod\limits_{x,\mu}\exp\bof{\beta\,\sabs{z^{\dagger}\of{x}\cdot z\of{x+\hat{\mu}}}^2}\ .\label{eq:lpartf2}
\end{multline}
The two partition functions $Z_{A}$ and $Z_{Q}$ are obviously not identical but closely related: by expanding the link weights appearing on the last lines of \eqref{eq:lpartf1} and \eqref{eq:lpartf2} in power series, we find for the one in \eqref{eq:lpartf1},
\begin{multline}
I_{0}\sof{2\,\beta\,\sabs{z^{\dagger}\of{x}\cdot z\of{x+\hat{\mu}}}}\e^{-2\,\beta}\,=\,\sum\limits_{n_{x,\mu}=0}^{\infty}\frac{\beta^{2\,n_{x,\mu}}\,\sabs{z^{\dagger}\of{x}\cdot z\of{x+\hat{\mu}}}^{2\,n_{x,\mu}}}{\of{n_{x,\mu}!}^2}\,e^{-2\,\beta}\\
\,=\,\sum\limits_{n_{x,\mu}=0}^{\infty}\frac{\beta^{n_{x,\mu}}\,\sabs{z^{\dagger}\of{x}\cdot z\of{x+\hat{\mu}}}^{2\,n_{x,\mu}}}{n_{x,\mu}!}\,w_{n_{x,\mu}}\of{\beta}\ ,\label{eq:lu1linkweight}
\end{multline}
with
\[
w_{n}\of{\beta}=\frac{\beta^{n}\e^{-2\,\beta}}{n!}\ ,\label{eq:addu1weight}
\]
and for the one in \eqref{eq:lpartf2} respectively
\[
\exp\of{\beta\,\sabs{z^{\dagger}\of{x}\cdot z\of{x+\hat{\mu}}}^{2}}\,=\,\sum\limits_{n_{x,\mu}=0}^{\infty}\frac{\beta^{n_{x,\mu}}\,\sabs{z^{\dagger}\of{x}\cdot z\of{x+\hat{\mu}}}^{2\,n_{x,\mu}}}{n_{x,\mu}!}\ .\label{eq:lquarticlinkweight}
\]
So, \eqref{eq:lpartf1} differs from \eqref{eq:lpartf2} only by the presence of an extra weight factor \eqref{eq:addu1weight} for each term of the power-series expansion of the link weights.\\

As expected from universality, and as argued in \cite{Vecchia2} on the basis of a large $N$ study: in the continuum limit $\of{\beta\rightarrow \infty}$, in $\of{1+1}$ dimensions, the two partition functions \eqref{eq:lpartf1} and \eqref{eq:lpartf2} seem to give rise to the same physics. In the strong coupling regime however, the two lattice formulations can behave differently: the system described by \eqref{eq:lpartf2} (still in $\of{1+1}$ dimensions) develops a first-order transition in the limit $\of{N\rightarrow \infty}$, which separates the strong coupling phase $\of{\beta/N \lesssim 1}$ from the weak coupling one $\of{\beta/N \gtrsim 1}$\cite{Vecchia2}. For the partition function \eqref{eq:lpartf1} on the other hand, the transition is absent in the limit $\of{N\rightarrow \infty}$ and only the weak coupling phase survives in the infinite $N$ limit \cite{Vecchia2}. We will verify these statements in our numerical simulations in Sec.~\ref{sec:results} .

\section{Dual Formulation}\label{sec:dualformulation}
After having introduced the two most-widely used lattice formulations of the $\CPn{N-1}$ model, \eqref{eq:lpartf1} and \eqref{eq:lpartf2}, we continue in Sec.~\ref{ssec:fluxreppartf} by giving a quick review on how the flux-variable representation of \eqref{eq:lpartf2}, as introduced in \cite{Chandrasekharan}, can be obtained, and how this representation can also be used for the partition function \eqref{eq:lpartf1}. In Sec.~\ref{ssec:conservedcurrnts} we briefly discuss how chemical potentials can be incorporated into the flux-variable representations of \eqref{eq:lpartf1} and \eqref{eq:lpartf2} without giving rise to a sign problem. Finally, in Sec.~\ref{ssec:observables} we define some observables that will be used later on.     

\subsection{Flux-Variable Representation of the Partition Function}\label{ssec:fluxreppartf}
For simplicity, we describe the dualization procedure on the example of the quartic action version \eqref{eq:lpartf2} of the lattice $\CPn{N-1}$ partition function which was also the one that was used in \cite{Chandrasekharan}. We follow the derivations in \cite{Chandrasekharan,Vetter} and start by explicitly writing out all sums in the exponential of \eqref{eq:lpartf2} :
\[
Z_{Q}\,=\,\int\DD{z^{\dagger},z}\,\exp\bof{\beta\,\sum\limits_{x}\sum\limits_{\mu=1}^{d}\sum\limits_{a,b=1}^{N}\sof{\bar{z}_{a}\of{x}z_{b}\of{x}}\sof{\bar{z}_{b}\of{x+\hat{\mu}}z_{a}\of{x+\hat{\mu}}}}\ .\label{eq:partf0}
\]
Now we write the exponential of the summed terms in \eqref{eq:partf0} as the product of exponentials of the individual terms and then use the power series representation for each of these exponentials to find:
\[
Z_{Q}\,=\,\int\DD{z^{\dagger},z}\,\prod\limits_{x}\prod\limits_{\mu=1}^{d}\prod\limits_{a,b=1}^{N}\,\sum\limits_{n_{x,\mu}^{a\,b}=0}^{\infty}\,\bcof{\frac{\beta^{n_{x,\mu}^{a\,b}}}{n_{x,\mu}^{a\,b}!}\of{\sof{\bar{z}_{a}\of{x}z_{b}\of{x}}\sof{\bar{z}_{b}\of{x+\hat{\mu}}z_{a}\of{x+\hat{\mu}}}}^{n_{x,\mu}^{a\,b}}}\ ,\label{eq:partf1}
\]
which can be written as
\[
Z_{Q}\,=\,\mathcal{N}\cdot\sum\limits_{\cof{n}}\bof{\prod\limits_{x}\prod\limits_{\mu=1}^{d}\prod\limits_{a,b=1}^{N}\frac{\beta^{n_{x,\mu}^{a\,b}}}{n_{x,\mu}^{a\,b}!}}\bof{\prod\limits_{x}\,F\of{q\of{x},p\of{x}}}\ ,\label{eq:cpnpartf2pre}
\]
with $\mathcal{N}$ an unimportant normalization constant and $F\of{q,p}$ being the result of the integration over the $z_{a}\of{x}=r^{a}_{x}\e^{\ii \phi_{x}^{a}}$ on each site, i.e.
\[
F\of{q,p}\,=\,\int\idd{\bar{z}}{N}\,\idd{z}{N}\,\delta\sof{\abs{z}^2-1}\prod\limits_{a=1}^{N}\of{\bar{z}_{a}}^{q_{a}}\of{z_{a}}^{p_{a}}\,=\,\frac{\pi^{N}\prod\limits_{a=1}^{N}\delta\of{q_{a}-p_{a}}\,q_{a}!}{\sof{N-1+\sum\limits_{a=1}^{N}\,q_{a}}!}\ ,\label{eq:fweight}
\]
where 
\[
q_{a}\of{x}\,=\,\sum\limits_{\mu=1}^{d}\sum\limits_{b=1}^{N}\sof{n_{x,\mu}^{a\,b}+n_{x-\hat{\mu},\mu}^{b\,a}}\quad,\qquad p_{a}\of{x}\,=\,\sum\limits_{\mu=1}^{d}\sum\limits_{b=1}^{N}\sof{n_{x,\mu}^{b\,a}+n_{x-\hat{\mu},\mu}^{a\,b}}\ .\label{eq:qpvars}
\]
The path integral has now turned into an infinite sum of terms which are labeled by different values of the $d\cdot V\cdot N^2$ non-negative, integer-valued $n_{x,\mu}^{a\,b}$ variables. The $n_{x,\mu}^{a\,b}$ are called \emph{flux-variables} as they live on the links of the lattice and, according to \eqref{eq:partf1}, can be interpreted as enumerating the number of hoppings of pairs (mesons) $\bar{z}_{a}\,z_{b}$ between the sites $x$ and $x+\hat{\mu}$.\\ 

The partition function \eqref{eq:partf1} is the one that was used in \cite{Vetter}. In the given form, all the $n_{x,\mu}^{a\,b}$ variables in \eqref{eq:partf1} are subject to the discrete delta-function\footnote{What we call a "discrete delta-function" is in fact just a Kronecker delta, i.e. $\delta\of{x}\,:=\,\delta_{0,x}$ .} constraints in \eqref{eq:fweight} (via \eqref{eq:qpvars}): each of the $n_{x,\mu}^{a\,b}$ appears in two such constraints on each site that it touches. This, in combination with the large number of $d\cdot V\cdot N^2$ distinct $n_{x,\mu}^{a\,b}$ variables, makes it difficult to recognize the true structure of the constraints and correspondingly, what freedom is left in the choice of values for the $n_{x,\mu}^{a\,b}$ .\\

To reduce the number of constrained variables, we decompose the $N\times N$ matrices $n_{x,\mu}^{a\,b}$ into symmetric and anti-symmetric pieces, parametrized by new variables $k_{x,\mu}^{a\,b}\in\mathbb{Z}$ and $l_{x,\mu}^{a\,b}\in\mathbb{N}_{0}$, such that:
\[
k_{x,\mu}^{a\,b}\,=\,n_{x,\mu}^{a\,b}-n_{x,\mu}^{b\,a}\quad,\qquad 2\,l_{x,\mu}^{a\,b}\,+\,\sabs{k_{x,\mu}^{a\,b}}\,=\,n_{x,\mu}^{a\,b}+n_{x,\mu}^{b\,a},
\]
i.e.
\[
n_{x,\mu}^{a\,b}\,=\,{\tfrac{1}{2}}\sof{\sabs{k_{x,\mu}^{a\,b}}+k_{x,\mu}^{a\,b}}+l_{x,\mu}^{a\,b}\ .\label{eq:nfromkl}
\]
After carrying out the substitutions, we find:
\begin{multline}
Z_{Q}\,=\,\sum\limits_{\cof{k,l}}\prod\limits_{x}\bcof{\bof{\prod\limits_{\mu=1}^{d}\prod\limits_{a,b=1}^{N}\frac{\beta^{\frac{1}{2}\ssof{\ssabs{k_{x,\mu}^{a\,b}}+k_{x,\mu}^{a\,b}}+l_{x,\mu}^{a\,b}}}{\of{\frac{1}{2}\of{\ssabs{k_{x,\mu}^{a\,b}}+k_{x,\mu}^{a\,b}}+l_{x,\mu}^{a\,b}}!}}\\
\frac{\prod\limits_{a}^{N}\bof{\delta\sof{\sum\limits_{\mu=1}^{d}\sum\limits_{b=1}^{N}\sof{k_{x,\mu}^{a\,b}-k_{x-\hat{\mu},\mu}^{a\,b}}}\sof{\sum\limits_{\mu=1}^{d}\sum\limits_{b=1}^{N}\sof{\frac{1}{2}\ssof{\ssabs{k_{x,\mu}^{a\,b}}+\ssabs{k_{x-\hat{\mu},\mu}^{a\,b}}}+l_{x,\mu}^{a\,b}+l_{x-\hat{\mu},\mu}^{a\,b}}}!}}{\sof{N-1+\sum\limits_{\mu=1}^{d}\sum\limits_{c,b=1}^{N}\sof{\frac{1}{2}\ssof{\ssabs{k_{x,\mu}^{c\,b}}+\ssabs{k_{x-\hat{\mu},\mu}^{c\,b}}}+l_{x,\mu}^{c\,b}+l_{x-\hat{\mu},\mu}^{c\,b}}}!}}\ ,\label{eq:cpnpartf2a}
\end{multline}
where now only the $N\of{N-1}/2$ independent components of the anti-symmetric $k_{x,\mu}^{a\,b}$ are still subject to the delta-function constraints while the $N\of{N+1}/2$ independent components of the symmetric $l_{x,\mu}^{a\,b}$ are free. By using the symmetry properties of the $k_{x,\mu}^{a\,b}$ and $l_{x,\mu}^{a\,b}$, \eqref{eq:cpnpartf2a} can also be written as:
\begin{multline}
Z_{Q}\,=\,\sum\limits_{\cof{k,l}}\prod\limits_{x}\bcof{\bof{\prod\limits_{\mu=1}^{d}\prod\limits_{a=1}^{N}\frac{\beta^{l_{x,\mu}^{a\,a}}}{l_{x,\mu}^{a\,a}!}\prod\limits_{b=a+1}^{N}\frac{\beta^{\ssabs{k_{x,\mu}^{a\,b}}+2\,l_{x,\mu}^{a\,b}}}{\of{\ssabs{k_{x,\mu}^{a\,b}}+l_{x,\mu}^{a\,b}}!\,l_{x,\mu}^{a\,b}!}}\\
\frac{\prod\limits_{a=1}^{N}\bof{\delta\sof{\sum\limits_{\mu=1}^{d}\sum\limits_{b=1}^{N}\sof{k_{x,\mu}^{a\,b}-k_{x-\hat{\mu},\mu}^{a\,b}}}\sof{\sum\limits_{\mu=1}^{d}\sum\limits_{b=1}^{N}\sof{\frac{1}{2}\ssof{\ssabs{k_{x,\mu}^{a\,b}}+\ssabs{k_{x-\hat{\mu},\mu}^{a\,b}}}+l_{x,\mu}^{a\,b}+l_{x-\hat{\mu},\mu}^{a\,b}}}!}}{\sof{N-1+\sum\limits_{\mu=1}^{d}\sum\limits_{c=1}^{N}\sof{l_{x,\mu}^{c\,c}+l_{x-\hat{\mu},\mu}^{c\,c}\,+\,\sum\limits_{b=c+1}^{N}\sof{\ssabs{k_{x,\mu}^{c\,b}}+\ssabs{k_{x-\hat{\mu},\mu}^{c\,b}}+2\ssof{l_{x,\mu}^{c\,b}+l_{x-\hat{\mu},\mu}^{c\,b}}}}}!}}\\
=\,\sum\limits_{\cof{k,l}}\prod\limits_{x}\bcof{\bof{\prod\limits_{\mu=1}^{d}\prod\limits_{a=1}^{N}\frac{\beta^{l_{x,\mu}^{a\,a}}}{l_{x,\mu}^{a\,a}!}\prod\limits_{b=a+1}^{N}\frac{\beta^{\ssabs{k_{x,\mu}^{a\,b}}+2\,l_{x,\mu}^{a\,b}}}{\of{\ssabs{k_{x,\mu}^{a\,b}}+l_{x,\mu}^{a\,b}}!\,l_{x,\mu}^{a\,b}!}}\hspace{130pt}\\
\bof{\prod\limits_{a=1}^{N}\delta\sof{\sum\limits_{\mu=1}^{d}\sfof{\sum\limits_{b=a+1}^{N}\sof{k_{x,\mu}^{a\,b}-k_{x-\hat{\mu},\mu}^{a\,b}}-\sum\limits_{b=1}^{a-1}\sof{k_{x,\mu}^{b\,a}-k_{x-\hat{\mu},\mu}^{b\,a}}}}}\\
\frac{\prod\limits_{a=1}^{N}\sof{\sum\limits_{\mu=1}^{d}\sum\limits_{b=1}^{N}\sof{\frac{1}{2}\ssof{\ssabs{k_{x,\mu}^{a\,b}}+\ssabs{k_{x-\hat{\mu},\mu}^{a\,b}}}+l_{x,\mu}^{a\,b}+l_{x-\hat{\mu},\mu}^{a\,b}}}!}{\sof{N-1+\sum\limits_{\mu=1}^{d}\sum\limits_{c=1}^{N}\sof{l_{x,\mu}^{c\,c}+l_{x-\hat{\mu},\mu}^{c\,c}\,+\,\sum\limits_{b=c+1}^{N}\sof{\ssabs{k_{x,\mu}^{c\,b}}+\ssabs{k_{x-\hat{\mu},\mu}^{c\,b}}+2\ssof{l_{x,\mu}^{c\,b}+l_{x-\hat{\mu},\mu}^{c\,b}}}}}!}}\ ,\label{eq:cpnpartf2}
\end{multline}
in which form the dependency of the weights on the flux variables, as well as the constraints that are imposed on the $k$ variables, are most easily recognized (the constraints will be discussed further in Sec. \ref{ssec:constraints}). Note that \eqref{eq:cpnpartf2} possesses global $\mathbb{Z}_{2}$ and $\mathbb{Z}_{N}$ symmetries, as all weights are both, invariant under a collective sign-flip $\of{k\rightarrow-k}$ of the $k$-variables, and under collective cyclic shifts of the internal space indices $\of{a\rightarrow a+1}$ of all the $k$ and $l$-variables.\\

To obtain also $Z_{A}$ from \eqref{eq:laction1} in terms of $k$ and $l$-variables, we note that using the multinomial expansion,
\begin{multline}
\frac{\beta^{2\,n_{x,\mu}}}{{n_{x,\mu}!}^{2}}\sabs{z^{\dagger}\of{x}\cdot z\of{x+\hat{\mu}}}^{2\,n_{x,\mu}}\,=\,\frac{\beta^{2\,n_{x,\mu}}}{{n_{x,\mu}!}^{2}}\bof{\sum\limits_{a,b=1}^{N}\underbrace{\sof{\bar{z}_{a}\of{x}z_{b}\of{x}}\sof{\bar{z}_{b}\of{x+\hat{\mu}}z_{a}\of{x+\hat{\mu}}}}_{q_{\mu}^{a\,b}\of{x}}}^{n_{x,\mu}}\\
\,=\,\frac{\beta^{2\,n_{x,\mu}}}{{n_{x,\mu}!}}\,\sum\limits_{\cof{n_{x,\mu}^{a\,b}\big|\sum_{a,b=1}^{N}n_{x,\mu}^{a\,b}=n_{x,\mu}}}\,\prod\limits_{a,b,=1}^{N}\frac{\sof{q_{\mu}^{a\,b}\of{x}}^{n_{x,\mu}^{a\,b}}}{n_{x,\mu}^{a\,b}!}\ ,\label{eq:multinomialexp}
\end{multline} 
and therefore, since according to \eqref{eq:nfromkl}, we have that $n_{x,\mu}^{a\,b}\,=\,\frac{1}{2}\sof{\sabs{k_{x,\mu}^{a\,b}}+k_{x,\mu}^{a\,b}}+l_{x,\mu}^{a\,b}$, a configuration of $k$ and $l$-variables represents on each link the contribution of a monomial that is part of a multinomial \eqref{eq:multinomialexp} with 
\[
n_{x,\mu}=\sum\limits_{a,b=1}^{N}\of{{\tfrac{1}{2}}\sabs{k_{x,\mu}^{a\,b}}+l_{x,\mu}^{a\,b}}\,=\,\sum\limits_{a=1}^{N}\bof{l_{x,\mu}^{a\,a}\,+\,\sum\limits_{b=a+1}^{N}\of{\sabs{k_{x,\mu}^{a\,b}}+2\,l_{x,\mu}^{a\,b}}}\ ,\label{eq:nkmap}
\]
and picks up the corresponding link weight $\,w_{n_{x,\mu}}\of{\beta}$, given by \eqref{eq:addu1weight} when changing from the quartic action \eqref{eq:caction2} to the auxiliary $\Un{1}$ action \eqref{eq:caction1}. We therefore find for $Z_{A}$ :
\begin{multline}
Z_{A}\,=\,\sum\limits_{\cof{k,l}}\prod\limits_{x}\bcof{\bof{\prod\limits_{\mu=1}^{d}\frac{\e^{-2\,\beta}}{\sof{\sum\limits_{a,b=1}^{N}\of{\frac{1}{2}\abs{k_{x,\mu}^{a\,b}}+l_{x,\mu}^{a\,b}}}!}\bof{\prod\limits_{a,b=1}^{N}\frac{\beta^{\ssof{\ssabs{k_{x,\mu}^{a\,b}}+k_{x,\mu}^{a\,b}}+2\,l_{x,\mu}^{a\,b}}}{\of{\frac{1}{2}\of{\ssabs{k_{x,\mu}^{a\,b}}+k_{x,\mu}^{a\,b}}+l_{x,\mu}^{a\,b}}!}}}\\
\frac{\prod\limits_{a}^{N}\delta\sof{\sum\limits_{\mu=1}^{d}\sum\limits_{b=1}^{N}\sof{k_{x,\mu}^{a\,b}-k_{x-\hat{\mu},\mu}^{a\,b}}}\sof{\sum\limits_{\mu=1}^{d}\sum\limits_{b=1}^{N}\sof{\frac{1}{2}\ssof{\ssabs{k_{x,\mu}^{a\,b}}+\ssabs{k_{x-\hat{\mu},\mu}^{a\,b}}}+l_{x,\mu}^{a\,b}+l_{x-\hat{\mu},\mu}^{a\,b}}}!}{\sof{N-1+\sum\limits_{\mu=1}^{d}\sum\limits_{c,b=1}^{N}\sof{\frac{1}{2}\ssof{\ssabs{k_{x,\mu}^{c\,b}}+\ssabs{k_{x-\hat{\mu},\mu}^{c\,b}}}+l_{x,\mu}^{c\,b}+l_{x-\hat{\mu},\mu}^{c\,b}}}!}}\ ,\label{eq:cpnpartf1}
\end{multline}
which is also invariant under a collective sign-flip $\of{k\rightarrow -k}$ and under collective, cyclic shifts of the internal space indices $\of{a\rightarrow a+1}$ of all the $k$ and $l$-variables.\\

Before we continue with the discussion of the properties of the flux-representation of $Z_{Q}$ and $Z_{A}$, we introduce the following vocabulary or naming for the different indices of the flux-variables $k_{x,\mu}^{a\,b}$ and $l_{x,\mu}^{a\,b}$: as usual, $x$ refers to a site on the lattice and $\mu$ to a direction. The indices $a,\,b$, we call \emph{internal space indices}, but it should be kept in mind that "internal space" does not mean "flavor space": $a,\,b$ are not flavor indices, i.e. the matrix $k_{x,\mu}$ does not transform like $U\,k_{x,\mu}\,U^{\dagger}$ under a global flavor symmetry transformation $U$, as should be clear from the fact that $k_{x,\mu}$ is always a matrix of integers which cannot undergo smooth changes. Flavor space has already been integrated out completely in the flux representations \eqref{eq:cpnpartf2}, \eqref{eq:cpnpartf1} of the two partition functions $Z_{Q}$ and $Z_{A}$, and the fact that a particular combination of values for the $k^{a\,b}_{x,\mu}$ variables gives rise to a non-vanishing weight for the corresponding configuration in $Z_{A}$ or $Z_{Q}$, just means that the product of all the $z_{a}$ and $\bar{z}_{b}$ in the lattice, with the multiplicities that are given by the values of all the $k^{a\,b}_{x,\mu}$ matrices, contains a "singlet" which survives the integration over the flavors. 

\subsection{Conserved Currents and Chemical Potentials}\label{ssec:conservedcurrnts}
The classical $\CPn{N-1}$ model possesses $\of{N-1}$ conserved currents, generated by the $\of{N-1}$ diagonal generalized Gell-Mann matrices,
\[
\tilde\lambda_{i,a\,b}\,=\,\sum\limits_{j=1}^{i}\,\delta_{j,a}\,\delta_{j,b}\sqrt{\frac{2}{i\of{i+1}}}\,-\,\delta_{i+1,a}\,\delta_{i+1,b}\sqrt{\frac{2\,i}{\of{i+1}}}\quad,\quad i\in\cof{1,\ldots,N-1}\ ,
\]
and a chemical potential can be coupled to each of the corresponding conserved charges, as has already been shown in \cite{Gattringer1}. To see how these chemical potentials enter in our flux-variable representation (which is based on $\order{N^2}$ variables per link), we will first give an alternative derivation of the flux-variable formulation given in \cite{Gattringer1} (which is based on $\order{2\,N}$ variables per link), from which we can then deduce how also our $k_{x,\nu}^{a\,b}$ variables should couple to these chemical potentials.\\

The starting point is again the auxiliary $\Un{1}$ lattice partition function \eqref{eq:lpartf1}, to which we add in the usual way the coupling to the chemical potentials:
\begin{multline}
Z_{A}\,=\,\int\DD{z^{\dagger},z,U}\,\exp\bof{\beta\,\sum\limits_{x}\sum\limits_{\nu=1}^{d}\sof{z^{\dagger}\of{x}\e^{\mu_{i}\tilde{\lambda}_{i}\delta_{\nu,d}}U_{\nu}\of{x}z\of{x+\hat{\nu}}\\
\qquad\qquad+\,z^{\dagger}\of{x}\e^{-\mu_{i}\tilde{\lambda}_{i}\delta_{\nu,d}}U^{\dagger}_{\nu}\of{x-\hat{\nu}}z\of{x-\hat{\nu}}}}\\
\,=\,\int\DD{r,\phi,\theta}\prod\limits_{x,\nu}\e^{-2\,\beta}\prod\limits_{a=1}^{N}\bcof{\sum\limits_{k_{x,\nu}^{a}=-\infty}^{\infty}\e^{\sof{\ii\theta_{x,\nu}+\tilde{\mu}_{a}\delta_{\nu,d}+\ii\ssof{\phi^{a}_{x+\hat{\nu}}-\phi^{a}_{x}}}k_{x,\nu}^{a}}\,I_{k_{x,\nu}^{a}}\of{2\,\beta\,r^{a}_{x}r^{a}_{x+\hat{\nu}}}}\ ,\label{eq:lpartf3}
\end{multline}
where, after the second equality sign, we have again used the identity \eqref{eq:besselid1}, as well as that $z_{a}\of{x}=r^{a}_{x}\,e^{\ii\phi^{a}_{x}}$ and $U_{\nu}\of{x}=\e^{\ii\theta_{x,\nu}}$, with corresponding measure
\[
\DD{r,\phi,\theta}=\prod\limits_{x}\bof{\idd{\phi_{x}}{N}\,\idd{r_{x}}{N}\,\delta\sof{1-\sum\limits_{a=1}^{N}\ssof{r_{x}^{a}}^{2}}\,\prod\limits_{\nu=1}^{d}\idd{\theta_{x,\nu}}{}}\,,\ r_{x}^{a}\in\roint{0,\infty}\,,\,\phi_{x}^{a},\theta_{x,\nu}\in\roint{0,2\,\pi}\ .
\]
Furthermore we introduced the $N$ "single flavor" chemical potentials $\tilde{\mu}_{a}=\sum_{i}\mu_{i}\tilde{\lambda}_{i,a\,a}$. By using that\footnote{We interpret factorials in terms of Gamma functions, i.e. $n!=\Gamma\of{1+n}$, so that $\frac{1}{n!}=\frac{1}{\Gamma\of{1+n}}=0$ for integers $n<0$.}
\[
I_{k}\of{2\,x}\,=\,\sum\limits_{l=0}^{\infty}\frac{x^{k+2\,l}}{\of{k+l}!\,l!}\,=\,\sum\limits_{l=0}^{\infty}\frac{x^{\abs{k}+2\,l}}{\of{\abs{k}+l}!\,l!}\ ,
\]
and by integrating out the angles $\theta_{x,\nu}$ and $\phi_{x}^{a}$, the partition function \eqref{eq:lpartf3} becomes (we add a tilde to $Z_{A}$ in order to distinguish it from \eqref{eq:cpnpartf1})
\begin{multline}
\tilde{Z}_{A}\,=\,\sum\limits_{\cof{k,\,l}}\bcof{\prod\limits_{x}\bof{\prod\limits_{\nu}\,\e^{-2\,\beta}\delta\sof{\sum\limits_{a}k_{x,\nu}^{a}}\prod\limits_{a}\e^{\tilde{\mu}_{a}\,k_{x,\nu}^{a}\delta_{\nu,d}}\,\frac{\beta^{\ssabs{k_{x,\nu}^{a}}+2\,l_{x,\nu}^{a}}}{\ssof{\ssabs{k_{x,\nu}^{a}}+l_{x,\nu}^{a}}!\,l_{x,\nu}^{a}!}}\\
\cdot\int\idd{r_{x}}{N}\,\delta\sof{1-\sum\limits_{a}\of{r^{a}_{x}}^{2}}\,\prod\limits_{a}\,\delta\sof{\sum\limits_{\nu}\sof{k_{x,\nu}^{a}-k_{x-\hat{\nu}}^{a}}}\of{r_{x}^{a}}^{1+\sum\limits_{\nu}\of{\ssabs{k_{x,\nu}^{a}}+\ssabs{k_{x-\hat{\nu},\nu}^{a}}+2\of{l_{x,\nu}^{a}+l_{x-\hat{\nu},\nu}^{a}}}}}\\
=\,\sum\limits_{\cof{k,\,l}}\bcof{\prod\limits_{x}\bof{\prod\limits_{\nu}\,\e^{-2\,\beta}\delta\sof{\sum\limits_{a}k_{x,\nu}^{a}}\prod\limits_{a}\e^{\tilde{\mu}_{a}\,k_{x,\nu}^{a}\delta_{\nu,d}}\,\frac{\beta^{\ssabs{k_{x,\nu}^{a}}+2\,l_{x,\nu}^{a}}}{\ssof{\ssabs{k_{x,\nu}^{a}}+l_{x,\nu}^{a}}!\,l_{x,\nu}^{a}!}}\\
\cdot\frac{\prod\limits_{a}\,\delta\sof{\sum\limits_{\nu}\sof{k_{x,\nu}^{a}-k_{x-\hat{\nu},\nu}^{a}}}\sof{\sum\limits_{\nu}\sof{\frac{1}{2}\sof{\ssabs{k_{x,\nu}^{a}}+\ssabs{k_{x-\hat{\nu},\nu}^{a}}}+l_{x,\nu}^{a}+l_{x-\hat{\nu},\nu}^{a}}}!}{\sof{N-1+\sum\limits_{a}\sum\limits_{\nu}\sof{\frac{1}{2}\sof{\ssabs{k_{x,\nu}^{a}}+\ssabs{k_{x-\hat{\nu},\nu}^{a}}}+l_{x,\nu}^{a}+l_{x-\hat{\nu},\nu}^{a}}}!}}\ ,\label{eq:cpnpartf3}
\end{multline}
which is our desired equation.\\

As in the previous section, we can use the relation between \eqref{eq:lu1linkweight} and \eqref{eq:lquarticlinkweight} to write down an expression for $\tilde{Z}_{Q}$ (a version of $Z_{Q}$ with $2\,N$ instead of $N^2$ degrees of freedom per link): we simply have to divide each link-weight in \eqref{eq:cpnpartf3} by
\[
w_{n_{x,\nu}}\,=\,\frac{\beta^{n_{x,\nu}}\,\e^{-2\,\beta}}{n_{x,\nu}!}\ ,\label{eq:addu1weighttilted}
\]
where this time, in terms of the $k_{x,\nu}^{a}$ and $l_{x,\nu}^{a}$ we have
\[
n_{x,\nu}=\sum_{a}\sof{{\textstyle\frac{1}{2}}\ssabs{k_{x,\nu}^{a}}\,+\,l_{x,\nu}^{a}}\ .
\]
The resulting expression for $\tilde{Z}_{Q}$ then reads:
\begin{multline}
\tilde{Z}_{Q}=\sum\limits_{\cof{k,\,l}}\bcof{\prod\limits_{x}\bof{\prod\limits_{\nu}\delta\sof{\sum\limits_{a}k_{x,\nu}^{a}}\sof{\sum\limits_{a}\sof{{\textstyle\frac{1}{2}}\ssabs{k_{x,\nu}^{a}}+l_{x,\nu}^{a}}}!\prod\limits_{a}\e^{\tilde{\mu}_{a}\,k_{x,\nu}^{a}\delta_{\nu,d}}\frac{\beta^{\frac{1}{2}\ssabs{k_{x,\nu}^{a}}+l_{x,\nu}^{a}}}{\ssof{\ssabs{k_{x,\nu}^{a}}+l_{x,\nu}^{a}}!\,l_{x,\nu}^{a}!}}\\
\cdot\frac{\prod\limits_{a}\,\delta\sof{\sum\limits_{\nu}\sof{k_{x,\nu}^{a}-k_{x-\hat{\nu},\nu}^{a}}}\sof{\sum\limits_{\nu}\sof{\frac{1}{2}\sof{\ssabs{k_{x,\nu}^{a}}+\ssabs{k_{x-\hat{\nu},\nu}^{a}}}+l_{x,\nu}^{a}+l_{x-\hat{\nu},\nu}^{a}}}!}{\sof{N-1+\sum\limits_{a}\sum\limits_{\nu}\sof{\frac{1}{2}\sof{\ssabs{k_{x,\nu}^{a}}+\ssabs{k_{x-\hat{\nu},\nu}^{a}}}+l_{x,\nu}^{a}+l_{x-\hat{\nu},\nu}^{a}}}!}}\ .\label{eq:cpnpartf4}
\end{multline}
   
In contrast to \eqref{eq:cpnpartf2a} and \eqref{eq:cpnpartf1} which contain only one type of delta-function constraints, namely for each site a product
\[
\prod\limits_{a=1}^{N}\,\delta\bof{\sum\limits_{\nu=1}^{d}\sum\limits_{b=1}^{N}\,\sof{k_{x,\nu}^{a\,b}-k_{x-\hat{\nu},\nu}^{a\,b}}}\ ,\label{eq:nsqconstr}
\]
the new partition functions \eqref{eq:cpnpartf3} and \eqref{eq:cpnpartf4} contain two different types of constraints: there are still $N$ \emph{on-site} constraints of the form
\[
\prod\limits_{a=1}^{N}\,\delta\bof{\sum\limits_{\nu=1}^{d}\,\sof{k_{x,\nu}^{a}-k_{x-\hat{\nu},\nu}^{a}}}\ ,\label{eq:twonconstrsite}
\]
which implement conservation laws for the fluxes associated with the $N$ different $k^{a}$-variables, $a\in\cof{1,\ldots,N}$, but there are now also the \emph{on-link} constraints
\[
\delta\bof{\sum\limits_{a=1}^{N}\,k_{x,\nu}^{a}}\ ,\label{eq:twonconstrlink}
\]
which require the $k^{a}$ variables that live on the same link, to add up to zero, showing that only $\of{N-1}$ of the $N$ conserved fluxes are independent. The constraints therefore implement precisely the conservation of the $\of{N-1}$ classically conserved currents of the $\CPn{N-1}$ model and the corresponding conserved charges are precisely the ones to which the $\tilde{\mu}_{a}$ couple.\\

By comparing \eqref{eq:nsqconstr} with \eqref{eq:twonconstrsite} and \eqref{eq:twonconstrlink}, we notice that we can relate the flux-variables $k_{x,\nu}^{a\,b}$ and $k_{x,\nu}^{a}$ from \eqref{eq:cpnpartf2a},\eqref{eq:cpnpartf1} and \eqref{eq:cpnpartf3},\eqref{eq:cpnpartf4} respectively, by setting $k_{x,\nu}^{a}\,=\,\sum\limits_{b}\,k_{x,\nu}^{a\,b}$. The on-link constraints \eqref{eq:twonconstrlink} are then automatically satisfied due to the anti-symmetry of the $k_{x,\nu}^{a\,b}$ in the internal-space indices $\of{a,b}$, which is the reason why this constraint is absent in the formulations \eqref{eq:cpnpartf2a} and \eqref{eq:cpnpartf1}. The $k_{x,\nu}^{a\,b}$ in \eqref{eq:cpnpartf1} should therefore couple to $\tilde{\mu}_{a}$ through an extra weight factor
\[
\prod\limits_{a}\e^{\tilde{\mu}_{a}\sum\limits_{b}k_{x,d}^{a\,b}}
\]
for temporal links. This also applies to the partition function $Z_{Q}$ from \eqref{eq:cpnpartf2a}. So, to conclude: all partition functions, \eqref{eq:cpnpartf2a}, \eqref{eq:cpnpartf1} as well as \eqref{eq:cpnpartf3} and \eqref{eq:cpnpartf4}, remain sign-problem free when introducing chemical potentials that couple to the classically conserved charges.\\

Note that for $\tilde{\mu}_{a}=0$, the partition functions \eqref{eq:cpnpartf3} and \eqref{eq:cpnpartf4} are also invariant under a collective flip of the signs of all $k$-variables or a collective, cyclic rotation of the internal space indices, just as \eqref{eq:cpnpartf2a} and \eqref{eq:cpnpartf1}. But a non-zero value of one of the chemical potentials in general breaks this global $\mathbb{Z}_{2}$ and $\mathbb{Z}_{N}$ symmetries explicitly in all four versions.\\ 

\subsection{Observables}\label{ssec:observables}
When working with a dual, flux-variable representation of a partition function, the definition of meaningful observables that can be measured during a Monte Carlo simulation is much less intuitive than when working with the original configuration variables. The only safe way to get correct expressions for physical observables in terms of flux-variables is either to define and insert the observable before the dualization process and dualize directly the resulting expression (see the derivation of the two-point function \eqref{eq:cpntwopointfunc3} in Sec.~\ref{ssec:intspacesubwormalgo}), or to define the observable in terms of derivatives of the logarithm of the partition function.\\

The only observables we are interested in, which cannot be defined directly in terms of derivatives of $\log\of{Z}$, are related to the two-point function\footnote{The two-point function \eqref{eq:cpntwopointfunc3} and the magnetic susceptibility could of course be defined in terms of derivatives of $Z$ or $\log\of{Z}$ if we would add appropriate source terms $\sim \sum\limits_{i}J_{a\,b}\of{x}\,\bar{z}_{a}\of{x}z_{b}\of{x}$ to the action, or alternatively the adjoint form $\sum\limits_{i}J_{i}\of{x}\,z^{\dagger}\of{x}\lambda_{i} z\of{x}$, where the $\lambda_{i}$ are the $\of{N^2-1}$ $\SU{N}$ generators. In the latter case the magnetic susceptibility reads $\chi_{m}=\frac{1}{2\,V}\sum\limits_{x,y,i}\left.\spartd{\log\of{Z}}{J_{i}\of{x}}{J_{i}\of{y}}\right|_{J=0}=\frac{1}{2\,V}\sum\limits_{x,y,i}\sof{\avof{z^{\dagger}\of{x}\lambda_{i}z\of{x}\,z^{\dagger}\of{y}\lambda_{i}z\of{y}}-\avof{z^{\dagger}\of{x}\lambda_{i}z\of{x}}\avof{z^{\dagger}\of{y}\lambda_{i}z\of{y}}}$, which, according to the Fierz identities for the generators of $\SU{N}$ and because $z^{\dagger}\of{x}\cdot z\of{x}=1$, is the same as \eqref{eq:magsusc}, provided that $\avof{\phi^{a\,b}}$ vanishes.} \eqref{eq:cpntwopointfunc3} defined below in Sec.~\ref{ssec:ordworm}, namely the magnetic susceptibility,
\[
\chi_{m}\,=\,\frac{1}{V}\sum\limits_{x,y}\bof{\sum\limits_{a,b}\avof{\phi^{a\,b}\of{x}\phi^{b\,a}\of{y}}-\frac{1}{N}}\ ,\label{eq:magsusc}
\]
where $\phi^{a\,b}\of{x}=z_{a}\of{x} \bar{z}_{b}\of{x}$, and the so-called \emph{second moment correlation length} (see e.g. \cite{Campostrini}),
\[
\xi_{G}\,=\,\frac{1}{2\,\sin\of{\frac{\pi}{L}}}\of{\tfrac{\sum\limits_{x,y}\sof{\sum\limits_{a,b}\avof{\phi^{a\,b}\of{x}\phi^{b\,a}\of{y}}-\frac{1}{N}}}{\sum\limits_{\bar{x},\bar{y},t_{x},t_{y}}\e^{\frac{2\,\pi\,\ii\,\ssof{t_y-t_x}}{L}}\sof{\sum\limits_{a,b}\avof{\phi^{a\,b}\of{\bar{x},t_{x}}\phi^{b\,a}\of{\bar{y},t_{y}}}-\frac{1}{N}}}-1}^{1/2}\ ,\label{eq:smcorrlen}
\]
where it should be noted that in this form, \eqref{eq:magsusc} and \eqref{eq:smcorrlen} are only valid as long as no non-trivial condensate develops, as otherwise the corresponding disconnected pieces would have to be subtracted from the two-point function first. In any case, as we are interested in comparing our results with those from the literature \cite{Vecchia,Vecchia2,Flynn} where the magnetic susceptibility $\chi_{m}$ and the second moment correlation length $\xi_{G}$ have been defined as \eqref{eq:magsusc} and \eqref{eq:smcorrlen} respectively, we will do so as well.\\

The remaining observables that will be considered are the average energy per site,
\[
\avof{E}\,=\,-\frac{1}{V}\partd{\log\of{Z}}{\beta}\ ,\label{eq:avenergy}
\]
and the specific heat,
\[
C_{E}\,=\,\beta^2\sof{\savof{E^2}-\savof{E}^2}\,=\,\frac{\beta^2}{V}\partdm{\log\of{Z}}{\beta}{2}\ ,\label{eq:specificheat}
\]
as well as the $\of{N-1}$ charge densities,
\[
\savof{n_{i}}\,=\,\frac{1}{V}\partd{\log\of{Z}}{\mu_{i}}\,=\,\frac{1}{V}\sum\limits_{a}\tilde{\lambda}_{i,a\,a}\partd{\log\of{Z}}{\tilde{\mu}_{a}}\ ,\label{eq:chargedens}
\]
and the corresponding $\of{N-1}^2$ covariances,
\[
V\of{\savof{n_{i}\,n_{j}}-\savof{n_{i}}\savof{n_{j}}}\,=\,\frac{1}{V}\spartd{\log\of{Z}}{\mu_{i}}{\mu_{j}}\,=\,\frac{1}{V}\sum\limits_{a,b}\tilde{\lambda}_{i,a\,a}\,\tilde{\lambda}_{j,b\,b}\,\spartd{\log\of{Z}}{\tilde{\mu}_{a}}{\tilde{\mu}_{b}}\ .\label{eq:chargecov}
\]  

\section{Simulation Methods}\label{sec:simulationmethods}
Because of the constraints imposed on the $k$-variables in the dual versions \eqref{eq:cpnpartf2a}, \eqref{eq:cpnpartf1}, \eqref{eq:cpnpartf3} and \eqref{eq:cpnpartf4} of the $\CPn{N-1}$-partition functions \eqref{eq:lpartf1} and \eqref{eq:lpartf2}, a worm algorithm has to be used to generate the configurations required for Monte Carlo estimates for expectation values of observables.\\ 
In this section, we first give an explanation for the apparent ergodicity problem \cite{Vetter} that occurs when using a "naive" worm algorithm to simulate the $\CPn{N-1}$ model with $N>2$ in the flux-variable representation from \cite{Chandrasekharan} (i.e. our $Z_{Q}$ from eq.~\eqref{eq:cpnpartf2a}) and then introduce our "internal space sub-worm algorithm" which solves the problem.

\subsection{Constraints}\label{ssec:constraints}
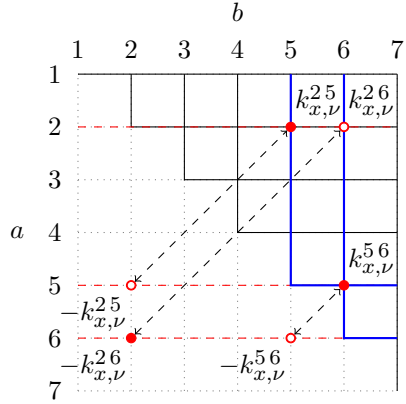
\begin{figure}[h]
\centering
\begin{minipage}[t]{\linewidth}
\centering
\input{tikzimg/constraints.tex}
\end{minipage}
\caption{Each grid point in the figure represents an $\of{a,b}$-component of the flux-variable $k_{x,\nu}^{a\,b}$ that points from the site $x$ in $\nu$-direction. If we increase for example the component $k_{x,\nu}^{5\,6}$ (indicated by a filled red circle) this implies, due to the anti-symmetry of the $k_{x,\nu}^{a\,b}$-variables in the $\of{a,b}$-indices, that we decrease at the same time the component $k_{x,\mu}^{6\,5}=-k_{x,\nu}^{5\,6}$ (indicated by an empty red circle). In order to satisfy all the delta-function constraints in \eqref{eq:cpnpartf2} (without changing a $k$-variable that points in another space-time direction) we have to compensate the changes in $k_{x,\nu}^{5\,6}$ and $k_{x,\nu}^{6\,5}$ so that the total sums of changes done to all components with the same index $a$ (i.e. along the horizontal lines) are zero: we can therefore decrease for example $k_{x,\nu}^{5\,2}=-k_{x,\nu}^{2\,5}$ to compensate for the increase in $k_{x,\nu}^{5\,6}$. This however also increases $k_{x,\nu}^{2\,5}$ which has again to be compensated by decreasing any of the $k_{x,\nu}^{2,b}$. For simplicity, we choose to decrease $k_{x,\nu}^{2\,6}$, such that the corresponding increase in $k_{x,\nu}^{6\,2}=-k_{x,\nu}^{2\,6}$ compensates not only the increase in $k_{x,\nu}^{2\,5}$ but also the decrease in $k_{x,\nu}^{6\,5}=-k_{x,\nu}^{5\,6}$ from the initial update, rendering all delta-functions non-zero again. The minimal length for such an \emph{internal space update cycle} is three, i.e. one has to update at last three different components of the $k_{x,\nu}$ matrix, but there is no upper bound on the length of such update cycles.}
  \label{fig:worm1}
\end{figure}

The structure of the constraints imposed on the $k$-variables in our flux-variable formulation \eqref{eq:cpnpartf2} of the $\CPn{N-1}$ model is more involved than what one encounters for example in the flux-variable formulation of the $\On{N}$\cite{Gattringer1} or the principal chiral $\SU{2}$\cite{Rindlisbacher1} model. The $k$-variables in \eqref{eq:cpnpartf2} are on each site subject to the following discrete delta-function constraints:
\[
\prod\limits_{a}^{N}\delta\bof{\sum\limits_{\mu=1}^{d}\sum\limits_{b=1}^{N}\of{k_{x,\mu}^{a\,b}-k_{x-\hat{\mu},\mu}^{a\,b}}}\ .\label{eq:deltaconstr}
\]
As the $k$-variables enter \eqref{eq:deltaconstr} always in the form of a sum over $\mu$ and $b$, it seems at first that the constraint is in fact not that restrictive and that a \emph{defect} in one of the delta functions in \eqref{eq:deltaconstr}, coming from a change in, say $k_{x-\hat{\mu},\mu}^{5\,6}$ could be compensated not only by changing $k_{x,\mu}^{5\,6}$ (i.e. by propagating the defect from site $x$ to site $x+\hat{\mu}$) but also by e.g. changing $k_{x-\hat{\mu},\mu}^{5,2}$ instead. However, as the $k_{x,\mu}^{a\,b}$ are anti-symmetric in the indices $\of{a\,b}$, each $k$-variable appears in fact in \emph{two} of the delta-function constraints for the two sites that it touches. This means that a change, for example an increase of $k_{x-\hat{\mu},\mu}^{5\,6}$, introduces in \eqref{eq:deltaconstr} not just a defect in the constraint for $a=5$ but also another one in the constraint for $a=6$. A decrease in e.g. $k_{x-\hat{\mu},\mu}^{5\,2}$ would therefore just remove the defect in the constraint for $a=5$ but not the one in the constraint for $a=6$, and instead would introduce another defect in the constraint for $a=2$. The net effect would just be to move the defect from the constraint for $a=5$ to the constraint for $a=2$. As illustrated in Fig. \ref{fig:worm1} the defects in the delta-functions for $a=2$ and $a=6$ are however of such a form that we can remove them simultaneously by updating a third $k$-variable, namely $k_{x-\hat{\mu},\mu}^{2,6}$! This trivially extends to arbitrarily long chains of $k$-variables that live on the same link. So, although it is not possible to freely update pairs of $k$-variables that live on the same link, one can update arbitrary cycles consisting of at least three such $k$-variables, which naturally gives rise to what we will call an \emph{internal space worm}.\\
In Sec.~\ref{ssec:intspacesubwormalgo} we will describe an algorithm which combines a conventional worm that propagates defects from site to site with an \emph{internal space sub-worm} which makes it possible that the same defect can be propagated by different types of flux-variables.\\

In the naive implementation of \cite{Vetter}, an ordinary worm was used (see next Sec.), where the internal-space index pair $\of{a,b}$ of the flux-variables that are updated are kept fixed while the worm propagates the corresponding defect on the lattice; only when the worm closes, the internal space index pair of the flux-variables that are to-be updated can change. This means that if the worm length increases as a function of $\beta$, the algorithm updates for increasingly long times always only one type of configuration-variables, i.e. the $k$-variables which have all the same internal space index pair $\of{a,b}$. When the worm finally closes, the algorithm can be faced with an energy barrier that has to be overcome in order to change this index pair. This becomes particularly clear when we consider the situation where the simulation has just started (we start with $k_{x,\nu}^{a\,b}=0$ $\forall a,b,x,\nu$) and the first worm has just finished updating the, say, $\of{a,b}=\of{2,3}$ components of the $n$-variables in \eqref{eq:cpnpartf2pre} (related to the $k$-variables through \eqref{eq:nkmap}). This means that the factors for $a=2,3$ in the numerators of the site-weights \eqref{eq:fweight} are already increased, while the factors for the remaining values of $a$ are still zero, and it therefore costs some energy (which can become quite large) to move the defect from $\of{a,b}=\of{2,3}$ to another internal-space index pair and it could therefore take a very long time for the system to thermalize. By extending the standard worm algorithm with the internal space worm as described below in Sec.~\ref{ssec:intspacesubwormalgo}, not just the $\of{a,b}=\of{2,3}$ components of the $k$-variables are updated while the defect for $\of{a,b}=\of{2,3}$ is propagated on the lattice, but also the other components of the $k$-variables. The values of the different factors in the site-weights \eqref{eq:fweight} therefore grow more homogeneously and the energy barrier for a change of the internal space indices of the defect after the worm has closed, does not form.\\
Another difficulty arises: how can an algorithm that uses an ordinary worm to update the $k$-variables achieve that the $\of{a,b}=\of{2,3}$ component of a single $k$-variable is non-zero while the $\of{a,b}=\of{2,3}$ components of all its neighboring $k$-variables are zero? As with the standard worm algorithm, just one type of $k$-variable can be updated during a single worm update, and because worms always produce connected strings of updated variables, it would require not just one but several worm-moves along a special trajectory to produce such a situation. With the internal space sub-worm algorithm (see Sec.~\ref{ssec:intspacesubwormalgo}) on the other hand, such situations are produced all the time during individual worm steps! The internal space sub-worm algorithm is therefore able to take shortcuts in configuration space between configurations, which would be hard to connect with the ordinary worm algorithm.\\

Note that if $N\leq 2$, the internal space sub-worm algorithm is identical to the ordinary worm, as non-trivial internal space update cycles are possible only if $N\geq 3$. This explains why in \cite{Vetter} the ergodicity problem has only been observed for $N>2$.  

\subsection{Ordinary Worm Algorithm}\label{ssec:ordworm}
The general idea behind a worm algorithm\cite{Prokofev} is to update configuration variables for a partition function $Z$, which are subject to a so-called \emph{closed loop constraint}, by generating configurations that contribute to some partition functions $Z_{2}\of{x,y}$ instead of $Z$. Each of the $Z_{2}\of{x,y}$ is a partition function for the same system that is described by $Z$ itself, but in the presence of an external source at $x$ and an external sink at $y$, so that at these two sites, the \emph{closed loop constraint} can be violated. The constrained configuration variables are then updated by moving around either the source or the sink and whenever source and sink meet again, a new configuration that contributes to $Z$ can be obtained by dropping the source/sink pair.\\

As already mentioned, a naive application of this concept to the flux-variable representations \eqref{eq:cpnpartf2a} and \eqref{eq:cpnpartf1} of the $\CPn{N-1}$ partition function leads to wrong results for $N>2$ \cite{Vetter}. However, for the flux-variable representations \eqref{eq:cpnpartf3} and \eqref{eq:cpnpartf4}, this naive, standard worm algorithm works. Therefore, and because our internal space sub-worm algorithm is a generalization of the standard worm, it makes sense to describe first the standard worm here on the example of \eqref{eq:cpnpartf3}.\\

Before we can explain how the worm works, we first need to know how an external source or sink appears in our flux-variable representation \eqref{eq:cpnpartf3}. As in the $\CPn{N-1}$ model, the fields $z$ and $z^{\dagger}$ cannot appear on their own due to local $\Un{1}$ gauge symmetry, we have to consider $\Un{1}$ gauge-invariant pairs (mesons) $z_{a}\of{x}\,\bar{z}_{b}\of{x}$ as sources and sinks, i.e. we define a source at a point $x$ as 
\[
\phi^{a\,b}\of{x}\,=\,z_{a}\of{x}\bar{z}_{b}\of{x}\ ,\label{eq:srcpair}
\]
and the corresponding sink as
\[
\bar{\phi}^{a\,b}\of{x}\,=\,\phi^{b\,a}\of{x}\,=\,z_{b}\of{x}\bar{z}_{a}\of{x}\ .\label{eq:snkpair}
\]
We are therefore interested in the flux-representation of the following partition function:
\begin{multline}
Z_{A,2}^{a_0\,b_0}\of{x,y}\,=\,\int\DD{z^{\dagger},z,U}\,z_{a_0}\of{x}\bar{z}_{b_0}\of{x}\bar{z}_{a_0}\of{y}z_{b_0}\of{y}\\
\cdot\exp\bof{\beta\,\sum\limits_{x,\nu}\,\sof{z^{\dagger}\of{x}\e^{\mu_{i}\tilde{\lambda}_{i}\delta_{\nu,d}}U_{\nu}\of{x}z\of{x+\hat{\nu}}\\
+\,z^{\dagger}\of{x}\e^{-\mu_{i}\tilde{\lambda}_{i}\delta_{\nu,d}}U^{\dagger}_{\nu}\of{x-\hat{\nu}}z\of{x-\hat{\nu}}}}\ .\label{eq:z2cpnlpartf3a}
\end{multline}
By going through the same steps that led us from \eqref{eq:lpartf3} to \eqref{eq:cpnpartf3}, the partition function \eqref{eq:z2cpnlpartf3a} can be turned into:
\begin{multline}
\tilde{Z}_{A,2}^{{\color{red}a\,b}}{\color{red}\of{x,y}}\,=\,\sum\limits_{\cof{k,\,l}}\bcof{\prod\limits_{z}\bof{\prod\limits_{\nu}\e^{-2\,\beta}\delta\sof{\sum\limits_{c}k_{z,\nu}^{c}}\prod\limits_{c}\e^{\tilde{\mu}_{c}\,k_{z,\nu}^{c}\delta_{\nu,d}}\,\frac{\beta^{\abs{k_{z,\nu}^{c}}+2\,l_{z,\nu}^{c}}}{\of{\abs{k_{z,\nu}^{c}}+l_{z,\nu}^{c}}!\,l_{z,\nu}^{c}!}}\\
\cdot\bof{\prod\limits_{c}\,\delta\sof{{\color{red}\sof{\delta^{c,b}-\delta^{c,a}}\sof{\delta_{x,z}-\delta_{y,z}}}+\sum\limits_{\nu}\sof{k_{z,\nu}^{c}-k_{z-\hat{\nu}}^{c}}}}\\
\cdot\frac{\prod\limits_{c}\,\sof{\sum\limits_{\nu}\sof{{\color{red}\frac{1}{2}\sof{\delta^{c,a}+\delta^{c,b}}\sof{\delta_{x,z}+\delta_{y,z}}}+\frac{1}{2}\sof{\sabs{k_{z,\nu}^{c}}+\sabs{k_{z-\hat{\nu},\nu}^{c}}}+l_{z,\nu}^{c}+l_{z-\hat{\nu},\nu}^{c}}}!}{\sof{N-1{\color{red}+\delta_{x,z}+\delta_{y,z}}+\sum\limits_{c}\sum\limits_{\nu}\sof{\frac{1}{2}\sof{\sabs{k_{z,\nu}^{c}}+\sabs{k_{z-\hat{\nu},\nu}^{c}}}+l_{z,\nu}^{c}+l_{z-\hat{\nu},\nu}^{c}}}!}}\ ,\label{eq:z2cpnpartf3}
\end{multline}
where we marked in red the changes caused in \eqref{eq:cpnpartf3} by the insertion of the external source/sink pair \eqref{eq:srcpair} and \eqref{eq:snkpair}. Equation \eqref{eq:z2cpnpartf3} is of course closely related to the two-point function for the field $\phi^{a\,b}\of{x}$ from \eqref{eq:srcpair}, which reads:
\[
\avof{\phi^{a\,b}\of{x}\phi^{b\,a}\of{y}}\,=\,\frac{\tilde{Z}_{A,2}^{a\,b}\of{x,y}}{\tilde{Z}_{A}}\ ,\label{eq:cpntwopointfunc3}
\]
and can therefore be measured \cite{Prokofev} by recording how often the worm's tail is located at site $x$ and its head at site $y$, such that the corresponding configuration contributes $\tilde{Z}_{A,2}^{a\,b}\of{x,y}$, as well as how often the external source-sink pair is removed, such that a configuration that contributes to $\tilde{Z}_{A}$ is produced.\\ 

\begin{figure}[!h]
\centering
\input{tikzimg/ordinaryworm.tex}
\caption{The figure illustrates the working principle of an ordinary worm algorithm used to update the $k$-variables in $\tilde{Z}_{A}$ in \eqref{eq:cpnpartf3}: in the upper-left drawing, the worm update starts in a configuration that contributes to the partition sum $\tilde{Z}_{A}$ and proposes to insert at some site $x=x_0$ an external source/sink pair $\phi^{a\,b}\,\phi^{b\,a}$. If this insertion is accepted by a Metropolis acceptance test, one has a configuration that contributes to $\tilde{Z}_{A,2}^{a\,b}\of{x_0,x_0}$ (see \eqref{eq:z2cpnpartf3}). Now, as depicted in the next picture, one can propose to move the sink $\phi^{b\,a}$ (which can be thought of as the head of a worm) in a randomly chosen direction $\nu$ from site $x$ to the neighboring site $y=x+\hat{\nu}$, compensating for the charge displacement by changing $k_{x,\nu}^{a}\rightarrow k_{x,\nu}^{a}+1$ and $k_{x,\nu}^{b}\rightarrow k_{x,\nu}^{b}-1$ if $\nu$ is a positive direction and vice versa if $\nu$ is a negative direction. If the proposed move is accepted, one obtains a configuration that contributes to the partition function $\tilde{Z}_{A,2}^{a\,b}\of{x_0,y}$. One can then set $x=y$, update $y=x+\hat{\nu}$ for a new randomly chosen direction $\nu$ and again propose to move the head of the worm to this new site $y$. In this manner the worm's head continues to move to new sites $y\rightarrow y+\hat{\nu}$ (where $\nu$ is always chosen randomly) until $x=x_{0}$, so that, as depicted in the bottom-left drawing, the external sink $\phi^{b\,a}$ sits again on the same site $x_{0}$ as the source $\phi^{a\,b}$. If this happens it can be proposed to remove again the source/sink pair $\phi^{a\,b}$, $\phi^{b\,a}$, and if this proposal is accepted, one ends up in a new configuration that contributes to the original partition function $\tilde{Z}_{A}$ . One can then pick a new location $x_{0}$ and continue with a new worm update at $x=x_0$.}
\label{fig:ordworm}
\end{figure}
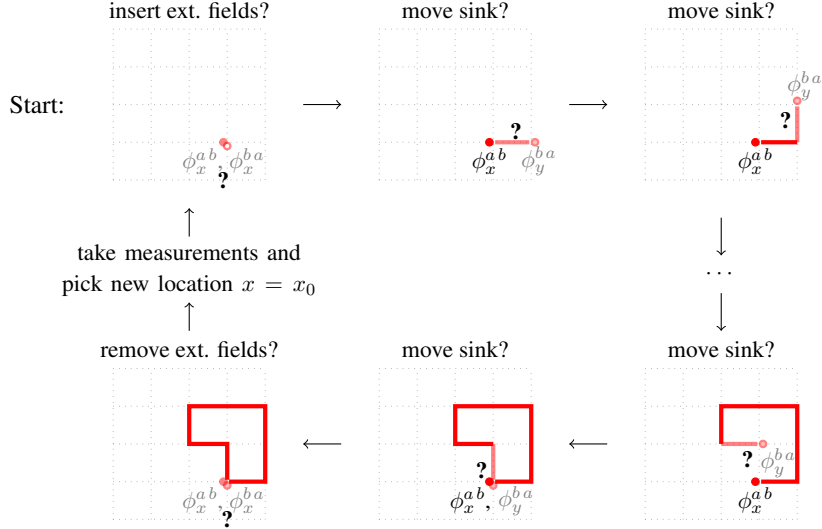

The basic working principle of the ordinary worm-algorithm is the following (see Fig.~\ref{fig:ordworm}): a worm update starts in a configuration that contributes to the partition sum $\tilde{Z}_{A}$ and proposes to insert at some site $x=x_0$ an external source/sink pair $\phi^{a\,b}\,\phi^{b\,a}$. If this insertion is accepted by a Metropolis acceptance test\cite{Metropolis}, one has a configuration that contributes to $\tilde{Z}_{A,2}^{a\,b}\of{x_{0},x_{0}}$. Now one can propose to move the sink (which can be thought of as the head of a worm) in a randomly chosen direction $\nu$ from site $x$ to the neighboring site $y=x+\hat{\nu}$, compensating for the charge displacement by updating appropriate flux-variables. Due to the on-link constraint $\delta\ssof{\sum_{c}\,k^{c}_{x,\nu}}$ in \eqref{eq:z2cpnpartf3} and because source and sink are mesons, one always has to update two $k$-variables simultaneously; in our case, if $\nu$ is a positive direction, the displacement of the sink would require to update $k_{x,\nu}^{a}\rightarrow k_{x,\nu}^{a}+1$ and $k_{x,\nu}^{b}\rightarrow k_{x,\nu}^{b}-1$, and vice versa if $\nu$ is a negative direction. If the proposed move is accepted, one obtains a configuration that contributes to the partition function $\tilde{Z}_{A,2}^{a\,b}\of{x_0,y}$. One can then set $x=y$, update $y=x+\hat{\nu}$ for a new randomly chosen direction $\nu$ and again propose to move the head of the worm to this new site $y$. In this manner the worm's head continues to move to new sites $y\rightarrow y+\hat{\nu}$ (where $\nu$ is always chosen randomly) until $x=x_{0}$ so that the external sink $\phi^{b\,a}$ hits again the site $x_{0}$ where the source $\phi^{a\,b}$ is located. If this happens, it can be proposed to remove again the source/sink pair $\phi^{a\,b}\,\phi^{b\,a}$, and if this proposal is accepted, the worm update ends and one ends up in a new configuration that contributes to the original partition function $\tilde{Z}_{A}$ . One can then take measurements for observables that depend on configurations of $\tilde{Z}_{A}$, if necessary, and then pick a new random location $x_{0}$ to start the next worm update. 

\subsection{Internal Space Sub-Worm Algorithm}\label{ssec:intspacesubwormalgo}
As explained in the previous section, for our choice of flux variables, there is not just one, but an infinite number of ways how a defect in the delta-function constraints (as introduced for example by the displacement of an external source or sink) can be compensated by an update of a sequence of appropriate flux variables. Our \emph{internal space sub-worm algorithm} (ISSW algorithm) takes this into account and thereby ensures ergodicity and ensures that the contribution of different configurations to the entropy is correctly taken into account.\\
 
We discuss the ISSW algorithm on the example of the quartic partition function $Z_{Q}$ form \eqref{eq:cpnpartf2a}. So this time, we are interested in the corresponding flux-representation of:
\[
Z_{Q,2}^{a_0\,b_0}\of{x,y}\,=\,\int\DD{z^{\dagger},z}\,z_{a_0}\of{x}\bar{z}_{b_0}\of{x}\bar{z}_{a_0}\of{y}z_{b_0}\of{y}\prod\limits_{x,\mu}\exp\of{\beta\,\sabs{z^{\dagger}\of{x}\cdot z\of{x+\hat{\mu}}}^2}\ .\label{eq:z2cpnpartf0}
\]
By going through the same steps that turned \eqref{eq:partf0} into \eqref{eq:cpnpartf2a}, the partition function \eqref{eq:z2cpnpartf0} yields:
\begin{multline}
Z_{Q,2}^{{\color{red}a_0\,b_0}}{\color{red}\of{x,y}}\,=\,\sum\limits_{\cof{k,l}}\prod\limits_{z}\bcof{\bof{\prod\limits_{\mu=1}^{d}\prod\limits_{a,b=1}^{N}\frac{\beta^{\frac{1}{2}\ssof{\ssabs{k_{z,\mu}^{a\,b}}+k_{z,\mu}^{a\,b}}+l_{z,\mu}^{a\,b}}}{\of{\frac{1}{2}\of{\ssabs{k_{z,\mu}^{a\,b}}+k_{z,\mu}^{a\,b}}+l_{z,\mu}^{a\,b}}!}}\\
\cdot\prod\limits_{a}^{N}\delta\sof{{\color{red}\sof{\delta^{a,b_{0}}-\delta^{a,a_{0}}}\sof{\delta_{x,z}-\delta_{y,z}}}+\sum\limits_{\mu=1}^{d}\sum\limits_{b=1}^{N}\sof{k_{z,\mu}^{a\,b}-k_{z-\hat{\mu},\mu}^{a\,b}}}\\
\cdot\frac{\prod\limits_{a}^{N}\sof{{\color{red}\frac{1}{2}\sof{\delta^{a,a_{0}}+\delta^{a,b_{0}}}\sof{\delta_{x,z}+\delta_{y,z}}}+\sum\limits_{\mu=1}^{d}\sum\limits_{b=1}^{N}\sof{\frac{1}{2}\ssof{\ssabs{k_{z,\mu}^{a\,b}}+\ssabs{k_{z-\hat{\mu},\mu}^{a\,b}}}+l_{z,\mu}^{a\,b}+l_{z-\hat{\mu},\mu}^{a\,b}}}!}{\sof{N-1\,{\color{red}+\,\delta_{x,z}+\delta_{y,z}}+\sum\limits_{\mu=1}^{d}\sum\limits_{c,b=1}^{N}\sof{\frac{1}{2}\ssof{\ssabs{k_{z,\mu}^{c\,b}}+\ssabs{k_{z-\hat{\mu},\mu}^{c\,b}}}+l_{z,\mu}^{c\,b}+l_{z-\hat{\mu},\mu}^{c\,b}}}!}}\ ,\label{eq:z2cpnpartf2a}
\end{multline}
where we marked again in red the changes caused in \eqref{eq:z2cpnpartf2a} by the insertion of the external source/sink pairs \eqref{eq:srcpair} and \eqref{eq:snkpair} into \eqref{eq:cpnpartf2a}.\\

The internal space sub-worm algorithm now starts exactly like the ordinary worm described in the previous section: one starts in a configuration that contributes to the partition sum $Z_{Q}$ and proposes to insert at some site $x=x_0$ an external source/sink pair $\phi^{a\,b}\,\phi^{b\,a}$. If this insertion is accepted by a Metropolis acceptance test, the system is in a configuration that contributes to the partition function $Z_{Q,2}^{a\,b}\of{x,x}$. One then proposes to move the sink in a randomly chosen direction $\nu$ from site $x$ to the neighboring site $y=x+\hat{\nu}$ and compensates for the charge displacement by updating appropriate flux-variables. But instead of just updating $k_{x,\nu}^{a\,b}$, i.e. the $k$-variable with the same internal space indices as the source and sink (which would be the analog to the simultaneous update of $k_{x,\nu}^{a}$ and $k_{x,\nu}^{b}$ in the description of the ordinary worm above in Sec.~\ref{ssec:ordworm}), one now runs an \emph{internal space sub-worm cycle} which explores all possibilities by which the defects introduced by the displacement of the external sink could be compensated; this includes the possibility of just updating $k_{x,\nu}^{a\,b}$, but also the possibility of doing so by updating a sequence of $k$-variables of length $n$ instead, e.g. $k_{x,\nu}^{a\,c_1}\,k_{x,\nu}^{c_1\,c_2}\,\cdots\,k_{x,\nu}^{c_n\,b}$ with $n>1$. Such a sequence is set up element by element in a worm-like manner: a random internal space index $c_1$ is chosen and one proposes to update the $k_{x,\nu}^{a\,c_1}$ variable and to temporarily replace the original sink $\phi^{b\,a}\of{y}$ by a source/sink pair $\phi^{b\,c_1}\of{x}$ and $\phi^{c_1\,a}\of{y}$ in order to compensate for the temporary new defects. If this update is accepted, one continues by choosing a new random internal space index $c_2$ and proposes to update $k_{x,\nu}^{c_1\,c_2}$ while replacing $\phi^{b\,c_1}\of{x}$ and $\phi^{c_1\,a}\of{y}$ by $\phi^{b\,c_2}\of{x}$ and $\phi^{c_2\,a}\of{y}$ respectively, and so on, until the new randomly chosen internal space index $c_{n}$ coincides either with $a$ or $b$. If $c_n$ coincides with $a$, the internal space sub-worm cycle ends by bringing the original external sink $\phi^{b\,a}\of{y}$ back to site $x$ while having updated a closed cycle of $k$-variables, $k_{x,\nu}^{a\,c_1}\,k_{x,\nu}^{c_1\,c_2}\,\cdots\,k_{x,\nu}^{c_n\,a}$, which effectively propagates no defects, whereas if $c_n$ coincides with $b$, the sub-worm cycle ends by restoring the original external sink $\phi^{b\,a}\of{y}$ on site $y$ while having updated a sequence of $k$-variables, $k_{x,\nu}^{a\,c_1}\,k_{x,\nu}^{c_1\,c_2}\,\cdots\,k_{x,\nu}^{c_n\,b}$, that compensates for the defects that were introduced by moving $\phi^{b\,a}$ from $x$ to $y$. In the latter case, we now have a configuration that contributes to the partition sum $Z_{Q,2}^{a\,b}\of{x_{0},y}$. One can now set $x=y$ and update $y=x+\hat{\nu}$ for a new randomly chosen direction $\nu$ and start a new sub-worm cycle for the corresponding $k$-variables. In this manner the worm's head continues to move to new sites $y\rightarrow y+\hat{\nu}$ (where $\nu$ is always chosen randomly after every completed sub-worm cycle) until $y=x_{0}$ and the external sink $\phi^{b\,a}$ hits again the site $x_{0}$. If this happens, i.e. if the worm closes, it can be proposed to remove again the source sink pair $\phi^{a\,b}$, $\phi^{b\,a}$ from the system, and if this proposal is accepted, one ends up in a new configuration that contributes to the original partition function $Z_{Q}$ .\\

\begin{figure}[!h]
\centering
\begin{minipage}[t]{\linewidth}
\centering
\input{tikzimg/subworm.tex}
\end{minipage}
\caption{The figure illustrates (from left to right) how a sub-worm cycle of the algorithm described in the main text, works, assuming that at step \ref{it:choosedir} a positive direction $\nu$ is selected. The small grids represent the sum of all the $k_{x,\mu}$ matrices and sources/sinks that enter the delta-function constraints for $x$ and $x+\ohat{\nu}$ on the second line of \eqref{eq:z2cpnpartf2a}. Different rows/columns correspond to different $a$/$b$ coordinates of $k_{x,\mu}^{a\,b}$ respectively. Due to the anti-symmetry of $k_{x,\mu}^{a\,b}$, there is also an anti-symmetry between the rows and columns in the grids and we can choose to just focus on the rows: each row in a grid represents one of the $N$ constraints in the product on the second line of \eqref{eq:z2cpnpartf2a} and the net change done to each row has therefore to be zero. Individual changes can be caused by insertions of external source/sink pairs or by updates of flux-variables $k_{x,\nu}^{a\,b}$. Source/sink pairs are represented by pairs (\urcross,\ucross), joined by a dotted line, where a \urcross in row $a$, joined to a \ucross in row $b$ in the grid for site $x$ represents a pair $\obar{z}_{a}\of{x} z_{b}\of{x}\equiv \phi^{a\,b}\of{x}$. An update of a flux-variable $k_{x,\nu}^{a\,b}\rightarrow k_{x,\nu}^{a\,b}+1$ is represented by a joined pair (\uocirc,\ufcirc) in the grid for site $x$, where \uocirc is in row $a$ and \ufcirc in row $b$, and a pair (\ufcirc,\uocirc) in the grid for site $x+\ohat{\nu}$, where \ufcirc is in row $a$ and \uocirc in row $b$. In the last column, we have depicted the two possible ways in which the sub-worm cycle can end: either with $b=b_{0}$, in which case the head of the worm can move from site $x$ to site $x+\ohat{\nu}$ before a new direction $\nu$ is chosen, or with $b=a_{0}$, in which case the head of the worm remains on site $x$ and the worm just chooses a new direction $\nu$. Note that in the former case, the net effect of all the \ufcirc and \uocirc in the grid for the site $x+\ohat{\nu}$ is the same as that of a (\ucross,\urcross) pair representing $\obar{z}_{b}\of{x+\ohat{\nu}}z_{a}\of{x+\ohat{\nu}}$, such that we are again in the situation that we had at the beginning for the site $x$, but now this situation occurs on site $x+\ohat{\nu}$. So the next worm step can proceed in a completely analogous way. Note that if at step \ref{it:choosedir} in the algorithm-description in the main text, a negative direction $\nu$ is chosen, the sub-worm cycle has to start with $a=b_0$ instead of $a=a_0$ in the second column in the figure, and also for the two possibilities by which the sub-worm cycle can end, as depicted in the last column, the roles of $a_0$ and $b_0$ are interchanged. This is necessary in order to satisfy detailed balance between start and end of the sub-worm cycles.}
\label{fig:worm2}
\end{figure}
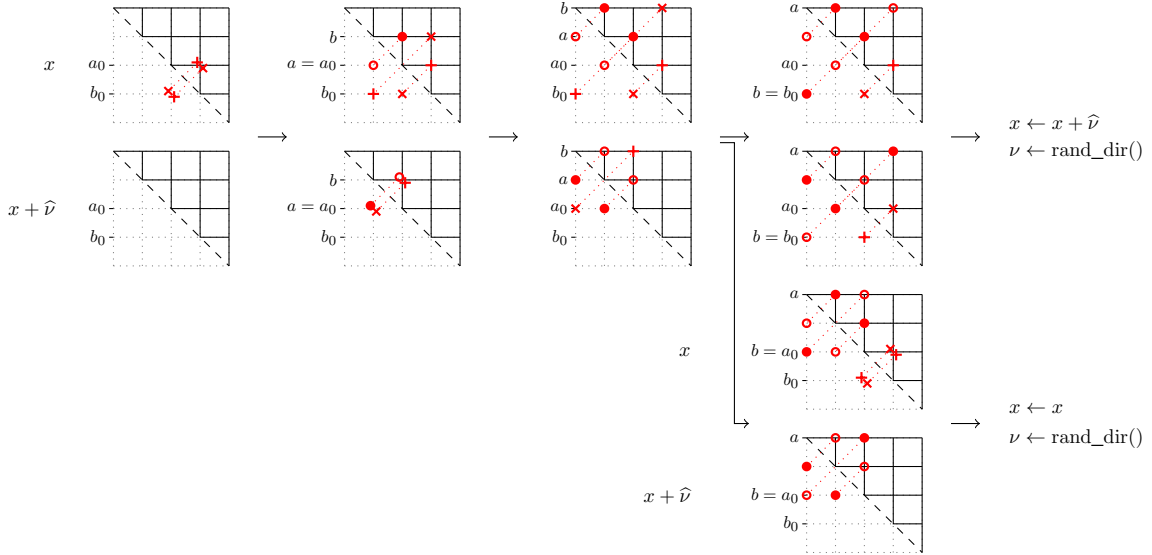

Note that whenever the worm chooses a negative direction $\nu$, the roles of the internal space indices $a$ and $b$ have to be interchanged for the corresponding sub-worm cycle. This is necessary in order to satisfy detailed balance for the start and end moves of the sub-worm cycles\footnote{In order to be able to set up detailed balance between two configurations $C$ and $C'$, the algorithm needs to be able to undergo direct transitions between the two configurations in both directions: from $C$ to $C'$ as well as from $C'$ to $C$. Stated differently: if the algorithm is currently in the configuration $C$ and is allowed to propose an update that leads from $C$ to another configuration $C'$, then, whenever the algorithm is in configuration $C'$, it needs to be able to propose also the inverse update that leads from $C'$ back to $C$.}.\\

If one would include appropriate non-zero meson-source terms (external background fields) in the $\CPn{N-1}$ action, the system would be allowed to produce dynamical meson sources and sinks which could replace some of the external, non-dynamical $\phi^{a\,b}$ that appear in the above description of our algorithm. For example, the worm could then change the internal space indices $\of{b,a}$ of its head at any time by replacing the third external source/sink $\phi^{b\,c_{n}}$ that appears temporarily during the sub-worm cycle, by a dynamical one, after which the worm's head would have internal space indices $\of{c_{n},a}$ instead of $\of{b,a}$. Furthermore, as is also the case in other models (see e.g. \cite{Rindlisbacher1}), the worm can become disconnected as at any time, the external sink at the worm's head can be replaced by a dynamical sink and a new head can be inserted at a new location, together with an appropriate dynamical source. Also here, the internal space indices of the head could of course be changed during this process.\\

A more detailed description of the ISSW algorithm (without external background fields) is given by the illustration in Fig.~\ref{fig:worm2} and the following step-by-step guide (the acceptance probabilities for the various moves will be discussed in Sec.~\ref{ssec:acceptanceprobabs}):
\begin{enumerate}[label=(\alph*)]\small
\item\label{it:isstartq} choose a random site $x_0\in\cof{1,\ldots,V}$, set $x=x_0$ and propose to insert at this site a source/sink pair $\phi^{a_{0}\,b_{0}}\of{x}\phi^{b_{0}\,a_{0}}\of{x}$ with randomly chosen $a_{0},b_{0}\in\cof{1,\ldots,N}$, $a_{0}\neq b_{0}$.
if the move is accepted: continue with \ref{it:isendq}, else: continue with \ref{it:isstartq},
\item\label{it:isendq}
\begin{itemize}
\item if $x=x_{0}$: 
\begin{itemize}
\item with probability $p_{t}$: continue with \ref{it:isendq2}, 
\item with parobability $\of{1-p_{t}}$: continue with \ref{it:choosedir},
\end{itemize}
\item else, if $x\neq x_{0}$: continue with \ref{it:choosedir},
\end{itemize}
\item\label{it:choosedir} choose a random direction $\nu\in\cof{\pm 1,\ldots,\pm d}$:
\begin{itemize}
\item if $\nu>0$: set $a=a_{0}$ and continue with \ref{it:posdir},
\item  else: set $a=b_{0}$ and continue with \ref{it:negdir},
\end{itemize}
\item\label{it:posdir} choose a random $b\in\cof{1,\ldots,N}\setminus\cof{a}$ and
\begin{itemize}
\item if $b\neq b_{0}$, propose to:
\begin{itemize}
\item update $k_{x,\nu}^{a\,b}\rightarrow k_{x,\nu}^{a\,b}+1$,
\item replace $\phi^{b_{0}\,a}\of{x}$ by $\phi^{b_{0}\,b}\of{x}$,
\item and insert $\phi^{b\,a_{0}}\of{x+\hat{\nu}}$,
\end{itemize}
if accepted: set $a=b$ and continue with \ref{it:posdircont}, else: continue with \ref{it:choosedir},
\item else if $b=b_{0}$, propose to:
\begin{itemize}
\item update $k_{x,\nu}^{a\,b}\rightarrow k_{x,\nu}^{a\,b}+1$,
\item remove $\phi^{b_{0}\,a}\of{x}$,
\item and insert $\phi^{b\,a_{0}}\of{x+\hat{\nu}}$,
\end{itemize}
if accepted: update $x\rightarrow x+\hat{\nu}$; continue with \ref{it:isendq}, else: continue with \ref{it:choosedir},
\end{itemize}
\item\label{it:posdircont} choose a random new $b\in\cof{1,\ldots,N}\setminus\cof{a}$ and
\begin{itemize}
\item if $b\neq b_{0}$ and $b\neq a_{0}$, propose to:
\begin{itemize}
\item update $k_{x,\nu}^{a\,b}\rightarrow k_{x,\nu}^{a\,b}+1$,
\item replace $\phi^{b_{0}\,a}\of{x}$ by $\phi^{b_{0}\,b}\of{x}$,
\item and replace $\phi^{a\,a_{0}}\of{x+\hat{\nu}}$ by $\phi^{b\,a_{0}}\of{x+\hat{\nu}}$,
\end{itemize}
if accepted: set $a=b$ and continue with \ref{it:posdircont}, else: continue with \ref{it:posdircont},
\item else if $b=b_{0}$ propose to:
\begin{itemize}
\item update $k_{x,\nu}^{a\,b}\rightarrow k_{x,\nu}^{a\,b}+1$,
\item remove $\phi^{b_{0}\,a}\of{x}$,
\item and replace $\phi^{a\,a_{0}}\of{x+\hat{\nu}}$ by $\phi^{b\,a_{0}}\of{x+\hat{\nu}}$,
\end{itemize}
if accepted: update $x\rightarrow x+\hat{\nu}$, continue with \ref{it:isendq}, else: continue with \ref{it:posdircont},
\item else if $b=a_{0}$ propose to:
\begin{itemize}
\item update $k_{x,\nu}^{a\,b}\rightarrow k_{x,\nu}^{a\,b}+1$,
\item replace $\phi^{b_{0}\,a}\of{x}$ by $\phi^{b_{0}\,b}\of{x}$,
\item and remove $\phi^{a\,a_{0}}\of{x+\hat{\nu}}$,
\end{itemize}
if accepted: continue with \ref{it:isendq}, else: continue with \ref{it:posdircont},
\end{itemize}
\item\label{it:negdir} choose a random $b\in\cof{1,\ldots,N}\setminus\cof{a}$ and
\begin{itemize}
\item if $b\neq a_{0}$ propose to:
\begin{itemize}
\item update $k_{x+\hat{\nu},\abs{\nu}}^{a\,b}\rightarrow k_{x+\hat{\nu},\abs{\nu}}^{a\,b}+1$,
\item replace $\phi^{a\,a_{0}}\of{x}$ by $\phi^{b\,a_{0}}\of{x}$,
\item and insert $\phi^{b_{0}\,b}\of{x+\hat{\nu}}$,
\end{itemize}
if accepted: set $a=b$ and continue with \ref{it:negdircont}, else continue with \ref{it:choosedir},
\item else if $b=a_{0}$ propose to:
\begin{itemize}
\item update $k_{x+\hat{\nu},\abs{\nu}}^{a\,b}\rightarrow k_{x+\hat{\nu},\abs{\nu}}^{a\,b}+1$,
\item remove $\phi^{a\,a_{0}}\of{x}$,
\item and insert $\phi^{b_{0}\,b}\of{x+\hat{\nu}}$,
\end{itemize}
if accepted: update $x\rightarrow x+\hat{\nu}$
, continue with \ref{it:isendq}, else: continue with \ref{it:choosedir},
\end{itemize}
\item\label{it:negdircont} choose a random new $b\in\cof{1,\ldots,N}\setminus\cof{a}$ and
\begin{itemize}
\item if $b\neq b_{0}$ and $b\neq a_{0}$, propose to:
\begin{itemize}
\item update $k_{x,\nu}^{a\,b}\rightarrow k_{x,\nu}^{a\,b}+1$,
\item replace $\phi^{a\,a_{0}}\of{x}$ by $\phi^{b\,a_{0}}\of{x}$,
\item and replace $\phi^{b_{0}\,a}\of{x+\hat{\nu}}$ by $\phi^{b_{0}\,b}\of{x+\hat{\nu}}$,
\end{itemize}
if accepted: set $a=b$ and continue with \ref{it:negdircont}, else: continue with \ref{it:negdircont},
\item else if $b=a_{0}$ propose to:
\begin{itemize}
\item update $k_{x,\nu}^{a\,b}\rightarrow k_{x,\nu}^{a\,b}+1$,
\item remove $\phi^{a\,a_{0}}\of{x}$,
\item and replace $\phi^{b_{0}\,a}\of{x+\hat{\nu}}$ by $\phi^{b_{0}\,b}\of{x+\hat{\nu}}$,
\end{itemize}
if accepted: update $x\rightarrow x+\hat{\nu}$, continue with \ref{it:isendq}, else: continue with \ref{it:negdircont},
\item else if $b=b_{0}$ propose to:
\begin{itemize}
\item update $k_{x,\nu}^{a\,b}\rightarrow k_{x,\nu}^{a\,b}+1$,
\item replace $\phi^{a\,a_{0}}\of{x}$ by $\phi^{b\,a_{0}}\of{x}$,
\item and remove $\phi^{b_{0}\,a}\of{x+\hat{\nu}}$,
\end{itemize}
if accepted: continue with \ref{it:isendq}, else: continue with \ref{it:negdircont},
\end{itemize}
\item\label{it:isendq2} Propose to remove the source/sink pair from the site $x$,\\
if this is accepted: continue with \ref{it:isstartq}, else: continue with \ref{it:isendq}.
\end{enumerate}

So far we haven't mentioned how the unconstrained $l$-variables are updated: alternating between worm-updates and update sweeps for the $l$-variable is inefficient if the simulation parameters are such that the average worm-length is large, as the worm then evolves for a long time in a quasi-fixed $l$-background. Furthermore, such an update strategy would strictly speaking break detailed balance. A better alternative is therefore to incorporate the update of the $l$-variables into the worm-update by inserting in the above algorithm, whenever an attempt is made to update a $k$-variable, a random choice whether the worm should really try to update a $k$-variable, or if rather a Metropolis update of a randomly chosen $l$-variable should be attempted instead.

\subsection{Detailed Balance}\label{ssec:detailedbalance}

In order for a Markov process to generate a sequence of configurations, 
\[
C_{1}\to C_{2}\to\ldots\to C_{n}\label{eq:markovchain1}
\]
so that for any observable $\obs$, it holds that 
\[
\avof{\obs}\,=\,\lim\limits_{n\to\infty}\frac{1}{n}\sum\limits_{i=1}^{n}\,\obs\of{C_{i}}\ ,
\]
a sufficient condition is, that transitions between neighboring configurations (configurations which can be turned into each other by single updates), e.g. $C$ and $C'$, occur with probabilities $P\of{C\to C'}$ and $P\of{C'\to C}$, which are in accordance with the detailed balance equation
\[
w\of{C}P\of{C\to C'}\,=\,w\of{C'}P\of{C'\to C}\ ,\label{eq:detailedbalance1}
\]
where $w\of{C}$ and $w\of{C'}$ are the weights for the configuration $C$ and $C'$, respectively. This means that if in \eqref{eq:markovchain1} we have $C_{n}=C$, and a transition to $C'$ is proposed so that $C_{n+1}=C'$, then this proposal is only accepted with probability $P\of{C\to C'}$, whereas otherwise $C_{n+1}=C$. If the set of possible candidate configurations is always the same throughout the simulation, appropriate \emph{transition probabilities} $P\of{C\to C'}$ and $P\of{C'\to C}$ can be obtained \cite{Metropolis} by setting
\[
P\of{C\to C'}\,=\,\min\bof{1,\frac{w\of{C'}}{w\of{C}}}\quad\text{and}\quad P\of{C'\to C}\,=\,\min\bof{1,\frac{w\of{C}}{w\of{C'}}}\ ,\label{eq:transprobab1}
\]
which obviously solves \eqref{eq:detailedbalance1}. However, for the algorithms described in the previous two sections, it often happens that the move $C\to C'$ is chosen with a different probability than its inverse move $C'\to C$, in which case one has to use a generalization of the Metropolis algorithm, which is known as \emph{Metropolis-Hastings} algorithm \cite{Hastings}. To do so, one thinks of each transition probability to consist of two factors: a \emph{move-choice} or \emph{selection probability}, $p_{s}$, and a so-called \emph{reduced transition probability}, $P_{r}$:
\[
P\of{C\to C'}\,=\,p_{s}\of{C\to C'}\,P_r\of{C\to C'}\ ,
\] 
and similarly
\[
P\of{C'\to C}\,=\,p_{s}\of{C'\to C}\,P_r\of{C'\to C}\ .
\]
Now, as the move-choice probability $p_{s}\of{C\to C'}$ is, as already mentioned, indeed the probability that if the system is currently in the configuration $C$, the algorithm chooses to propose the update that leads it into the configuration $C'$, this factor of the full transition probability $P\of{C\to C'}$ has already been taken into account at the moment where the final acceptance test for the transition is made. The final acceptance test should therefore rely only on the reduced transition probability $P_{r}\of{C\to C'}$, which, in order to satisfy the detailed balance equation \eqref{eq:detailedbalance1}, can be defined as 
\[
P_r\of{C\to C'}\,=\,\min\bof{1,\frac{p_{s}\of{C'\to C}\,w\of{C'}}{p_{s}\of{C\to C'}\,w\of{C}}}\ .\label{eq:redtransprobab1}
\]
Similarly, the corresponding inverse transition from $C'$ to $C$ would be carried out with probability
\[
P_r\of{C'\to C}\,=\,\min\bof{1,\frac{p_{s}\of{C\to C'}\,w\of{C}}{p_{s}\of{C'\to C}\,w\of{C'}}}\ .\label{eq:redtransprobab2}
\]
If the move-choice probabilities $p_{s}$ for a move and its inverse are the same, i.e. $p_{s}\of{C\to C'}=p_{s}\of{C'\to C}$, then they just cancel out in \eqref{eq:redtransprobab1} and \eqref{eq:redtransprobab2} and one recovers the ordinary Metropolis acceptance probabilities \eqref{eq:transprobab1}.\\

\subsection{Metropolis Acceptance Probabilities for the ISSW Algorithm}\label{ssec:acceptanceprobabs}
We are now in the position to discuss the acceptance probabilities (given by the reduced transition probabilities) for the different types of moves that occur in the internal space sub-worm algorithm described above in Sec.~\ref{ssec:intspacesubwormalgo}. While the move choice probabilities $p_{s}$ depend only on the algorithm, the transition probabilities are model-dependent and will therefore be different for simulations of $Z_{Q}$ and $Z_{A}$ from \eqref{eq:cpnpartf2a} and \eqref{eq:cpnpartf1}, respectively. We will only discuss the transition probabilities for the $Z_{Q}$-case but those for the $Z_{A}$ can be obtained in a completely analogous way. To simplify notation, we introduce the following definitions:
\begin{align}
& A_{x}^{a}\,=\,\sum\limits_{\nu=1}^{d}\,\sum\limits_{b=1}^{N}\,\of{\tfrac{1}{2}\sof{\ssabs{k_{x,\nu}^{a\,b}}+\ssabs{k_{x-\hat{\nu},\nu}^{a\,b}}}+l_{x,\nu}^{a\,b}+l_{x-\hat{\nu},\nu}^{a\,b}}\ ,\\
& w_{\beta}\of{k,l}\,=\,\frac{\beta^{\ssabs{k}+2\,l}}{\sof{\ssabs{k}+l}!\,l!}\ ,\\
& W_{N}\ssof{A}\,=\,\frac{\prod\limits_{a=1}^{N}\,A^{a}!}{\sof{N-1+\sum\limits_{a=1}^{N}\,A^{a}}!}\ ,
\end{align}
where $A=\of{A^{1},\ldots,A^{N}}$. To indicate a shift in the $a$-component of $A$, we will write $A+\hat{a}$, with
\[
\hat{a}=\of{\delta_{1,a},\ldots,\delta_{N,a}}\ .
\]
In the following description, $C$ always refers to the configuration that is currently realized in the system and $C'$ to the one that is obtained if the update under consideration is carried out.

\subsubsection{Start of an ISSW Update}\label{sssec:startissw}
The internal space sub-worm algorithm starts in a configuration that contributes to the partition function $Z_{Q}$. A particular move \ref{it:isstartq}. from the step-by-step description in Sec.~\ref{ssec:intspacesubwormalgo} is selected with probability
\[
p_{s}\of{C\to C'}\,=\,\frac{1}{V\,N\,\of{N-1}}
\]
and its inverse with probability
\[
p_{s}\of{C'\to C}\,=\,p_{t}\in\roint{0,1}\ ,
\]
where $p_{t}$ is a free parameter which we define to be $p_{t}=1/2$. According to \eqref{eq:redtransprobab1}, the reduced transition probability for the move $\of{C\to C'}$ can therefore be set to:
\[
P_{r}\of{C\to C'}\,=\,\min\of{1,\,p_{t}\,V\,N\,\of{N-1}\,\frac{W_{N}\ssof{A_{x}+\hat{a}_{0}+\hat{b}_{0}}}{W_{N}\ssof{A_{x}}}}\ .
\]
If the move is accepted, we set $C=C'$ and define $y=x=x_{0}$. The system is now in a configuration that contributes to the two-point partition function $Z_{Q,2}^{a_{0}\,b_{0}}\of{x,y}$, defined above in \eqref{eq:z2cpnpartf2a}.

\subsubsection{Start of Sub-Worm Cycle}\label{sssec:startsubwormcycle}
Whenever $x=y$, a new sub-worm cycle can be started by choosing a random direction $\nu\in\cof{\pm 1,\ldots,\pm d}$ (step \ref{it:choosedir}. from above) and setting $y=x+\hat{\nu}$ and
\[
\tilde{a}_{0}=\ucases{a_{0}\,,\ \nu>0\\b_{0}\,,\ \nu<0}\quad\text{and}\quad \tilde{b}_{0}=\ucases{b_{0}\,,\ \nu>0\\a_{0}\,,\ \nu<0}\ ,
\]
and finally $a=\tilde{a}_{0}$. Next (step \ref{it:posdir}. or \ref{it:negdir}.), we choose a random internal space index $b\in\cof{1,\ldots,N}\setminus\cof{a}$ and call $C'$ the configuration to which the corresponding move would lead. However, if $x=x_{0}$, the sub-worm cycle is proposed only with probability $\of{1-p_{t}}$, as in this case, with probability $p_{t}$, it can also be proposed instead to remove the source/sink pair from the system and terminate the ISSW update. The total move-choice probability $p_{s}\of{C\to C'}$ for starting the sub-worm cycle is therefore in general given by
\[
p_{s}\of{C\to C'}\,=\,\frac{1-p_{t}\,\delta_{x,x_{0}}}{2\,d\,\of{N-1}}\ .
\]
The move-choice probability $p_{s}\of{C'\to C}$ for the inverse move depends on whether $b=\tilde{b}_{0}$  is true or not: 
\begin{itemize}
\item if $b\neq\tilde{b}_{0}$, then $C'$ is an intermediate configuration from within a sub-worm cycle, which contains three external fields (sources/sinks), in which case we have (see Sec.~\ref{sssec:intermedsubcyclemove} below)
\[
p_{s}\of{C'\to C}\,=\,\frac{1}{N-1}\ ,
\] 
\item whereas if $b=\tilde{b}_{0}$, then the sub-worm cycle starts and ends at the same time (i.e. the move corresponds to an ordinary worm move), so that $C'$ is not an intermediate configuration from within a sub-worm cycle, and the move that brings the system back from $C'$ to the configuration $C$ has to be selected in a completely analogous way as the move from $C$ to $C'$, so that
\[
p_{s}\of{C'\to C}\,=\,\frac{1-p_{t}\,\delta_{y,x_{0}}}{2\,d\,\of{N-1}}\ .
\]
\end{itemize}
In total, $p_{s}\of{C'\to C}$ is given by
\[
p_{s}\of{C'\to C}\,=\,\frac{1}{N-1}\,\of{\frac{1-p_{t}\,\delta_{y,x_{0}}}{2\,d}}^{\delta_{b,\tilde{b}_{0}}}\ ,
\]
which is however not very useful, as also the configuration weights will be qualitatively different for the two cases $b=\tilde{b}_{0}$ and $b\neq\tilde{b}_{0}$, so that we will explain the corresponding reduced transition probabilities separately. Let us abbreviate
\[
\tilde{k}\,=\,\ucases{k_{x,\nu}^{a\,b}\,,\ \nu>0\\k_{y,\abs{\nu}}^{a\,b}\,,\ \nu<0}\quad,\quad \tilde{l}\,=\,\ucases{l_{x,\nu}^{a\,b}\,,\ \nu>0\\l_{y,\abs{\nu}}^{a\,b}\,,\ \nu<0}\ ,
\]
and
\[
\Delta\,=\,\sabs{\tilde{k}+1}-\sabs{\tilde{k}} .
\]
Then we find for the reduced transition probability $P_{r}\of{C\to C'}\,=\,\min\of{1,\,r}$, where
\begin{itemize}
\item if $b\neq\tilde{b}_{0}$:
\begin{multline}
r\,=\,\frac{2\,d}{1-p_{t}\,\delta_{x,x_{0}}}\,\frac{w_{\beta}\ssof{\tilde{k}+1,\tilde{l}}}{w_{\beta}\ssof{\tilde{k},\tilde{l}}}\\
\cdot\frac{W_{N}\sof{A_{x}+\hat{a}_{0}+\hat{b}_{0}+\frac{\Delta+1}{2}\cdot\hat{b}+\frac{\Delta-1}{2}\cdot\hat{a}}\,W_{N}\sof{A_{y}+\frac{1+\Delta}{2}\cdot\ssof{\hat{a}+\hat{b}}}}{W_{N}\sof{A_{x}+\hat{a}_{0}+\hat{b}_{0}}\,W_{N}\sof{A_{y}}}\ ,
\end{multline}
in which case, if the transition is accepted, we set $a=b$ and $C=C'$,

\item and if $b=\tilde{b}_{0}$:
\begin{multline}
r\,=\,\frac{1-p_{t}\,\delta_{y,x_{0}}}{1-p_{t}\,\delta_{x,x_{0}}}\,\frac{w_{\beta}\ssof{\tilde{k}+1,\tilde{l}}}{w_{\beta}\ssof{\tilde{k},\tilde{l}}}\\
\cdot\frac{W_{N}\sof{A_{x}+\frac{\Delta+1}{2}\cdot\ssof{\hat{a}_{0}+\hat{b}_{0}}}\,W_{N}\sof{A_{y}+\frac{\Delta+1}{2}\cdot\ssof{\hat{a}_{0}+\hat{b}_{0}}}}{W_{N}\sof{A_{x}+\hat{a}_{0}+\hat{b}_{0}}\,W_{N}\sof{A_{y}}}\ ,
\end{multline}
and if the transition is accepted, we set $x=y$ and $C=C'$.
\end{itemize}

\subsubsection{Intermediate and Final Sub-Worm Cycle Moves}\label{sssec:intermedsubcyclemove}
If a sub-worm cycle has successfully started, then $C$ is an intermediate configuration which contains three external fields, and the next move will therefore be either of type \ref{it:posdircont}. or type \ref{it:negdircont}. from the guide in Sec.~\ref{ssec:intspacesubwormalgo}. In any case, the only possible moves correspond to $\of{N-1}$ different choices for the internal space index $b\in\cof{1,\ldots,N}\setminus\cof{a}$, so that the move choice probability is given by
\[
p_{s}\of{C\to C'}\,=\,\frac{1}{N-1}\ .
\]
As before, $C'$ refers to the configuration that is obtained if the selected move is carried out. The move-choice probability for the inverse move is given by
\[
p_{s}\of{C'\to C}\,=\,\frac{1}{N-1}\,\of{\frac{1-p_{t}\,\delta_{y,x_{0}}}{2\,d}}^{\delta_{b,\tilde{b}_{0}}}\,\of{\frac{1-p_{t}\,\delta_{x,x_{0}}}{2\,d}}^{\delta_{b,\tilde{a}_{0}}}\ ,
\]
where the additional factors come from taking into account that if $b=\tilde{b}_{0}$ or $b=\tilde{a}_{0}$, then $C'$ will no longer be an intermediate sub-worm cycle configuration like $C$, and therefore a new sub-worm cycle would have to be started either from $x$ or from $y$, in order to get back to $C$. Defining again
\[
\tilde{k}\,=\,\ucases{k_{x,\nu}^{a\,b}\,,\ \nu>0\\k_{y,\abs{\nu}}^{a\,b}\,,\ \nu<0}\quad,\quad \tilde{l}\,=\,\ucases{l_{x,\nu}^{a\,b}\,,\ \nu>0\\l_{y,\abs{\nu}}^{a\,b}\,,\ \nu<0}\ ,
\]
and
\[
\Delta\,=\,\sabs{\tilde{k}+1}-\sabs{\tilde{k}}\ ,
\]
the reduced transition probability is in this case
\[
P_{r}\of{C\to C'}\,=\,\min\of{1,\,r}\ ,
\]
with:
\begin{itemize}
\item if $b\neq\tilde{b}_{0}$ and $b\neq\tilde{a}_{0}$:
\begin{multline}
r\,=\,\frac{w_{\beta}\ssof{\tilde{k}+1,\tilde{l}}}{w_{\beta}\ssof{\tilde{k},\tilde{l}}}\,\frac{W_{N}\sof{A_{x}+\hat{\tilde{b}}_{0}+\frac{\Delta+1}{2}\cdot\ssof{\hat{a}+\hat{b}}}}{W_{N}\sof{A_{x}+\hat{\tilde{b}}_{0}+\hat{a}}}\\
\cdot\frac{W_{N}\sof{A_{y}+\hat{\tilde{a}}_{0}+\frac{\Delta+1}{2}\cdot\ssof{\hat{a}+\hat{b}}}}{W_{N}\sof{A_{y}+\hat{\tilde{a}}_{0}+\hat{a}}}\ ,
\end{multline}
in which case, if the transition is accepted, we set $a=b$ and $C=C'$,

\item otherwise, if $b=\tilde{b}_{0}$:
\begin{multline}
r\,=\,\frac{1-p_{t}\,\delta_{y,x_{0}}}{2\,d}\,\frac{w_{\beta}\ssof{\tilde{k}+1,\tilde{l}}}{w_{\beta}\ssof{\tilde{k},\tilde{l}}}\,\frac{W_{N}\sof{A_{x}+\hat{\tilde{b}}_{0}+\hat{a}+\frac{\Delta-1}{2}\cdot\ssof{\hat{a}+\hat{b}}}}{W_{N}\sof{A_{x}+\hat{\tilde{b}}_{0}+\hat{a}}}\\
\cdot\frac{W_{N}\sof{A_{y}+\hat{\tilde{a}}_{0}+\frac{\Delta+1}{2}\cdot\ssof{\hat{a}+\hat{b}}}}{W_{N}\sof{A_{y}+\hat{\tilde{a}}_{0}+\hat{a}}}
\end{multline}
and if the transition is accepted, we set $x=y$ and $C=C'$,

\item and finally if $b=\tilde{a}_{0}$:
\begin{multline}
r\,=\,\frac{1-p_{t}\,\delta_{x,x_{0}}}{2\,d}\,\frac{w_{\beta}\ssof{\tilde{k}+1,\tilde{l}}}{w_{\beta}\ssof{\tilde{k},\tilde{l}}}\,\frac{W_{N}\sof{A_{x}+\hat{\tilde{b}}_{0}+\frac{\Delta+1}{2}\cdot\ssof{\hat{a}+\hat{b}}}}{W_{N}\sof{A_{x}+\hat{\tilde{b}}_{0}+\hat{a}}}\\
\cdot\frac{W_{N}\sof{A_{y}+\hat{\tilde{a}}_{0}+\hat{a}+\frac{\Delta-1}{2}\cdot\ssof{\hat{a}+\hat{b}}}}{W_{N}\sof{A_{y}+\hat{\tilde{a}}_{0}+\hat{a}}}\ ,
\end{multline}
where, if the transition is accepted, we set $y=x$ and $C=C'$.
\end{itemize}

\subsubsection{End of an ISSW Update}\label{sssec:endissw}
If $C$ is a configuration that contains a source/sink-pair where both external fields are located on the same site, so that $x=x_{0}$, it is proposed with probability $p_{t}$ that the pair is removed from the system and the ISSW update therefore terminates (step \ref{it:isendq2}. from the step-by-step description in Sec.~\ref{ssec:intspacesubwormalgo}). This is just the inverse move from the one described in Sec.~\ref{sssec:startissw}, and the reduced transition probability is therefore given by  
\[
P_{r}\of{C\to C'}\,=\,\min\of{1,\,\frac{1}{p_{t}\,V\,N\,\of{N-1}}\,\frac{W_{N}\ssof{A_{x}}}{W_{N}\ssof{A_{x}+\hat{a}_{0}+\hat{b}_{0}}}}\ .
\]
If the move is accepted, the system is again in a configuration that contributes to the ordinary partition function $Z_{Q}$, defined above in \eqref{eq:cpnpartf2a}.

\section{Results}\label{sec:results}
Here we present the results for some tests that we applied to our new algorithm, as well as a discussion of its efficiency. Although our algorithm works in arbitrary dimensions, all results presented here correspond to the $\of{1+1}$ dimensional case. 

\subsection{Crosscheck of Code}\label{ssec:crosscheck}
In order to test the correctness of our internal space sub-worm algorithm, we applied it to the dual formulation \eqref{eq:cpnpartf1} of the auxiliary $\Un{1}$ version of the $\CPn{N-1}$ model for $N=10$, and reproduced some of the results for  $E$, $\xi_{G}$ and $\chi_{m}$ (see \eqref{eq:avenergy}, \eqref{eq:smcorrlen}, \eqref{eq:magsusc} respectively), presented in \cite{Flynn}, where an over-heatbath algorithm was used to simulate \eqref{eq:laction1} in terms of the original configuration variables $z\of{x}$. The data is shown in Table~\ref{tbl:crosschecku1}. We also included corresponding results for the ordinary worm algorithm, applied to the alternative dual formulation \eqref{eq:cpnpartf3}, that was first proposed in \cite{Gattringer1}. As can be seen, the results for all three algorithms agree within error bounds (one-sigma).\\

\begin{table}[!h]
\centering
{
\small
    \begin{tabular}{ r l c l l l l }
    \hline
    $L$ & $\hspace{-1pt}\beta/N$ &  & \hspace{2pt}sweeps & $\hspace{12pt}E$ & $\hspace{10pt}\xi_{G}$ & $\hspace{10pt}\chi_{m}$\\ \hline
    $72$ & $0.8$ & \cite{Flynn} & $\phantom{1}80$M & $0.6670232(7)$ & $\phantom{1}4.5992(12)$ & $\phantom{1}28.0595(18)$\\
     & & $Z_{A}$, eq. \eqref{eq:cpnpartf1} & $\phantom{1}10$M & $0.6670204(79)$ & $\phantom{1}4.5847(71)$ & $\phantom{1}28.0408(144)$\\
     & & $\tilde{Z}_{A}$, eq. \eqref{eq:cpnpartf3} & $\phantom{1}10$M & $0.6670156(55)$ & $\phantom{1}4.5978(48)$ & $\phantom{1}28.0560(102)$\\ \hline
    $96$ & $0.85$ & \cite{Flynn} & $\phantom{1}80$M & $0.6222715(5)$ & $\phantom{1}6.3926(20)$ & $\phantom{1}46.863(4)$\\
     & & $Z_{A}$, eq. \eqref{eq:cpnpartf1} & $\phantom{1}10$M & $0.6222678(52)$ & $\phantom{1}6.3837(138)$ & $\phantom{1}46.862(35)$\\
     & & $\tilde{Z}_{A}$, eq. \eqref{eq:cpnpartf3} & $\phantom{1}10$M & $0.6222657(42)$ & $\phantom{1}6.3833(93)$ & $\phantom{1}46.854(24)$\\ \hline
    $136$ & $0.9$ & \cite{Flynn} & $\phantom{1}80$M & $0.5838365(3)$ & $\phantom{1}8.815(4)$ & $\phantom{1}78.202(8)$\\
     & & $Z_{A}$, eq. \eqref{eq:cpnpartf1} & $\phantom{10}5$M & $0.5838427(53)$ & $\phantom{1}8.792(32)$ & $\phantom{1}78.185(95)$\\
     & & $\tilde{Z}_{A}$, eq. \eqref{eq:cpnpartf3} & $\phantom{1}10$M & $0.5838345(28)$ & $\phantom{1}8.822(19)$ & $\phantom{1}78.220(51)$\\ \hline
    $184$ & $0.95$ & \cite{Flynn} & $100$M & $0.55026689(20)$ & $12.095(6)$ & $130.707(15)$\\
     & & $Z_{A}$, eq. \eqref{eq:cpnpartf1} & $\phantom{10}2.5$M & $0.55027714(527)$ & $12.028(82)$ & $130.414(283)$\\
     & & $\tilde{Z}_{A}$, eq. \eqref{eq:cpnpartf3} & $\phantom{10}8.5$M & $0.55026634(242)$ & $12.031(37)$ & $130.549(122)$\\
  \end{tabular}
}
  \caption{The table shows a comparison of some high precision results for the average energy $E$ from \eqref{eq:avenergy}, the correlation length $\xi_{G}$ from \eqref{eq:smcorrlen} and the magnetic susceptibility $\chi_{m}$ from \eqref{eq:magsusc} for $\CPn{9}$ on a $\of{1+1}$ dimensional $L\times L$ lattice, which were obtained in \cite[(see Tab. 6.3)]{Flynn} using the standard lattice action \eqref{eq:laction1} and an "over-heat bath algorithm", the corresponding results obtained with our "internal space sub-worm algorithm", using the partition function $Z_{A}$ from equation \eqref{eq:cpnpartf1}, and the ones obtained with the ordinary worm algorithm, using the partition function $\tilde{Z}_{A}$ from \eqref{eq:cpnpartf3} which was introduced in \cite{Gattringer1}.}
  \label{tbl:crosschecku1}
\end{table}

\begin{figure}[!h]
\centering
\begin{minipage}[t]{0.48\linewidth}
\centering
{\small $\avof{E}$ vs. $\beta/N$ for $Z_{Q}$}\\[-1pt]
\includegraphics[width=\linewidth]{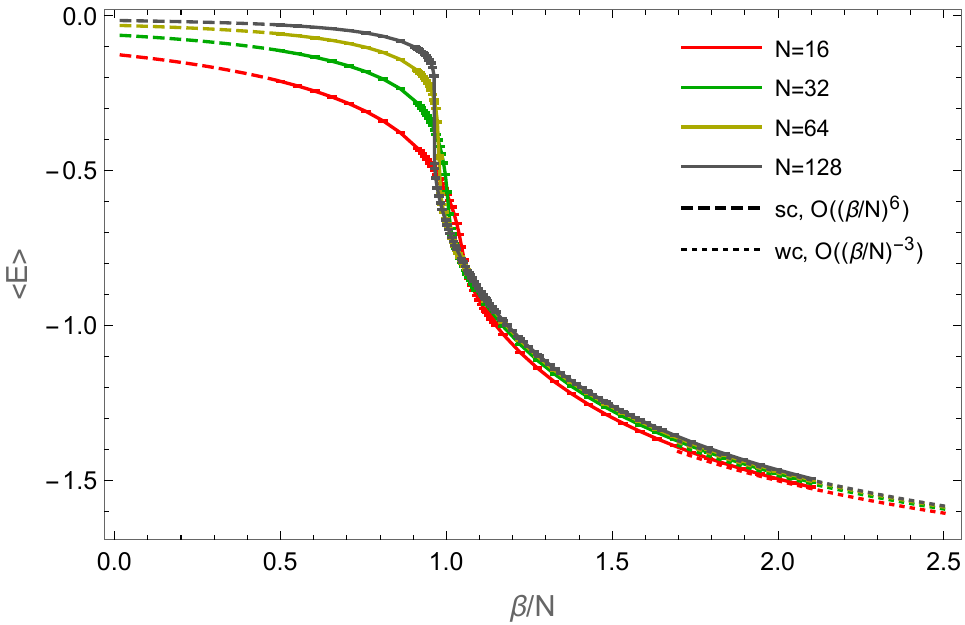}
\end{minipage}\hfill
\begin{minipage}[t]{0.495\linewidth}
\centering
{\small $C_{E}$ vs. $\beta/N$ for $Z_{Q}$}\\[-1pt]
\includegraphics[width=\linewidth]{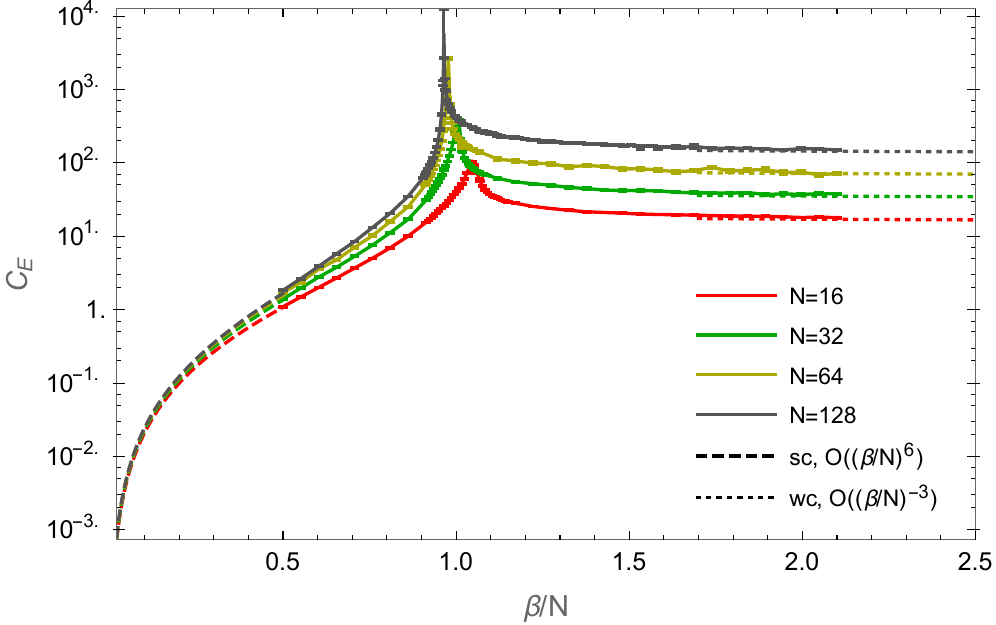}
\end{minipage}
\caption{The figure shows the average energy \eqref{eq:avenergy} (left) and specific heat \eqref{eq:specificheat} (right) for the $\CPn{N-1}$ model described by \eqref{eq:cpnpartf2a} (quartic action) in $\ssof{1+1}$ dimensions, for a system with volume $V=12^2$ and $N\in\cof{16,32,64,128}$. The dashed and dotted lines correspond to the analytic strong and weak coupling results respectively \cite{Vecchia}, which show excellent agreement with our numerical data generated with our internal space sub-worm algorithm (the partition function \eqref{eq:cpnpartf4} leads to exactly the same results with the algorithm described in Sec.~\ref{ssec:ordworm}).}
  \label{fig:energyvsbvarnquartic}
\end{figure}

\begin{figure}[!h]
\centering
\begin{minipage}[t]{0.48\linewidth}
\centering
{\small $\avof{E}$ vs. $\beta/N$ for $Z_{A}$}\\[-1pt]
\includegraphics[width=\linewidth]{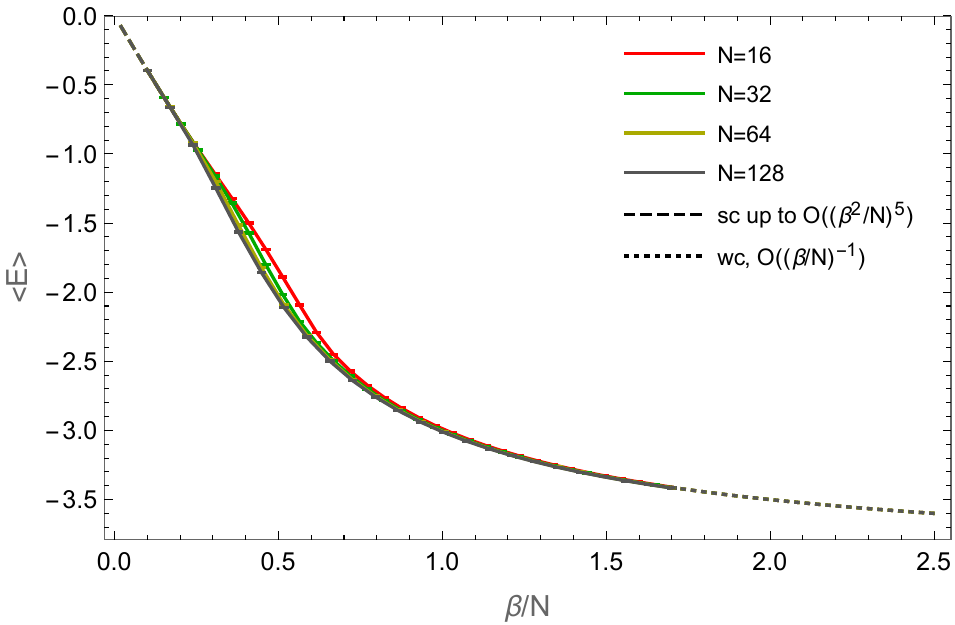}
\end{minipage}\hfill
\begin{minipage}[t]{0.495\linewidth}
\centering
{\small $C_{E}$ vs. $\beta/N$ for $Z_{A}$}\\[2pt]
\includegraphics[width=\linewidth]{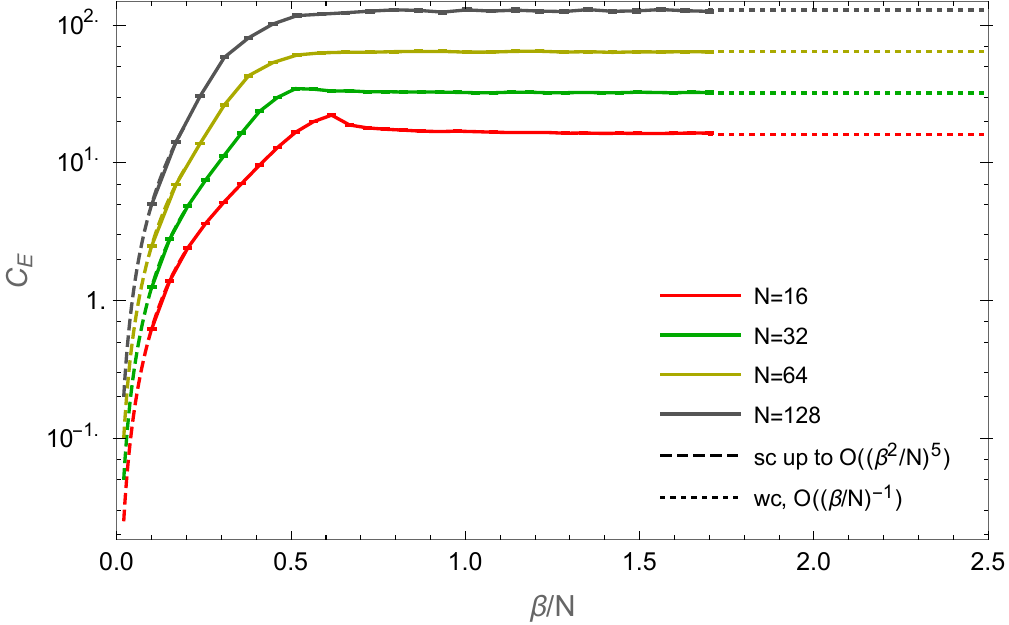}
\end{minipage}
\caption{As Fig. \ref{fig:energyvsbvarnquartic}, the figure shows the average energy density (left) and specific heat (right), but this time for the $\CPn{N-1}$ model described by \eqref{eq:cpnpartf1} (auxiliary $\Un{1}$ action). Again the system is $\ssof{1+1}$ dimensional, has volume $V=12^2$ and the number of flavors is $N\in\cof{16,32,64,128}$. The dashed and dotted lines correspond to the analytic strong and weak coupling results respectively \cite{Vecchia}, which show again excellent agreement with the numerical data generated with our internal space sub-worm algorithm (also here, the partition function \eqref{eq:cpnpartf3} leads to exactly the same results with the algorithm from Sec.~\ref{ssec:ordworm}, applied to the $\CPn{N-1}$ partition function of \cite{Gattringer1}).}
  \label{fig:energyvsbvarnauxu1}
\end{figure}

In Figures \ref{fig:energyvsbvarnquartic} and \ref{fig:energyvsbvarnauxu1} we show the average energy \eqref{eq:avenergy} and the corresponding specific heat \eqref{eq:specificheat} as a function of $\beta/N$ for different values of $N$ in a $\of{1+1}$ dimensional system of volume $V=12^2$, together with the analytic strong and weak coupling expansions provided in \cite{Vecchia}. Figure \ref{fig:energyvsbvarnquartic} shows the results for the quartic version $Z_{Q}$ from \eqref{eq:cpnpartf2a} and figure \ref{fig:energyvsbvarnauxu1} shows the ones for the auxiliary $\Un{1}$ version $Z_{A}$ from \eqref{eq:cpnpartf1}. In both cases, the numerical data nicely interpolates between the strong and weak coupling predictions (and perfectly matches them in the corresponding regions). Note the dramatic peak in the specific heat with the quartic version $Z_{Q}$.

\subsection{Efficiency}\label{ssec:efficiency}
A measure of the efficiency of an algorithm is given by its critical dynamical exponent, usually called $z$. This $z$ is observable-dependent and tells one how the integrated auto-correlation time $\tau_{int}$ of this observable scales as a function of the correlation length $\xi_{G}$, where one assumes a dependency of the form $\tau_{int}\propto \xi_{G}^{z}$. For local, Metropolis-type algorithms one usually finds $z\sim 2$. A value of $z>0$  means that the algorithm suffers from critical slowing down, i.e. if one changes the lattice size while keeping the physical size of the system fixed, the computational cost for achieving a predefined accuracy for the measurements will not just depend linearly on the lattice volume $V=L^d$ ($\propto$ number of degrees of freedom), but grow even faster like $\propto L^{d+z}$. However, as long as $z\lesssim 1$ the slowing down is usually considered to be weak.\\

\begin{figure}[!h]
\centering
\begin{minipage}[t]{0.49\linewidth}
\centering
\includegraphics[width=\linewidth]{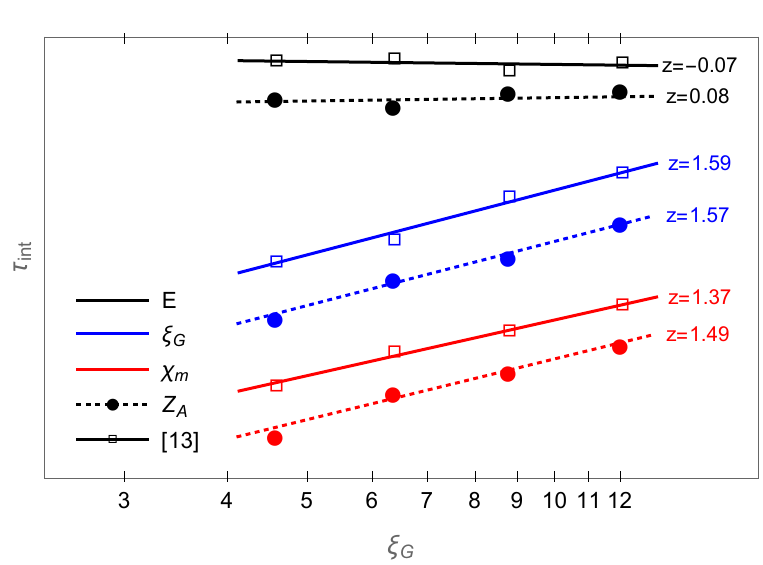}
\end{minipage}\hfill
\begin{minipage}[t]{0.49\linewidth}
\centering
\includegraphics[width=\linewidth]{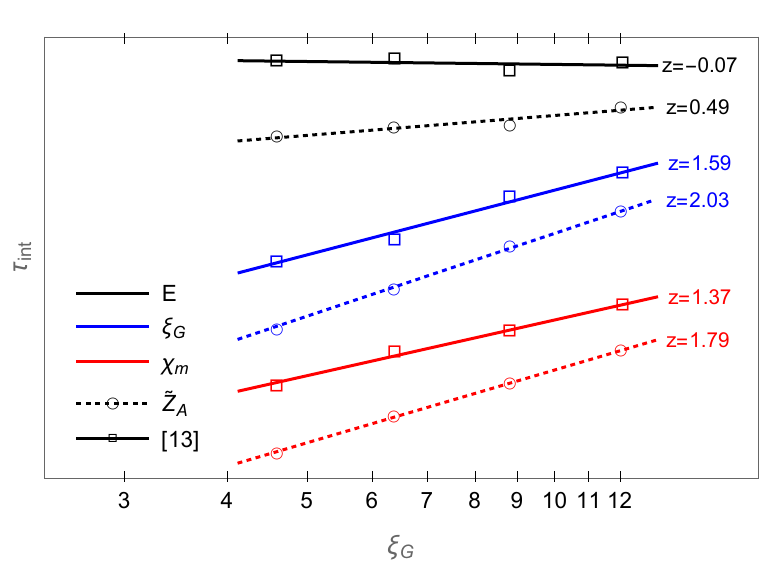}
\end{minipage}
\caption{Log-log plots of $\sim\tau_{int}$ vs. $\xi_{G}$ at fixed $L/\xi_{G}\approx 15$ for the three observables $E$, $\xi_{G}$ and $\chi_{m}$ and the three different algorithms from Table~\ref{tbl:crosschecku1} for the $\CPn{9}$ model. The $\tau_{int}$ values for the different observables are re-scaled by arbitrary constants, to fit in a common figure. The straight lines correspond to fits of the form $\tau_{int}\sim \xi_{G}^{z}$, where $z$ is the dynamical critical exponent. The left-hand figure shows the data for our ISSW algorithm applied to $Z_{A}$ from \eqref{eq:cpnpartf1}, together with the data for the over-heat bath algorithm from \cite{Flynn}. The right-hand figure shows the data for the ordinary worm algorithm applied to $\tilde{Z}_{A}$ from \cite{Gattringer1}, again together with the over-heat bath results from \cite{Flynn}. All three algorithms show similar behavior.}
  \label{fig:dyncritexp}
\end{figure}

In Fig.~\ref{fig:dyncritexp} we compare the dynamical critical exponents $z$ for the three observables $E$, $\xi_{G}$ and $\chi_{m}$ for the internal space sub-worm algorithm, the ordinary worm algorithm and the over-heat bath algorithm from \cite{Flynn} for the $\CPn{9}$ model at fixed $L/\xi_{G}\approx 15$. The values of $z$ were obtained from the data in Table~\ref{tbl:crosschecku1}, knowing that for an observable $\mathcal{O}$, 
\[
\sigma_{\bar{\mathcal{O}}}^{2}\,\approx\,\frac{\sigma_{\mathcal{O}}^{2}}{N_{m}}\,2\,\tau_{\mathcal{O}}^{int}\ ,\label{eq:sigmatau}
\]
where $\bar{\mathcal{O}}=\frac{1}{N_{m}}\sum\limits_{i=1}^{N_{m}}\mathcal{O}_{i}$, with $\mathcal{O}_{i}$ being the $i$-th out of $N_{m}$ measurements, $\sigma_{\mathcal{O}}^{2}=\frac{1}{N_{m}}\sum\limits_{i=1}^{N_{m}}\of{\mathcal{O}_{i}-\bar{\mathcal{O}}}^{2}$ and $\sigma^{2}_{\bar{\mathcal{O}}}=\savof{\bar{\mathcal{O}}^{2}}-\savof{\bar{\mathcal{O}}}^{2}$, so that $\epsilon_{\mathcal{O}}=\sqrt{\sigma_{\bar{\mathcal{O}}}^2}$ is the statistical error in $\bar{\mathcal{O}}$. If one now divides both sides of \eqref{eq:sigmatau} by $\bar{\mathcal{O}}^2$ in order to get rid of the explicit dependency of $\mathcal{O}$ on the lattice volume, and assumes that for fixed $L/\xi_{G}$, the resulting $\hat{\sigma}_{\mathcal{O}}^{2}\,=\,\sigma_{\mathcal{O}}^{2}/\bar{\mathcal{O}}^{2}$ becomes approximately independent of the lattice size for sufficiently large lattices, one finds that
\[
\tau^{int}_{\mathcal{O}}\,\propto\,\frac{\epsilon_{\mathcal{O}}^{2}\,\tilde{N}_{m}}{\bar{\mathcal{O}}^{2}}\ ,
\]
where $\tilde{N}_{m}$ is the effective number of measurements, which for non-local observables (like $\xi_{G}$ and $\chi_{m}$) is equal to $N_{m}$ but for local observables (like $E$) picks up an extra volume-factor, i.e. $\tilde{N}_{m}=V\cdot N_{m}$. As can be seen, the three algorithms give rise to very similar critical exponents for the three observables under consideration.\\

\begin{figure}[!h]
\centering
\begin{minipage}[t]{0.465\linewidth}
\centering
{\small $\tau_{int}\fof{E}$ vs. $\xi_{G}/L$ for $Z_{Q}$}\\[-5pt]
\includegraphics[width=\linewidth]{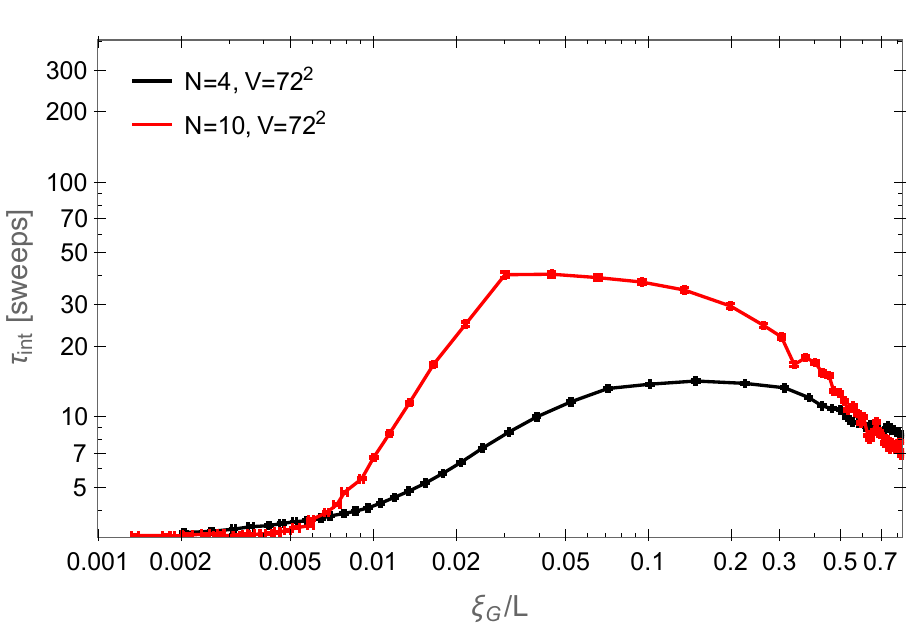}
\end{minipage}\hfill
\begin{minipage}[t]{0.475\linewidth}
\centering
{\small $\tau_{int}\fof{\chi_{m}}$ vs. $\xi_{G}/L$ for $Z_{Q}$}\\[-5pt]
\includegraphics[width=\linewidth]{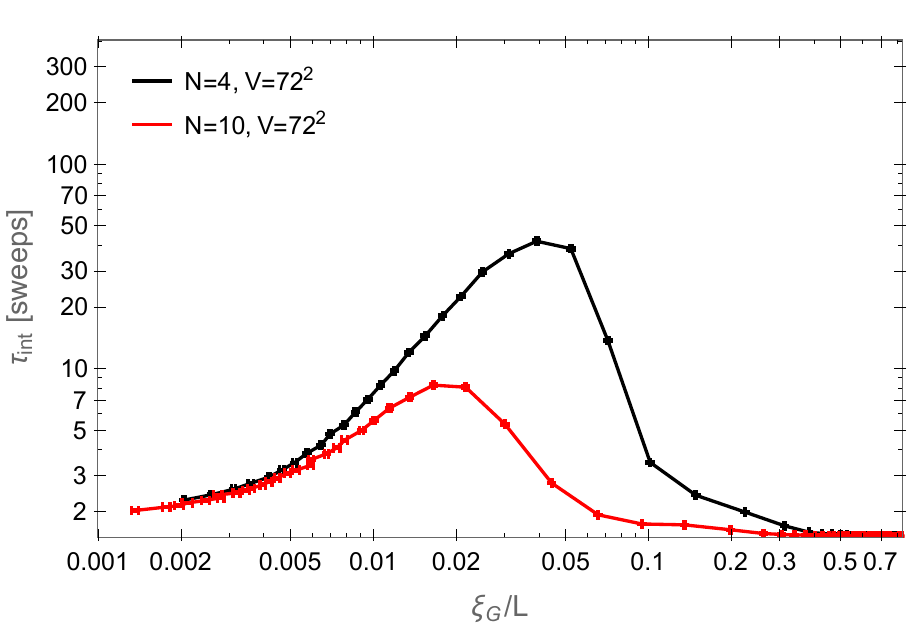}
\end{minipage}\\[10pt]
\begin{minipage}[t]{0.475\linewidth}
\centering
{\small $\tau_{int}\fof{n}$ vs. $\xi_{G}/L$ for $Z_{Q}$}\\[-5pt]
\includegraphics[width=\linewidth]{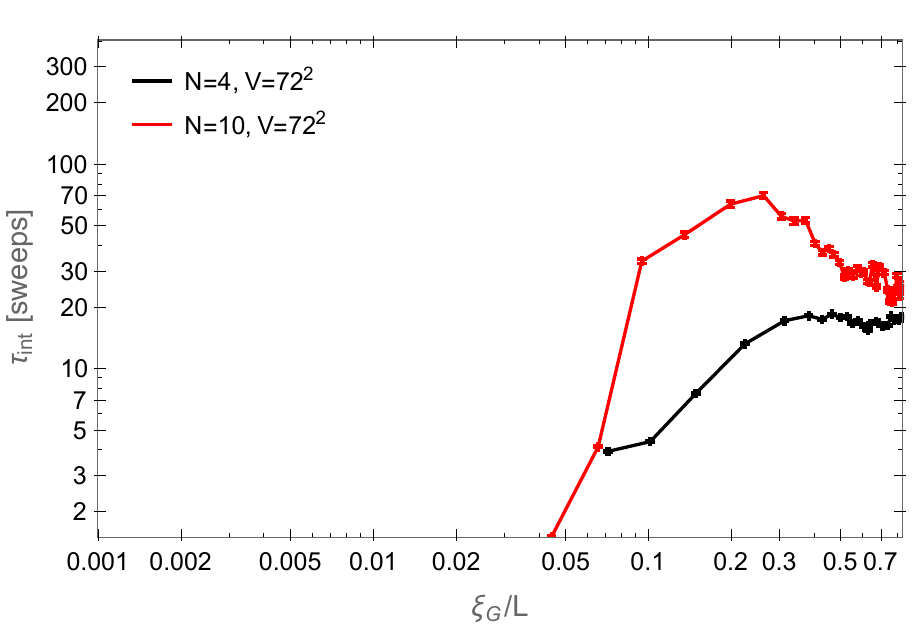}
\end{minipage}\hfill
\begin{minipage}[t]{0.465\linewidth}
\centering
{\small $\xi_{G}$ vs. $\beta/N$ for $Z_{Q}$}\\[3pt]
\includegraphics[width=\linewidth]{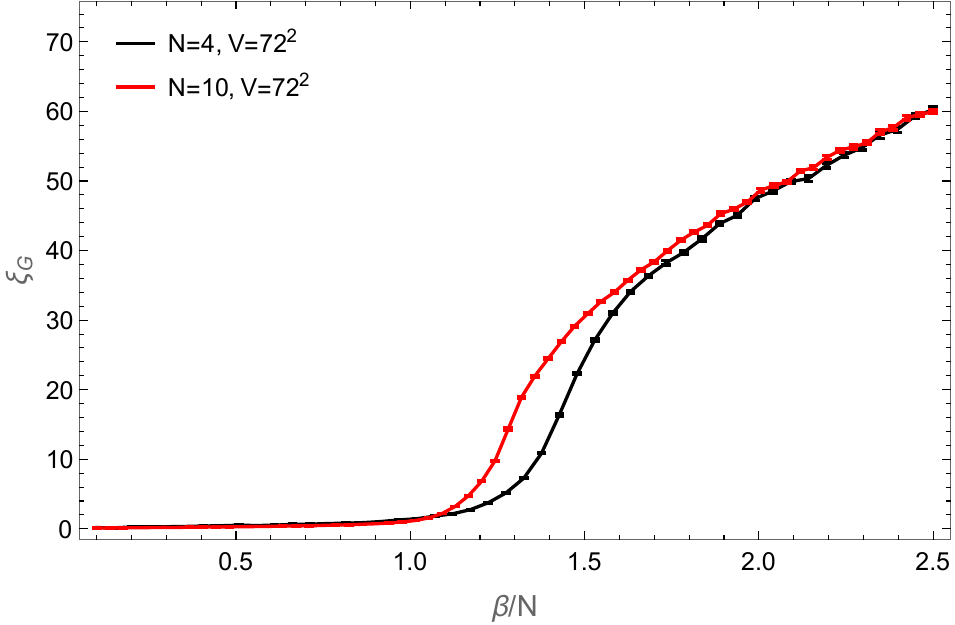}
\end{minipage}
\caption{The top-left figure shows the integrated auto-correlation time for the average energy \eqref{eq:avenergy} as a function of the second moment correlation length \eqref{eq:smcorrlen} (divided by the linear system size $L$) for the quartic version of the $\CPn{N-1}$ model, described by \eqref{eq:cpnpartf2a}. The systems have volume $V=72^2$ and $N\in\cof{4,10}$. As can be seen, the auto-correlation time quickly develops a plateau or even decreases as function of increasing correlation length as soon as one leaves the short-distance regime. The top-right and bottom-left figures show in a similar manner the integrated auto-correlation times for the magnetic susceptibility \eqref{eq:magsusc} and for one of the conserved charges \eqref{eq:chargedens}. The bottom-right figure shows how the second moment correlation length $\xi_{G}$ changes as a function of the coupling $\beta$.}
  \label{fig:autocorrtime2}
\end{figure}

\begin{figure}[!h]
\centering
\begin{minipage}[t]{0.465\linewidth}
\centering
{\small $\tau_{int}\fof{E}$ vs. $\xi_{G}/L$ for $Z_{A}$}\\[-5pt]
\includegraphics[width=\linewidth]{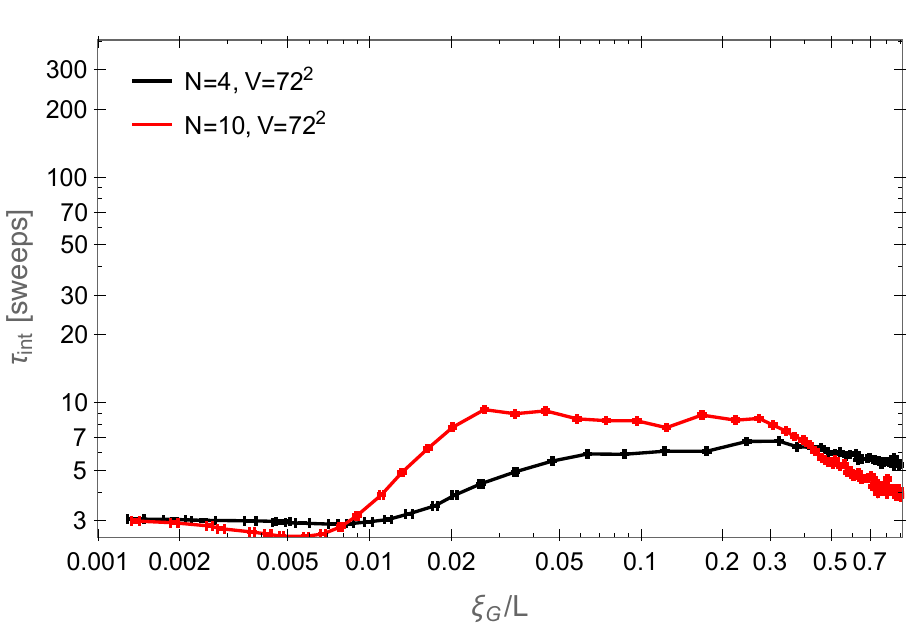}
\end{minipage}\hfill
\begin{minipage}[t]{0.475\linewidth}
\centering
{\small $\tau_{int}\fof{\chi_{m}}$ vs. $\xi_{G}/L$ for $Z_{A}$}\\[-5pt]
\includegraphics[width=\linewidth]{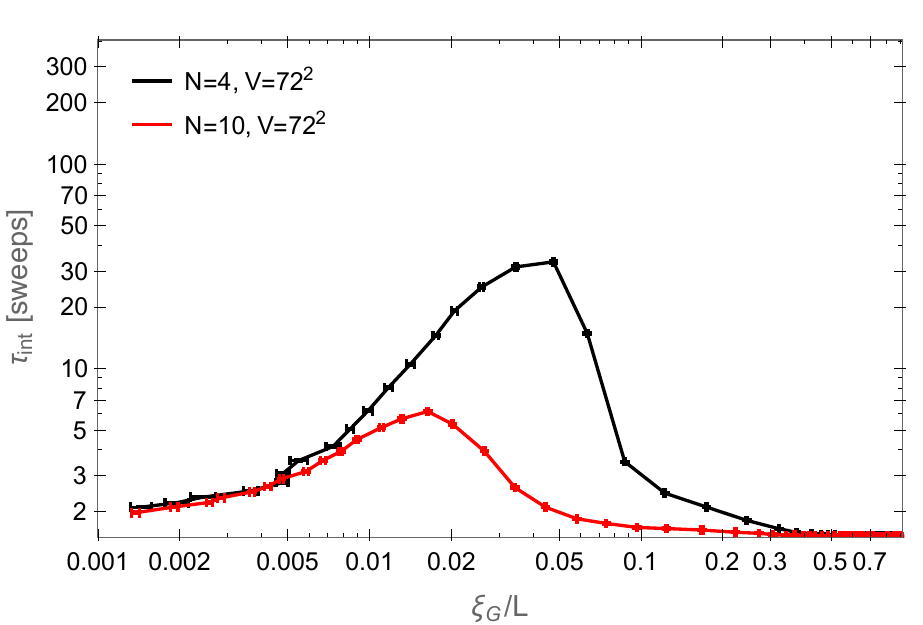}
\end{minipage}\\[10pt]
\begin{minipage}[t]{0.475\linewidth}
\centering
{\small $\tau_{int}\fof{n}$ vs. $\xi_{G}/L$ for $Z_{A}$}\\[-5pt]
\includegraphics[width=\linewidth]{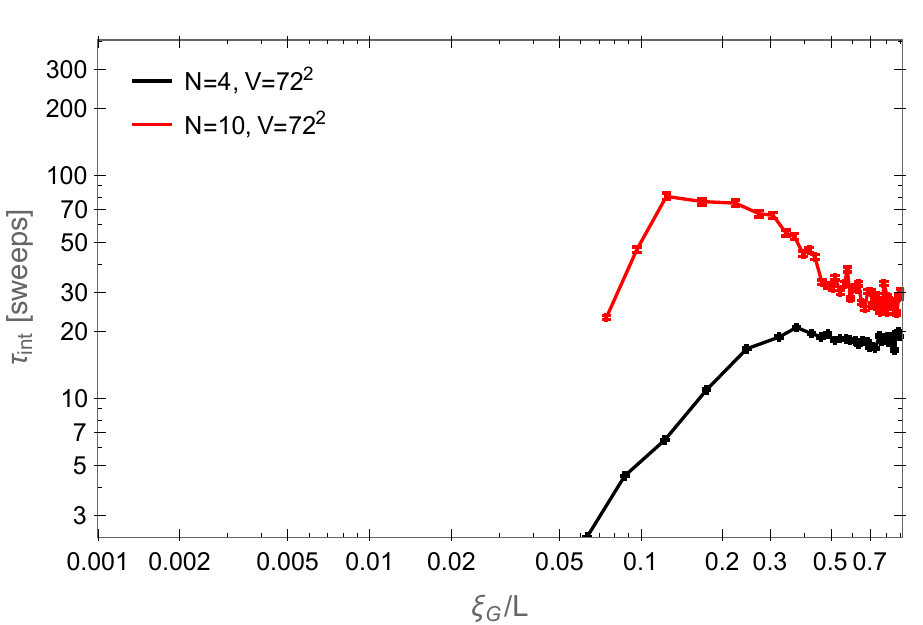}
\end{minipage}\hfill
\begin{minipage}[t]{0.465\linewidth}
\centering
{\small $\xi_{G}$ vs. $\beta/N$ for $Z_{A}$}\\[3pt]
\includegraphics[width=\linewidth]{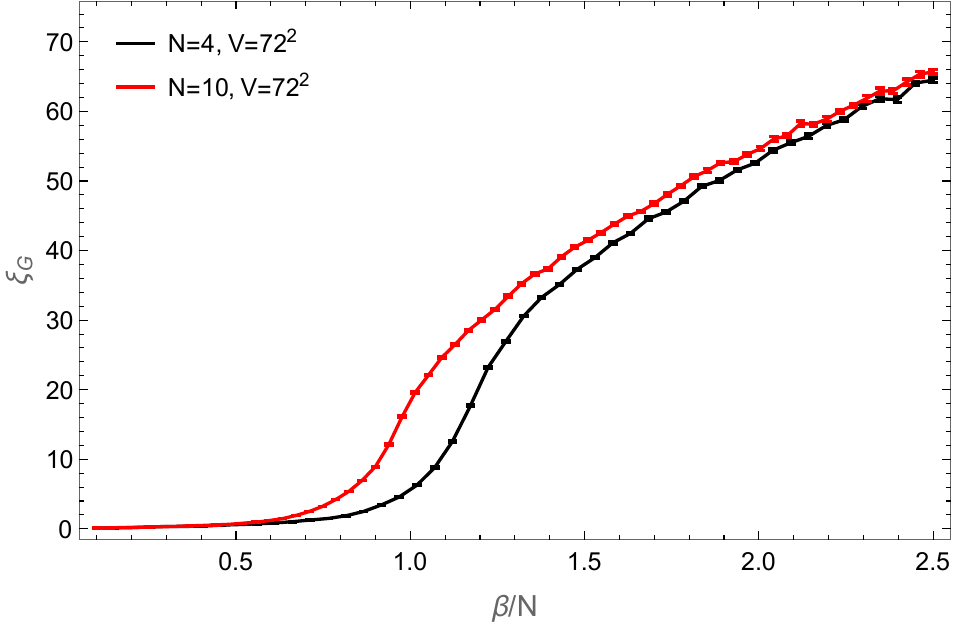}
\end{minipage}
\caption{Same as Fig.~\ref{fig:autocorrtime2} but for the auxiliary $\Un{1}$ version \eqref{eq:cpnpartf1} of the $\CPn{N-1}$ partition function. The top-left figure shows the integrated auto-correlation time for the average energy \eqref{eq:avenergy} as a function of the second moment correlation length \eqref{eq:smcorrlen} (divided by the linear system size $L$). Again, the systems have volume $V=72^2$ and $N\in\cof{4,10}$. As can be seen, also here the auto-correlation time quickly develops a plateau or even decreases as function of increasing correlation length as soon as one leaves the short-distance regime. The top-right and bottom-left figures show in a similar manner the integrated auto-correlation times for the magnetic susceptibility \eqref{eq:magsusc} and for one of the conserved charges \eqref{eq:chargedens}. The bottom-right figure shows how the second moment correlation length $\xi_{G}$ changes as a function of the coupling $\beta$.}
  \label{fig:autocorrtime2u1}
\end{figure}

We also determined the integrated auto-correlation time $\tau_{int}$ directly for the average energy \eqref{eq:avenergy}, the magnetic susceptibility \eqref{eq:magsusc} and the charge densities \eqref{eq:chargedens} (as long as all $\mu_{i}$ are set to zero, all charge densities are equivalent), for the two systems described by \eqref{eq:cpnpartf2a} and \eqref{eq:cpnpartf1} when updated with the ISSW algorithm. The results are shown in Fig.~\ref{fig:autocorrtime2} and Fig.~\ref{fig:autocorrtime2u1}, respectively, as functions of $\xi_{G}/L$, where $L$ is the linear system size (here $L=72$) and for $N=4,10$. The auto-correlation times are given in units of sweeps, where we define a sweep, to consist of a fixed number of worms, so that the average number of local updates that are processed during these worms, equals the number of degrees of freedom in the system (in our case $\#\text{d.o.f}=d\,N^2\,V$, where $d$ is the number of space-time dimensions, $N$ is the number of flavors and $V$ the system volume). As can be seen, for both discretizations of the $\CPn{N-1}$ model, the maximal integrated auto-correlation time $\tau_{int}$ for both, the average energy and the average charge density, increases with increasing $N$, while for $\chi_{m}$, it decreases. However, for all three quantities, the value of $\xi_{G}/L$, at which this maximum in $\tau_{int}$ occurs, decreases with increasing $N$.\\

\begin{figure}[!h]
\centering
\begin{minipage}[t]{0.485\linewidth}
\centering
{\small $\tau_{int}\fof{E}$ vs. $\xi_{G}/L$ for $Z_{Q}$}\\[-5pt]
\includegraphics[width=\linewidth]{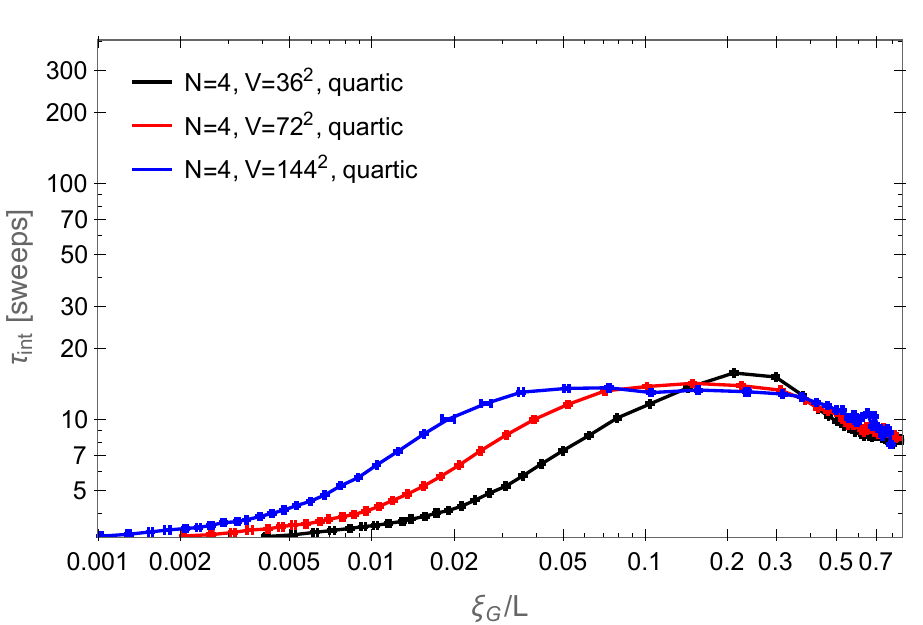}
\end{minipage}\hfill
\begin{minipage}[t]{0.495\linewidth}
\centering
{\small $\tau_{int}\fof{\chi_{m}}$ vs. $\xi_{G}/L$ for $Z_{Q}$}\\[-5pt]
\includegraphics[width=\linewidth]{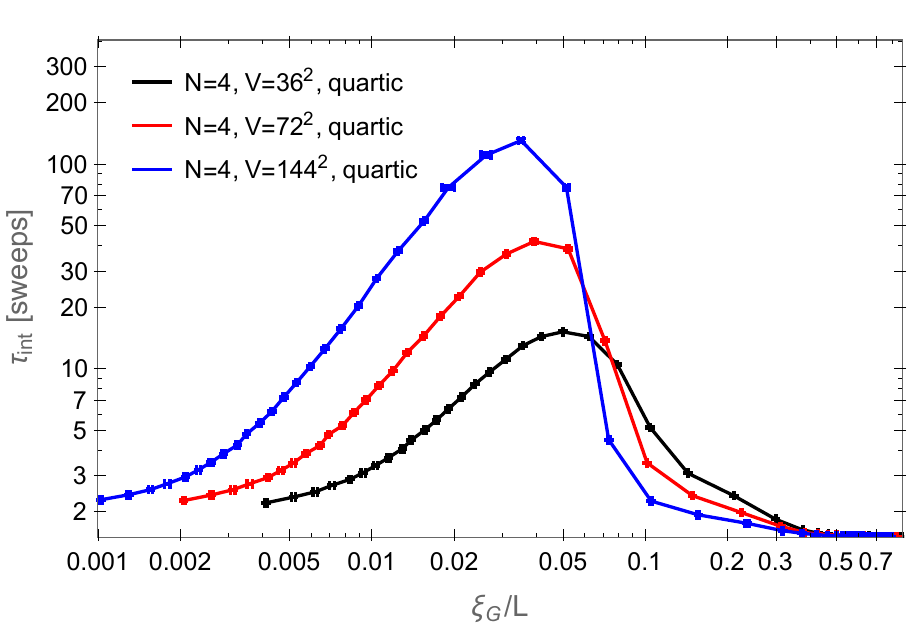}
\end{minipage}\\[10pt]
\begin{minipage}[t]{0.495\linewidth}
\centering
{\small $\tau_{int}\fof{n}$ vs. $\xi_{G}/L$ for $Z_{Q}$}\\[-5pt]
\includegraphics[width=\linewidth]{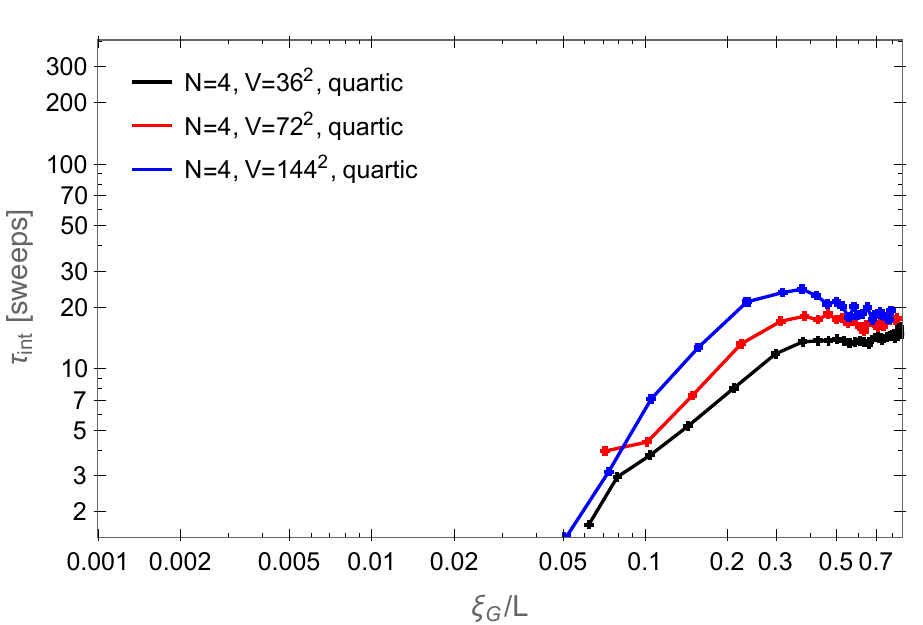}
\end{minipage}\hfill
\begin{minipage}[t]{0.485\linewidth}
\centering
{\small $\xi_{G}$ vs. $\beta/N$ for $Z_{Q}$}\\[3pt]
\includegraphics[width=\linewidth]{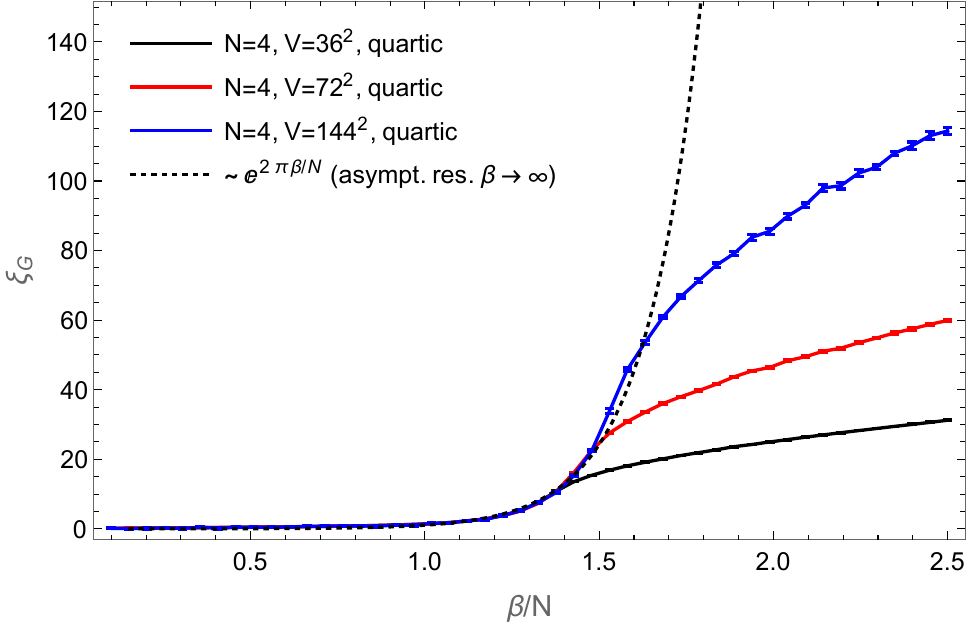}
\end{minipage}
\caption{Same as Fig.~\ref{fig:autocorrtime2} but for different volumes $V=36^2,72^2,144^2$ with $N=4$ kept fixed. The top-left figure shows again the integrated auto-correlation time for the average energy \eqref{eq:avenergy} as a function of $\xi_{G}/L$ where $\xi_{G}$ is the second moment correlation length \eqref{eq:smcorrlen}. As can be seen, for the average energy, the maximal auto-correlation time does not grow with increasing system size. However, for the magnetic susceptibility \eqref{eq:magsusc} (top-right) and the charge density \eqref{eq:chargedens} (bottom-left), this is obviously not true. The bottom-right figure shows how the second moment correlation length $\xi_{G}$ changes as a function of the coupling $\beta$ and the dotted black line corresponds to the expected asymptotic behavior for $\of{\beta\rightarrow \infty}$ in an infinite system. Of course, as our system is finite with periodic boundary conditions, the second moment correlation length $\xi_{G}$ cannot grow arbitrarily and stops following the asymptotic curve as soon as $\xi_{G}/L$ exceeds $\frac{1}{4}$ - $\frac{1}{2}$ .}
  \label{fig:autocorrtime3}
\end{figure}

\begin{figure}[!h]
\centering
\begin{minipage}[t]{0.495\linewidth}
\centering
{\small $\tau_{int}\fof{E}$ vs. $\xi_{G}/L$ for $\tilde{Z}_{Q}$}\\[-5pt]
\includegraphics[width=\linewidth]{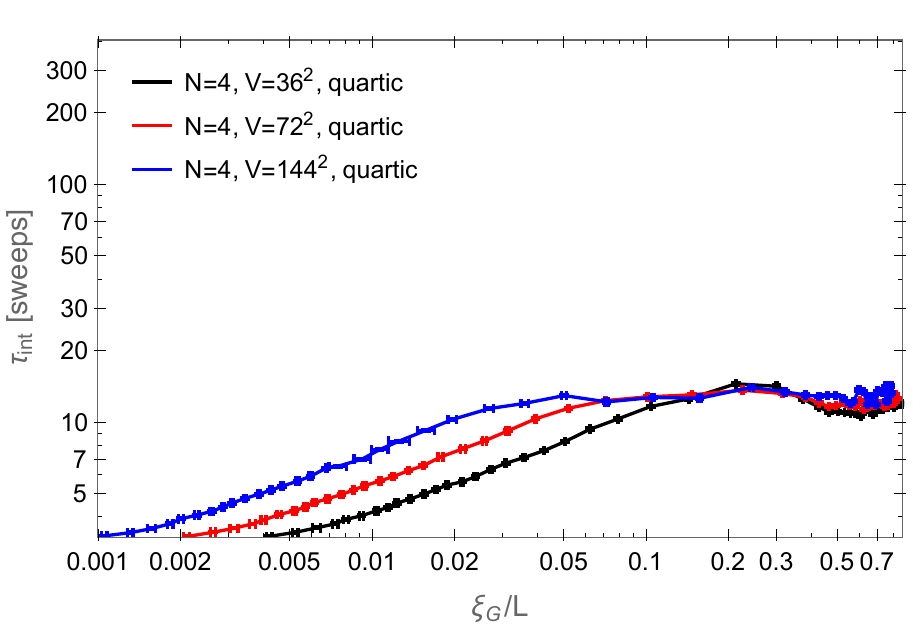}
\end{minipage}\hfill
\begin{minipage}[t]{0.485\linewidth}
\centering
{\small $\tau_{int}\fof{\chi_{m}}$ vs. $\xi_{G}/L$ for $\tilde{Z}_{Q}$}\\[-5pt]
\includegraphics[width=\linewidth]{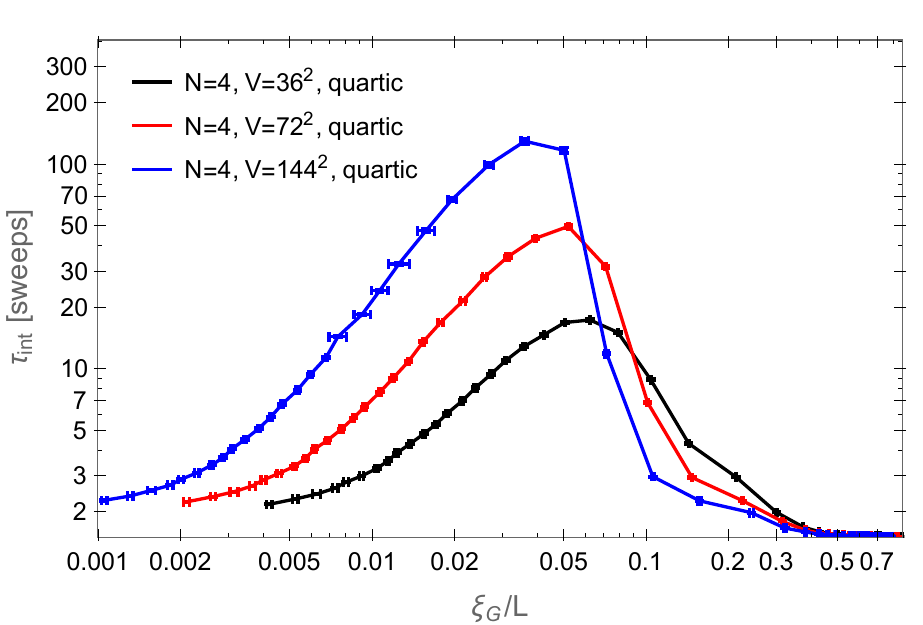}
\end{minipage}\\[10pt]
\begin{minipage}[t]{0.495\linewidth}
\centering
{\small $\tau_{int}\fof{n}$ vs. $\xi_{G}/L$ for $\tilde{Z}_{Q}$}\\[-5pt]
\includegraphics[width=\linewidth]{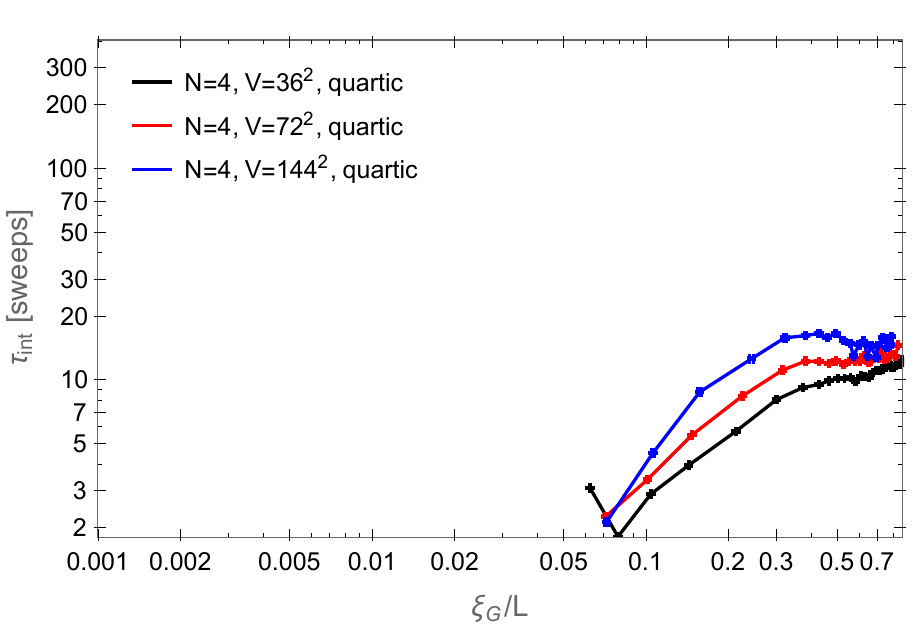}
\end{minipage}\hfill
\begin{minipage}[t]{0.485\linewidth}
\centering
{\small $\xi_{G}$ vs. $\beta/N$ for $\tilde{Z}_{Q}$}\\[3pt]
\includegraphics[width=\linewidth]{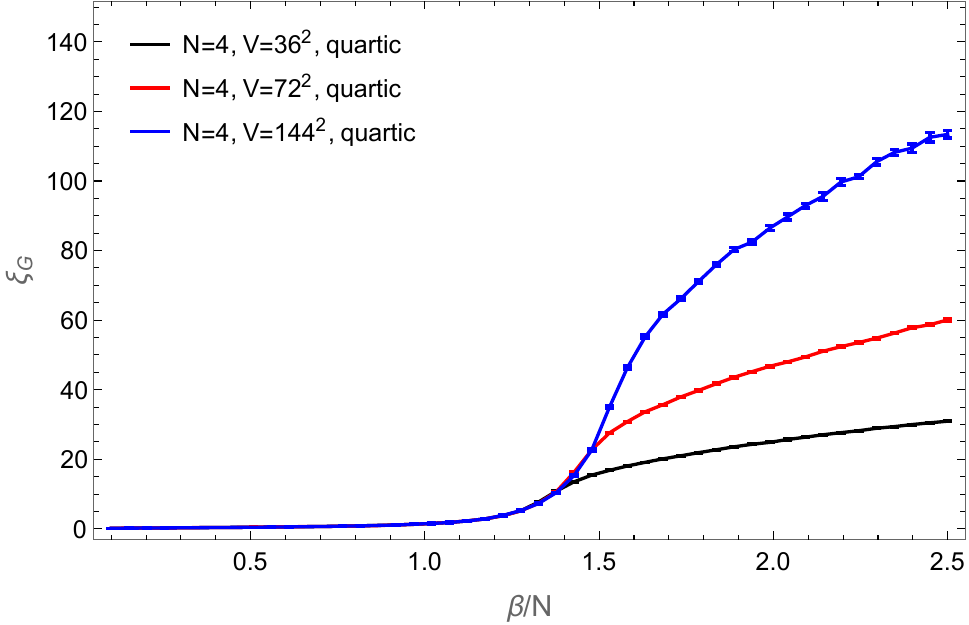}
\end{minipage}
\caption{Same as Fig.~\ref{fig:autocorrtime3} but for the flux variable representation $\tilde{Z}_{Q}$ from \eqref{eq:cpnpartf4}. The integrated auto-correlation times for the average energy (top-left), magnetic susceptibility (top-right) and the charge density (bottom-left), behave essentially in the same way as the corresponding auto-correlation times for the flux variable representation $Z_{Q}$ from \eqref{eq:cpnpartf2a}, shown in Fig.~\ref{fig:autocorrtime3}. Finally, the second moment correlation length $\xi_{G}$ (bottom-right) is identical to that obtained with $Z_{Q}$, as should be case since $\xi_{G}$ is a physical, algorithm-independent quantity.}
  \label{fig:autocorrtime4}
\end{figure}

\begin{figure}[!h]
\centering
\begin{minipage}[t]{0.485\linewidth}
\centering
{\small $\tau_{int}\fof{E}$ vs. $\xi_{G}/L$ for $\tilde{Z}_{Q}$}\\[-5pt]
\includegraphics[width=\linewidth]{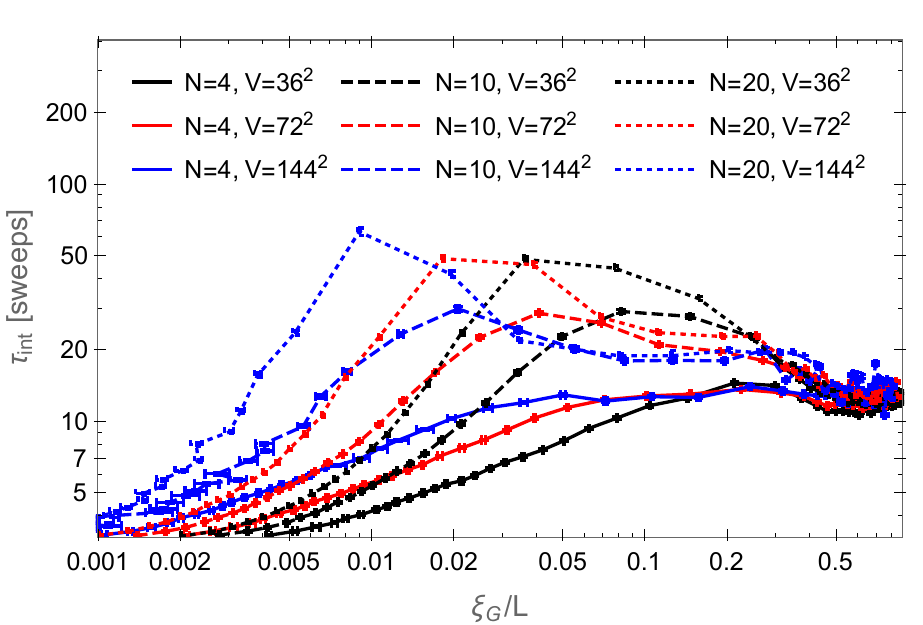}
\end{minipage}\hfill
\begin{minipage}[t]{0.495\linewidth}
\centering
{\small $\tau_{int}\fof{\chi_{m}}$ vs. $\xi_{G}/L$ for $\tilde{Z}_{Q}$}\\[-5pt]
\includegraphics[width=\linewidth]{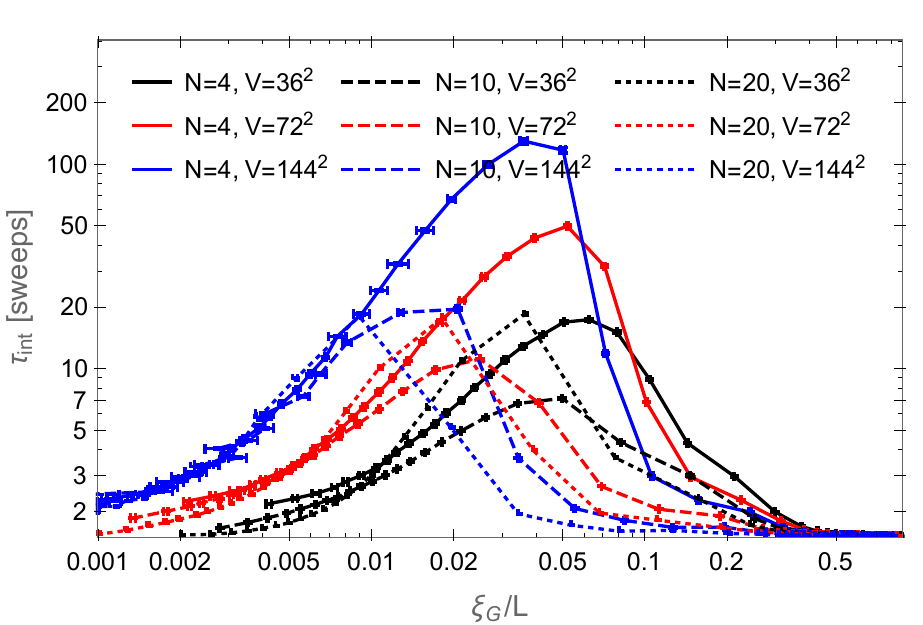}
\end{minipage}\\[10pt]
\begin{minipage}[t]{0.495\linewidth}
\centering
{\small $\tau_{int}\fof{n}$ vs. $\xi_{G}/L$ for $\tilde{Z}_{Q}$}\\[-5pt]
\includegraphics[width=\linewidth]{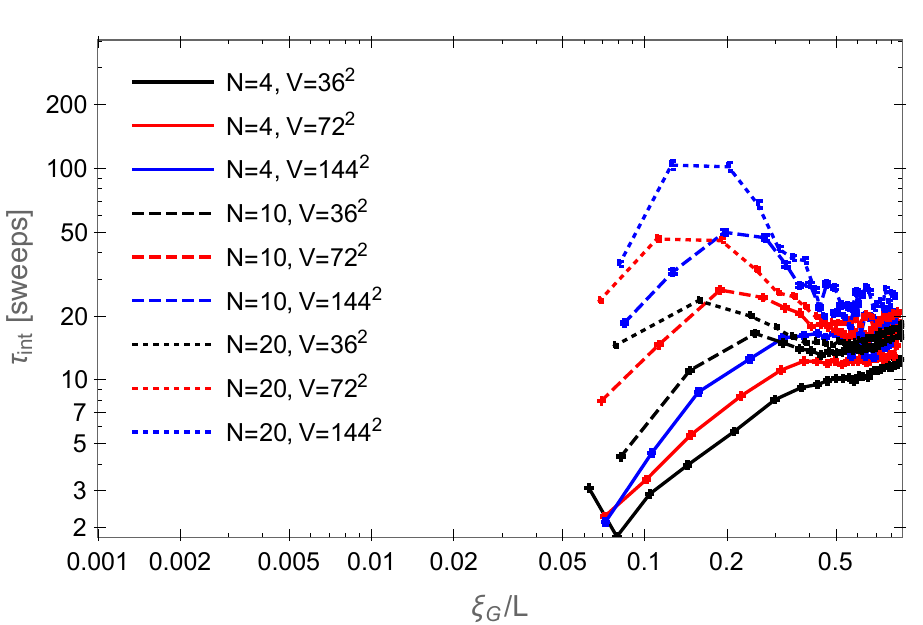}
\end{minipage}\hfill
\begin{minipage}[t]{0.485\linewidth}
\centering
{\small $\xi_{G}$ vs. $\beta/N$ for $\tilde{Z}_{Q}$}\\[3pt]
\includegraphics[width=\linewidth]{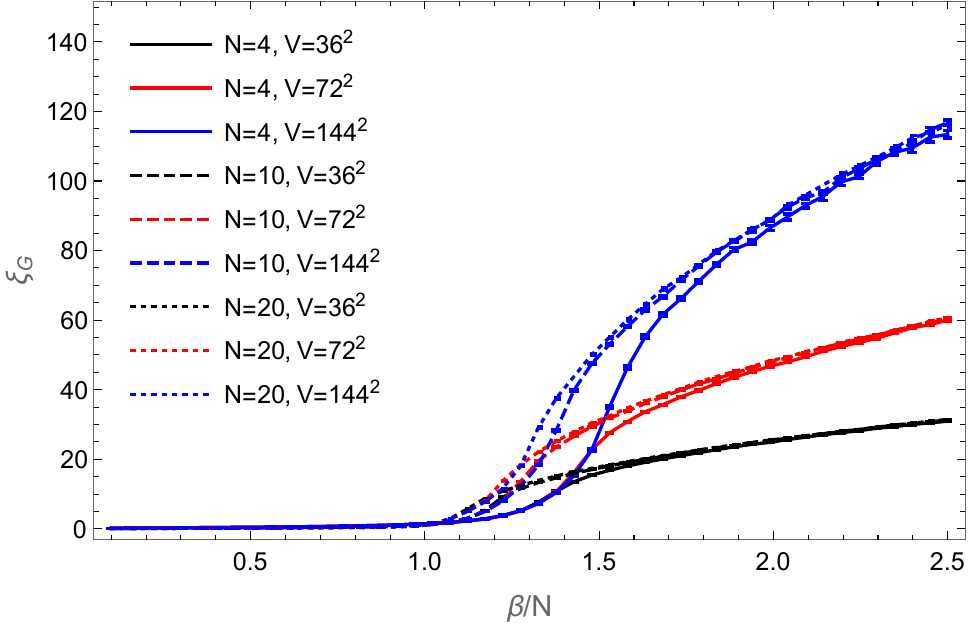}
\end{minipage}
\caption{Same as Fig.~\ref{fig:autocorrtime4} but including results also for $N=10,20$. For the average energy (top-left) and the magnetic susceptibility (top-right), the value of $\xi_{G}/L$ ($L$ being the linear system size) at which $\tau_{int}$ peaks, drops with increasing $N$. For the charge density (bottom-left), the value of $\xi_{G}/L$ after which the integrate auto-correlation time stops to increase with increasing $\xi_{G}$ also drops, but the the correlation time always grows with increasing system size.}
  \label{fig:autocorrtime5}
\end{figure}

Figures \ref{fig:autocorrtime3} and \ref{fig:autocorrtime4} show again, for the average energy \eqref{eq:avenergy}, the magnetic susceptibility \eqref{eq:magsusc} and the charge density \eqref{eq:chargedens}, the corresponding integrated auto-correlation times as functions of $\xi_{G}/L$, but this time for fixed $N=4$ but three different volumes $V=36^2,72^2,144^2$. As the Monte Carlo dynamics of the two systems described by $Z_{Q}$ and $Z_{A}$ from \eqref{eq:cpnpartf2a} and \eqref{eq:cpnpartf1}, respectively, are very similar when using the ISSW algorithm, we show in Fig.~\ref{fig:autocorrtime3} just the results for $Z_{Q}$ and in Fig.~\ref{fig:autocorrtime4} for comparison the corresponding results for $\tilde{Z}_{Q}$ from \eqref{eq:cpnpartf4}, simulated with the ordinary worm algorithm. As was already the case with $Z_{A}$ and $\tilde{Z}_{A}$ in Fig.~\ref{fig:dyncritexp}, the ISSW and the ordinary worm algorithm also perform very similarly when applied to $Z_{Q}$ and $\tilde{Z}_{Q}$, respectively. The figures also illustrate, that the value of $z$ can strongly depend on the value of $\xi_{G}/L$ at which $z$ is determined. Figure \ref{fig:autocorrtime5} is similar to Fig.~\ref{fig:autocorrtime4} but contains also data for $N=10,20$, showing that for $\tilde{Z}_{Q}$, the dependency of $\tau_{int}$ on $N$ is similar to what is shown in Fig.~\ref{fig:autocorrtime2} for the case of $Z_{Q}$.\\

For one quantity the behavior of the integrated auto-correlation time, as shown in Figs.~\ref{fig:autocorrtime2u1}-\ref{fig:autocorrtime5}, is a bit puzzling: for the magnetic susceptibility $\chi_{m}$, $\tau_{int}$ initially grows with increasing $\xi_{G}/L$, but then decreases again quite fast, well below $\xi_{G}/L\sim 0.25$ where finite size effects should start to become dominant. The same would happen with the integrated auto-correlation time for the correlation length $\xi_{G}$. The reason is that these two quantities are defined in terms of the two-point function \eqref{eq:cpntwopointfunc3}, and in contrast to the average energy or the charge density, which are just averages of properties of individual configurations that contribute to the partition function $Z$, an element $\savof{\phi^{a\,b}\of{x}\phi^{b\,a}\of{y}}$ of the two-point function is not simply the average of a property of configurations that contribute to $Z$, but rather a kind of reweighting factor between configurations that contribute to $Z$ and configurations that contribute to $Z^{a\,b}_{2}\of{x,y}$ (partition function for the system in the presence of an external source $\phi^{a\,b}$ at $x$ and an external sink $\phi^{b\,a}$ at $y$): it is related to, how "expensive" the changes required to configurations contributing to $Z$  are, in order to incorporate the external source-sink pair at $x$ and $y$. The problem is now that within our dual formalism, due to the strictly imposed conservation laws (see last part of Sec.~\ref{ssec:conservedcurrnts}), we cannot simply measure these reweighting factors on individual configurations that contribute to $Z$, but instead have to determine them stochastically by measuring the average frequency by which the worm algorithm reaches a configuration that contributes to $Z_{2}\of{x,y}$, when starting in a configuration that contributes to $Z$. This leads to some additional statistical noise in the measurements of the two-point function which has nothing to do with the true de-correlation rate for the observable. As long as the number of worms per sweep is large (as is the case for small $\beta$ where the correlation length on the lattice is small), the effect of this additional statistical noise is small, but as soon as the average worm length increases (together with the lattice correlation length), the number of worms that are required to process a sweep (as defined above) decreases and with it the statistics and therefore the accuracy for individual measurements of the two-point function. It then becomes impossible to determine the true integrated auto-correlation time for observables that depend on the two-point function by summing up the correlations between different measurements. On can also not just increase the number of worms per sweep in order to reduce the statistical noise in the individual measurements; as the worms do not just sample the two-point function but also update the system, this would also increase the number of sweeps between measurements and make it impossible to measure auto-correlation times which are shorter than this number of sweeps between measurements.\\

Another observable that would be of great interest is the topological susceptibility. In Monte Carlo simulations of the $\CPn{N-1}$ model in terms of the standard configuration variables $z\of{x}$, the system can tunnel only slowly between different topological sectors, which causes very long auto-correlation times for the topological charge and the topological susceptibility \cite{Flynn}. As in the dualization process that lead us to the flux-variable partition functions \eqref{eq:cpnpartf1}, \eqref{eq:cpnpartf2a}, \eqref{eq:cpnpartf3} and \eqref{eq:cpnpartf4}, the $z$-fields are integrated out analytically in order to obtain the weights for flux-variable configurations, and this integration runs over all possible configurations and covers therefore also all possible topological sectors, every single configuration in terms of flux-variables contains already contributions from all possible topological sectors. A tunneling between different sectors is therefore no longer necessary, which is why we think that if one could incorporate the measurement of the topological charge and topological susceptibility into our dual simulations, critical slowing down should also be absent for these observables. Unfortunately, such measurements are quite involved in our dual framework: as shown in appendix \ref{sec:thetaterm}, in order to measure the topological charge and its susceptibility, one has to introduce new plaquette degrees of freedom which then also enter the constraints imposed on the $k$-variables. This  would give rise to interesting effects, but as the weights for the plaquette variables can be negative\cite{Gattringer1}, one has to deal with a sign problem if one wants to sample the plaquette variables by Monte Carlo.\\

Also Wilson and Polyakov loops of the local $\Un{1}$ gauge field modify the constraints imposed on the $k$-variables, but they do not require the introduction of additional degrees of freedom. They can be defined as ordinary observables that are measured on closed-worm configurations (c.f. \cite{Irving} to see how Wilson loops can be defined in terms of the original configuration variables $z\of{x}$).\\

\subsection{Large $N$ Limit for $Z_{Q}$}\label{ssec:largenlimit}
As already mentioned at the end of Sec. \ref{ssec:latticeformulation}, the authors of \cite{Vecchia2} found that in the limit $\of{N\rightarrow\infty}$, the quartic action version of the lattice $\CPn{N-1}$, i.e. the one that in terms of our dual variables is described by $Z_{Q}$ from \eqref{eq:cpnpartf2a}, develops in $\of{1+1}$ dimensions a first order transition between the strong $\ssof{\beta\lesssim\beta_{cr}}$ and the weak coupling phase $\ssof{\beta\gtrsim\beta_{cr}}$, where $\beta_{cr}/N\approx 0.956$ (see \cite[Sec. 4]{Vecchia2}).\\

\begin{figure}[!h]
\centering
\begin{minipage}[t]{0.485\linewidth}
\centering
{\small $\avof{E}$ vs. $\beta/N$ for $Z_{Q}$}\\[-1pt]
\includegraphics[width=\linewidth]{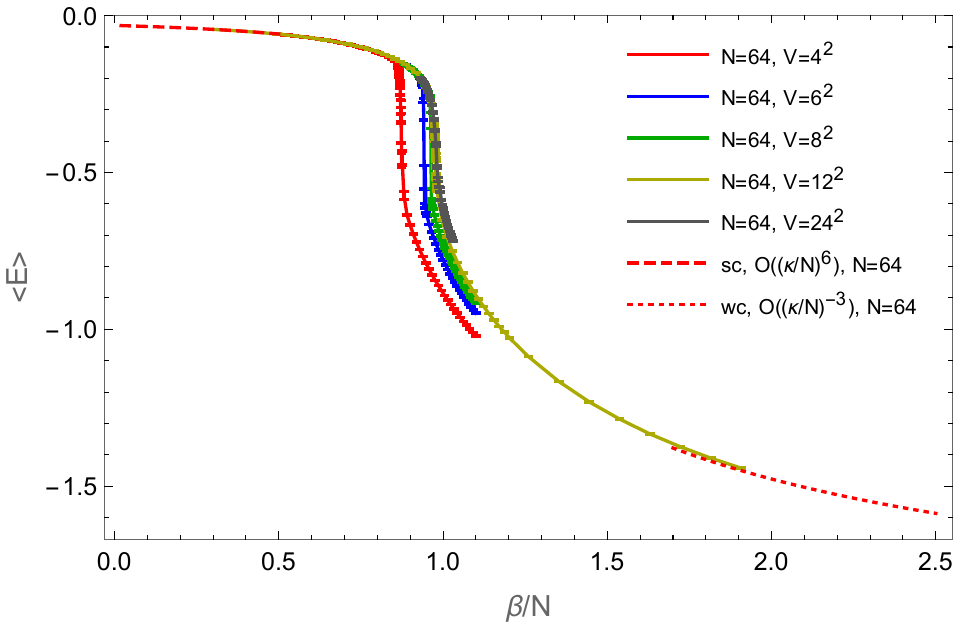}
\end{minipage}\hfill
\begin{minipage}[t]{0.495\linewidth}
\centering
{\small $C_{E}$ vs. $\beta/N$ for $Z_{Q}$}\\[1pt]
\includegraphics[width=\linewidth]{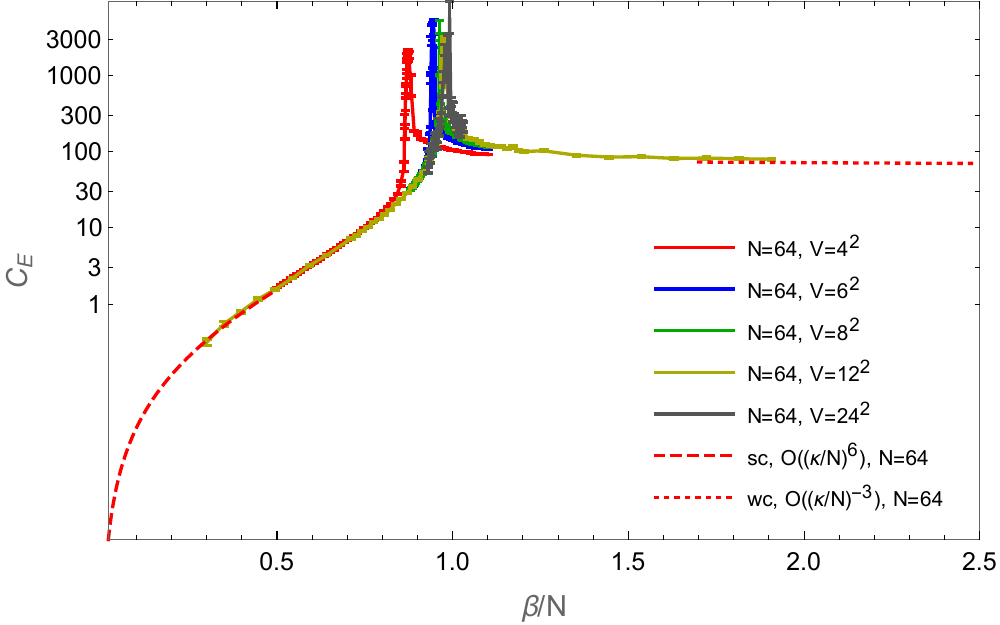}
\end{minipage}
\caption{The figure shows the average energy (left) and specific heat (right) for the $\CPn{N-1}$ model described by \eqref{eq:cpnpartf2a} (quartic action) in $\ssof{1+1}$ dimensions, with $N=64$ and for system sizes $V\in\cof{4^2,6^2,8^2,12^2,24^2}$. The red dashed and dotted lines correspond to the analytic strong and weak coupling results \cite{Vecchia}.}
  \label{fig:energyvsbvarsize}
\end{figure}

\begin{figure}[!h]
\centering
\begin{minipage}[t]{0.48\linewidth}
\centering
\includegraphics[width=\linewidth]{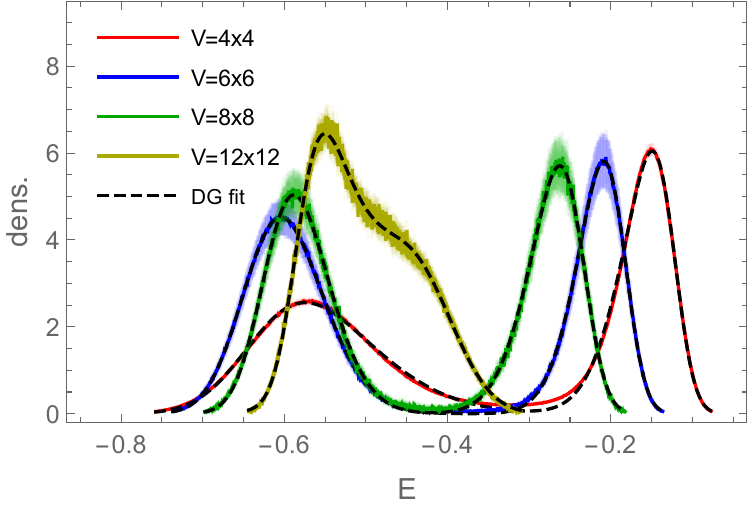}
\end{minipage}\hfill
\begin{minipage}[t]{0.48\linewidth}
\centering
\includegraphics[width=\linewidth]{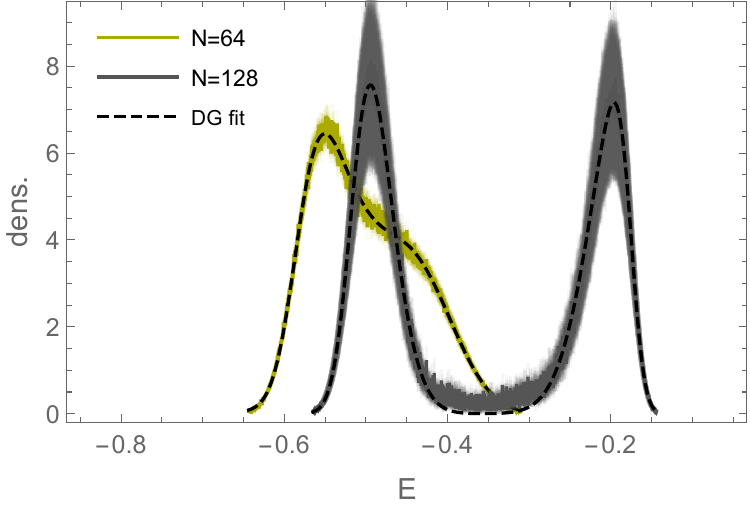}
\end{minipage}
\caption{The figures show the energy density distribution at the "pseudo-critical" point ($\beta/N$ near $1$) for the transition between the strong and the weak coupling phase for a $\ssof{1+1}$-dimensional system, described by the partition function \eqref{eq:cpnpartf2a} (quartic action). In the left-hand figure, the number $N$ of flavors is set to $N=64$ and the energy density is plotted for four different volumes between $V=4^2$ and $V=12^2$. The dotted black lines correspond to double-Gaussian fits. For $V<12^2$, the energy densities show a clear double-peak structure while for $V=12^2$, the two peaks start to merge. For even larger system sizes, the peaks would become indistinguishable. The right-hand figure shows the energy density for two systems, both of size $V=12^2$ but for different flavor numbers $N=64$ (yellow, same as yellow curve in the left-hand figure) and $N=128$ (gray). As can be seen, keeping the volume fixed but increasing the number of flavors enhances again the double-peak structure.}
  \label{fig:energydensdist}
\end{figure}

Also in our simulations of the system described by \eqref{eq:cpnpartf2a}, the "transition" between the weak and strong coupling regime becomes more pronounced when the number $N$ of flavors is increased, as can be seen in the left-hand part of Fig.~\ref{fig:energyvsbvarnquartic}, where the average energy is shown as a function of $\beta/N$ for a $\ssof{1+1}$ dimensional system of size $V=12^2$ for different $N$. However, if one keeps $N$ fixed and instead varies the system size, it turns out that the "transition" becomes \emph{smoother} if the system size is increased (see Figs.~\ref{fig:energyvsbvarsize} and \ref{fig:energydensdist}). Thus, the large $N$, thermodynamic behavior depends on the way the two limits are taken. A possible explanation for this behavior can be found by writing the Boltzmann factor in the following form:
\begin{multline}
\exp\bof{\beta\sum\limits_{x,\nu}\sabs{z^{\dagger}\of{x}\cdot z\of{x+\hat{\nu}}}^2}
\,=\\
\exp\bof{\beta\sum\limits_{x,\nu}\sum\limits_{a,b=1}^{N}r_{x}^{a}\,r_{x+\hat{\nu}}^{a}\,r_{x}^{b}\,r_{x+\hat{\nu}}^{b}\,\cos\of{\phi_{x}^{a}-\phi_{x+\hat{\nu}}^{a}-\phi_{x}^{b}+\phi_{x+\hat{\nu}}^{b}}}\ .\label{eq:qaction}
\end{multline}
In the strong coupling phase, $\beta$ is small and the entropy of the configuration variables dominates over the Boltzmann factor, no matter what the value of \eqref{eq:qaction} is. But with increasing $\beta$, the variation of the Boltzmann factor as a function of the configuration variables becomes more and more relevant and configurations which minimize the action become favored as soon as $\beta$ is large enough so that the Boltzmann factor can dominate over entropy for these configurations. In contrast to simple spin models, where the value of $\beta$ at which this change of dominance between entropy and Boltzmann weight happens marks the pseudo-critical point at which the system develops long range order (because the spatially ordered configurations are the ones that minimize the Euclidean lattice action), for the $\CPn{N-1}$ model (with quartic action) this is not necessarily the case. This can be seen by noting that there are two different (naive) ways to maximize the cosines in \eqref{eq:qaction} :
\begin{enumerate}[label=\roman*)]
\item\label{it:sporder} the first option is the usual one, where for each link $\of{x,\nu}$ and for each internal space index $a$ the angles $\phi_{x}^{a}$ and $\phi_{x+\hat{\nu}}^{a}$ are "in phase", so that all cosines in \eqref{eq:qaction} assume the value 1. In this case, the local $\Un{1}$ gauge-invariance gets "broken" and the system could develop true long-range order, provided the $r$-variables get ordered as well.
\item\label{it:intorder} The second possibility is to have on each site $x$ all the angles $\cof{\phi_{x}^{1}, \phi_{x}^{2}, \phi_{x}^{3}, \ldots}$ in phase, so that again all the cosines in \eqref{eq:qaction} assume the value $1$. This time however, the local $\Un{1}$ gauge-invariance is preserved as $\phi$-angles that correspond to different sites are still unrelated and there will be no true long-range order even if the $r$-variables get ordered.
\end{enumerate}
Depending on the system size and on the number of flavors $N$, it is either the "local internal space ordering" described in \ref{it:intorder} or the "real-space ordering" from \ref{it:sporder} which is cheaper, i.e. which requires a smaller change in entropy (coming from the reduction of the effective configuration space due to the ordering): for a small system with large $N$, the option \ref{it:sporder} will be cheaper in which the $\Un{1}$ gauge-invariance is broken (which is consistent with the finding in \cite{Vecchia2}), while for a large system with smaller $N$, the system will chose option \ref{it:intorder}, which preserves $\Un{1}$ gauge-invariance.\\
As for any fixed $N$, the transition between the strong and weak coupling region becomes smoother with increasing system size, it will not become a first order phase transition in the thermodynamic limit, so the infinite volume and infinite $N$ limits do not commute.\\

Of course \ref{it:sporder} and \ref{it:intorder} are two extremes: mixtures of the two orderings are also possible and in option \ref{it:sporder}, some permutations in the choice of the pairs $\of{a,b}$ for which $\phi_{x}^{a}$ is associated to $\phi_{x+\hat{\nu}}^{b}$ are allowed. But this heuristic description motivates our construction of a Monte Carlo algorithm decorrelating both types of order.\\

It should also be noted that the partition function, as it is a path integral, is always a sum over all possible orientations of the complex vectors $z\of{x}$ on each site $x$. By saying that the system puts some of the $\phi$-angles "in phase", we mean that configurations for which these angles are the same, give the dominant contribution to the integral over the $\phi$-angles.\\ 

\begin{figure}[!h]
\centering
\begin{minipage}[t]{0.46\linewidth}
\centering
{\small $\avof{E}$ vs. $\beta/N$ for $Z_{Q}$}\\[0pt]
\includegraphics[width=\linewidth]{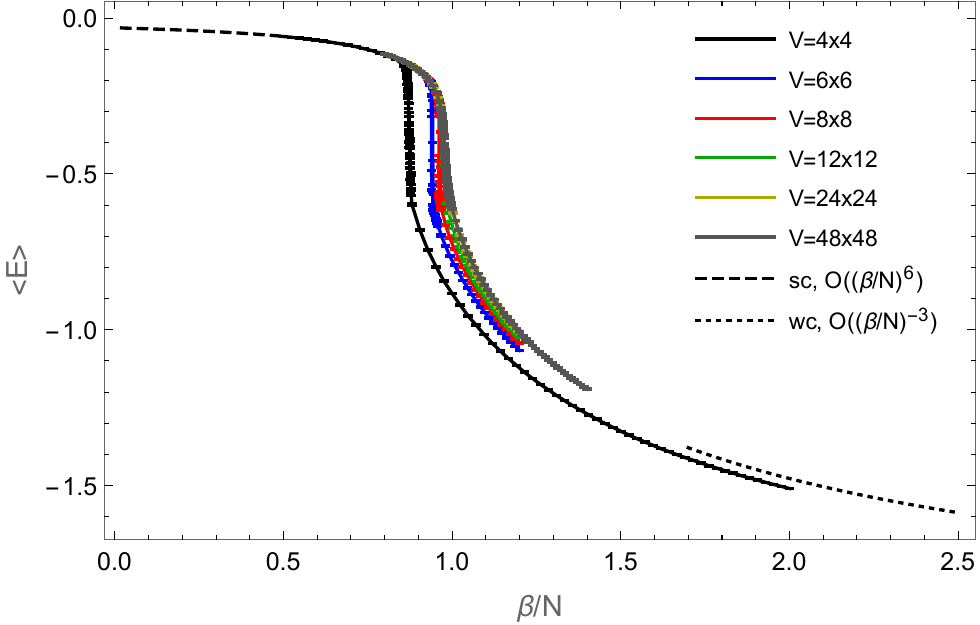}
\end{minipage}\hfill
\begin{minipage}[t]{0.47\linewidth}
\centering
{\small $C_{E}$ vs. $\beta/N$ for $Z_{Q}$}\\[0pt]
\includegraphics[width=\linewidth]{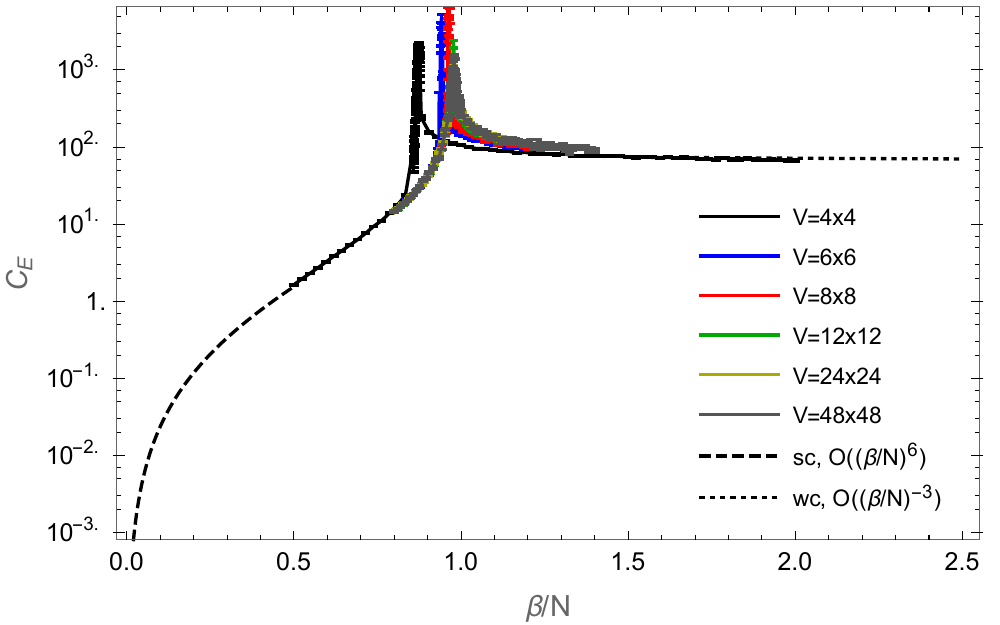}
\end{minipage}\\[0.25cm]
\begin{minipage}[t]{0.47\linewidth}
\centering
{\small $\chi_{m}$ vs. $\beta/N$ for $Z_{Q}$}\\[0pt]
\includegraphics[width=\linewidth]{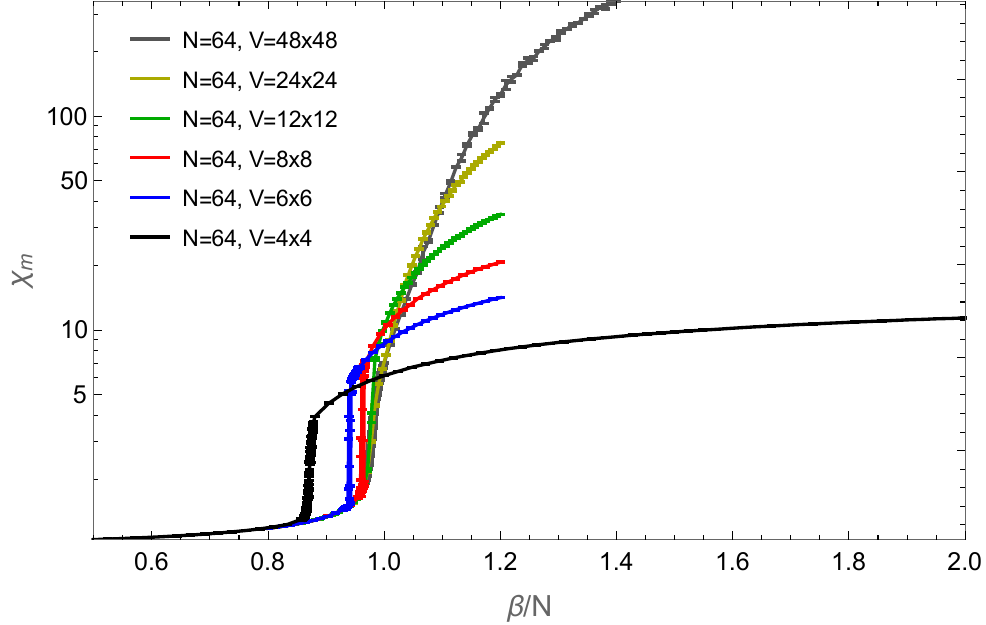}
\end{minipage}\hfill
\begin{minipage}[t]{0.46\linewidth}
\centering
{\small $V/N\ssof{\ssavof{n^2}-\ssavof{n}^{2}}$ vs. $\beta/N$ for $Z_{Q}$}\\[0pt]
\includegraphics[width=\linewidth]{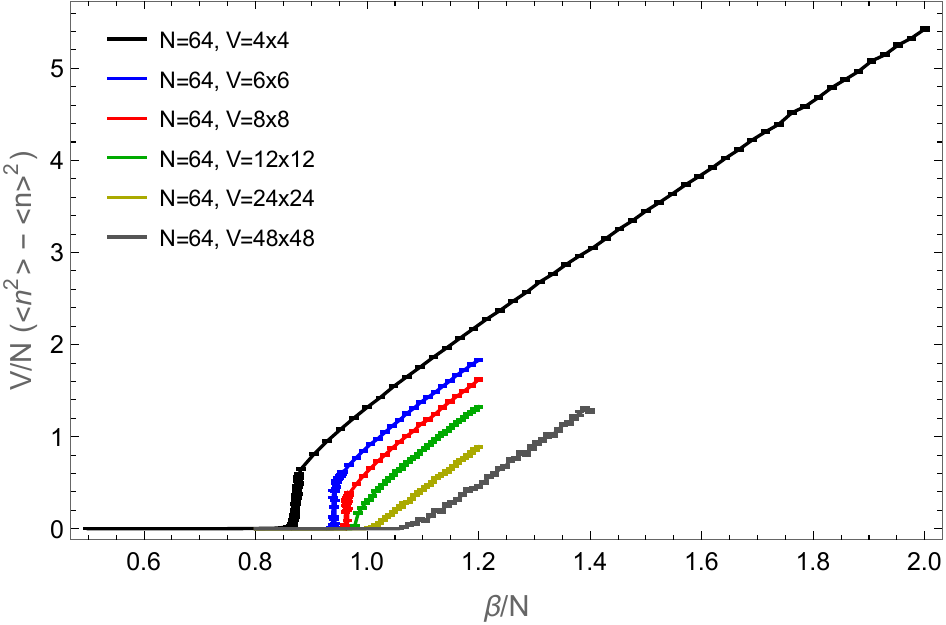}
\end{minipage}
\caption{The figure shows again the average energy (top-left) and specific heat (top-right) for the $\CPn{N-1}$ model described by \eqref{eq:cpnpartf2a} (quartic action) in $\ssof{1+1}$ dimensions, with $N=64$ and for system sizes $V\in\cof{4^2,6^2,8^2,12^2,24^2,48^2}$, just as in Fig.~\ref{fig:energyvsbvarsize}, but this time we also show the magnetic susceptibility $\chi_{m}$ (bottom-left) from \eqref{eq:magsusc} and the trace of the charge-density covariance matrix (bottom-right) from \eqref{eq:chargecov}. As can be seen, the abrupt changes in the average energy and the magnetic susceptibility, as well as the peak in the specific heat always occur at $\beta/N<1$ for all system sizes. For the trace of the covariance matrix of the charge densities, this is only true for $V\leq 12^2$, while for $V=24^2,48^2$ there is no longer an abrupt change, and it becomes non-zero only for $\beta/N>1$.}
  \label{fig:energyvsbvarsize2}
\end{figure}

\begin{figure}[!h]
\centering
\begin{minipage}[t]{0.465\linewidth}
\centering
{\small $\avof{E}$ vs. $\beta/N$ for $Z_{A}$}\\[0pt]
\includegraphics[width=\linewidth]{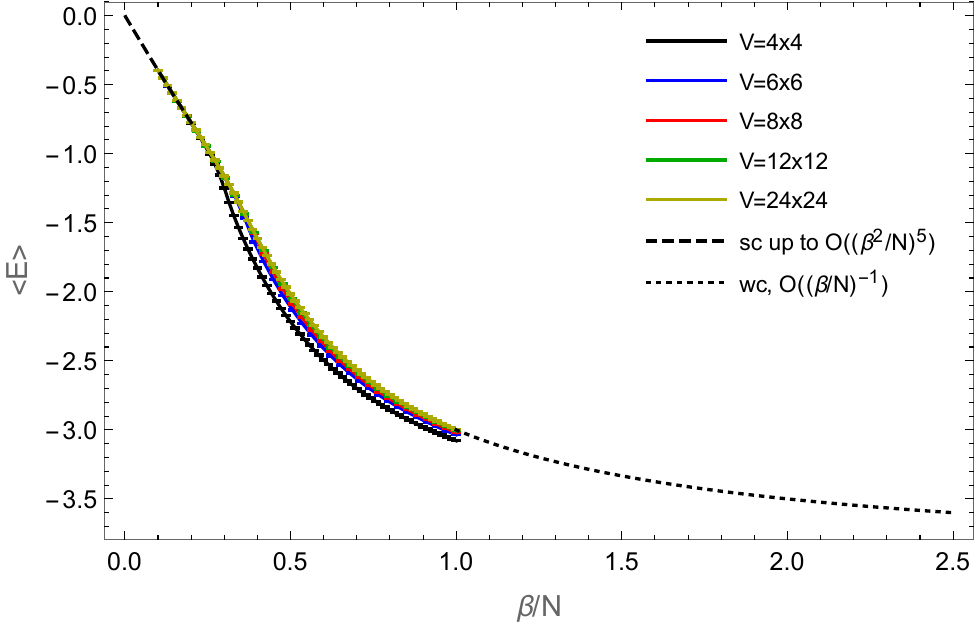}
\end{minipage}\hfill
\begin{minipage}[t]{0.475\linewidth}
\centering
{\small $C_{E}$ vs. $\beta/N$ for $Z_{A}$}\\[0pt]
\includegraphics[width=\linewidth]{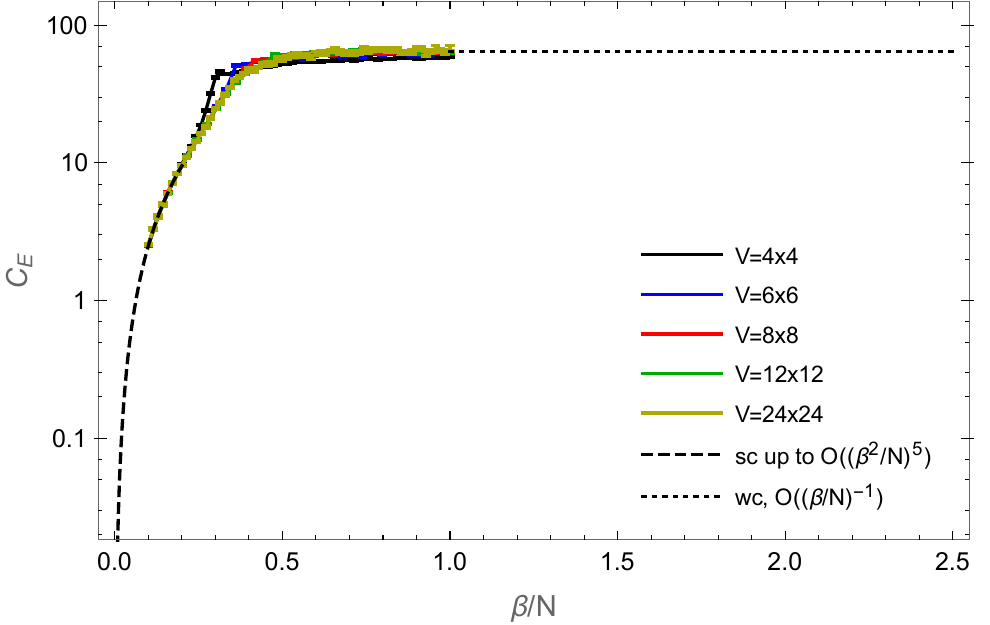}
\end{minipage}\\[0.25cm]
\begin{minipage}[t]{0.465\linewidth}
\centering
{\small $\chi_{m}$ vs. $\beta/N$ for $Z_{A}$}\\[0pt]
\includegraphics[width=\linewidth]{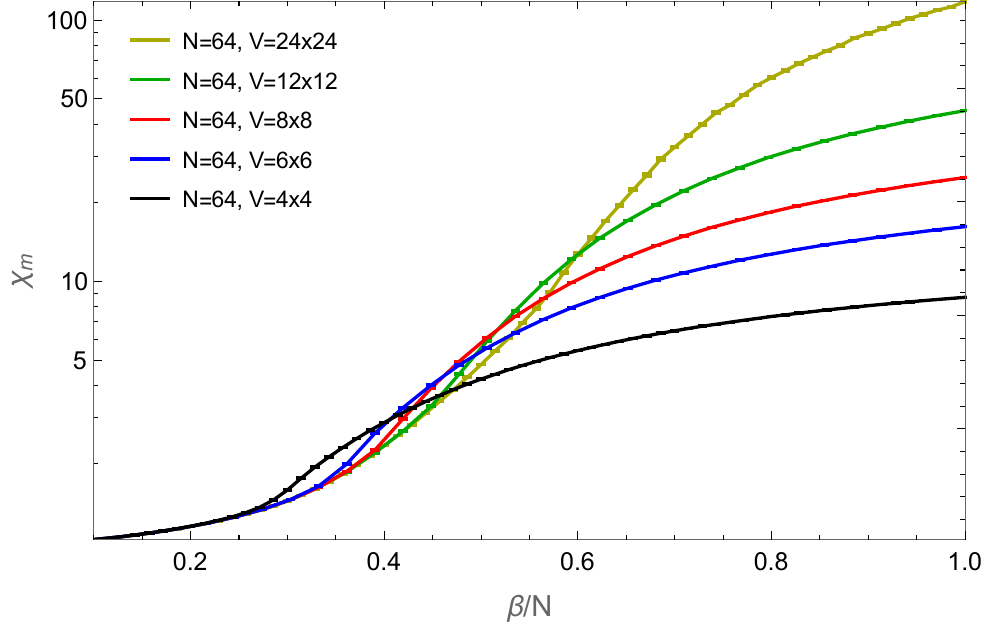}
\end{minipage}\hfill
\begin{minipage}[t]{0.465\linewidth}
\centering
{\small $V/N\ssof{\ssavof{n^2}-\ssavof{n}^{2}}$ vs. $\beta/N$ for $Z_{A}$}\\[0pt]
\includegraphics[width=\linewidth]{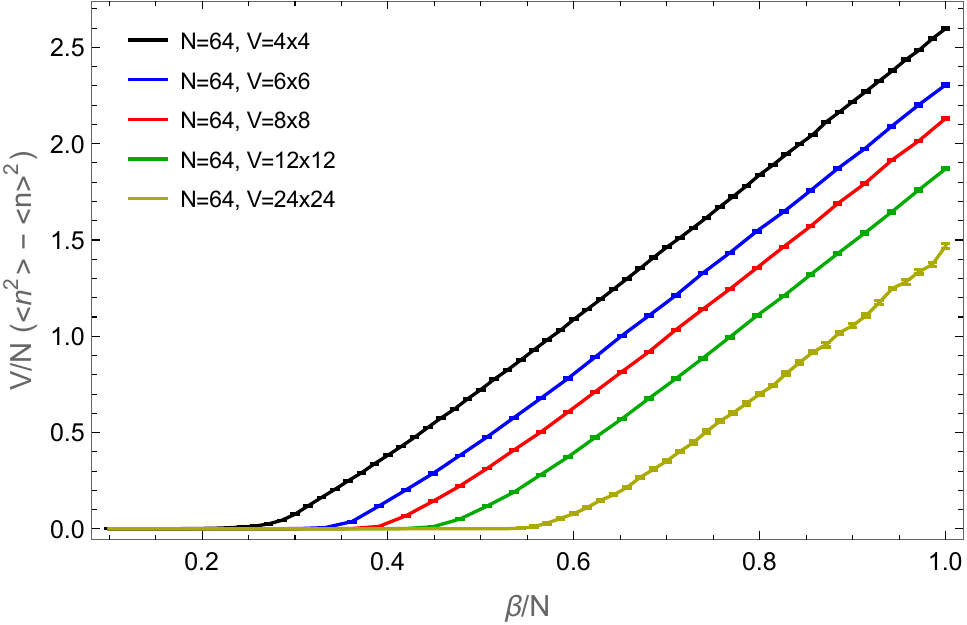}
\end{minipage}
\caption{Same as Fig.~\ref{fig:energyvsbvarsize2} but for comparison, this time for the version \eqref{eq:cpnpartf1} (auxiliary $\Un{1}$ action) of the $\CPn{N-1}$ partition function, again in $\ssof{1+1}$ dimensions, with $N=64$ and for system sizes $V\in\cof{4^2,6^2,8^2,12^2,24^2}$. As can be seen, for this version of the model, there are no abrupt jumps in any of the shown quantities and the pseudo-critical value of $\beta/N$ for the de-confinement transition in the bottom-right plot just increases steadily with increasing system size.}
  \label{fig:energyvsbvarsizeauxu1}
\end{figure}

Another thing to be mentioned is, that in the cases where the double-peak structure can be observed in the energy density, i.e. for small volumes and large $N$, the trace of the charge-density covariance matrix \eqref{eq:chargecov} becomes non-zero at the same value of $\beta/N$ at which the peak in the specific heat occurs, while as soon as the double-peak structure disappears for sufficiently large volumes, the trace of the charge density covariance matrix becomes non-zero only at a larger value of $\beta/N$ (see Fig.~\ref{fig:energyvsbvarsize2}), and the value of $\beta/N$ at which this happens increases even further with increasing volume. The reason why this is interesting is, that the trace of the charge density covariance matrix can be interpreted as an order parameter for the breaking of the global $\mathbb{Z}_{2}$ symmetry. As this quantity depends only on the $k$-variables directly, this indicates that the peak in the specific heat at $\beta/N \lesssim 1$ is caused purely by a reordering of the $l$-variables: for example by a breaking of the global $\mathbb{Z}_{N}$ symmetry in the form of a condensation of the $l^{a\,a}$-variables for some $a$ (on which also the magnetic susceptibility depends through the diagonal entries of the correlator). But this is so far only speculation and needs to be verified in the future. Also the meaning of these discrete symmetries should be clarified (remember that the internal space indices are related to, but not the same as flavor-space indices, so the breaking of these symmetries should not be in contradiction with the Mermin-Wagner theorem, which would forbid a spontaneous breaking of a subgroup of the continuous $\SU{N}$ flavor-symmetry group).\\
For comparison, Fig.~\ref{fig:energyvsbvarsizeauxu1} shows the same data for the auxiliary $\Un{1}$ version of the $\CPn{N-1}$ partition function from \eqref{eq:cpnpartf1} for which no jump in the energy density occurs. Clearly, the auxiliary $\Un{1}$ discretization shows a smoother approach to the continuum limit, and should be preferred over $Z_{Q}$ for that purpose.\\

\section{Summary \& Conclusion}\label{sec:summary}

We looked at the two most common lattice formulations of the $\CPn{N-1}$ model (referred to as quartic and auxiliary $\Un{1}$ version, respectively), and reviewed two possibilities how the corresponding partition functions can be expressed in terms of integer valued (dual) flux-variables by integrating out the original degrees of freedom. The two possibilities to dualize the $\CPn{N-1}$ partition functions differ by the number of independent degrees of freedom that are used to parametrize different configurations in the dual partition function: the first possibility \cite{Chandrasekharan} yields a system of configurations depending on $\order{N^2}$ independent flux variables per link, while configurations of the system obtained by the second possibility depend on just $\order{2\,N}$ flux variables per link\cite{Gattringer1}. It turns out that in terms of both sets of flux-variables, the partition functions for the quartic and the auxiliary $\Un{1}$ versions of the $\CPn{N-1}$ model differ only by an extra weight factor for each link. After having discussed the relation between the constraints in the $\order{N^2}$ and the $\order{2\,N}$ flux-variables per link versions of the $\CPn{n-1}$ partition function, which can be associated with conservation laws, we then found that not just the version from \cite{Gattringer1} but all four flux-variable representations (with/without $\Un{1}$ field, with $\order{2\,N}$ or $\order{N^2}$ flux variables) allow for the introduction of chemical potentials without giving rise to a sign problem.\\

It has previously been observed in \cite{Vetter} that a naive application of a worm algorithm to the flux-variable formulation of the $\CPn{N-1}$ model form \cite{Chandrasekharan} with $\order{N^2}$ d.o.f. per link, gives rise to an ergodicity problem when $N>2$. The problem can be solved by extending the ordinary worm algorithm by additional moves, which allow the worm to propagate a defect not just from site to site, but also to temporarily introduce another defect which can be used to move in internal space. This additional freedom allows the worm to take shortcuts in configuration space and directly relate configurations, which would require many intermediate updates in order to be connected by an ordinary worm. This is the basic idea underlying our \emph{internal space sub-worm} algorithm.\\
As to our knowledge, no algorithm has been tested so far for the $\order{2\,N}$ d.o.f. per link formulation of the $\CPn{N-1}$ model from \cite{Gattringer1}, we also presented a simulation algorithm for this system. Due to the simpler structure of the constraints imposed on the flux-variables, an ordinary worm algorithm is sufficient in this case.\\ 
Both, the ordinary worm for the $\order{2\,N}$ and the internal space sub-worm algorithm for the $\order{N^2}$ d.o.f. per link version of the $\CPn{N-1}$ partition function, have been tested and yield identical results, which also compare well with predictions from strong and weak coupling expansions as well as with previous results from the literature. It also turns out that both algorithms seem to perform equivalently well in the sense that they yield equal accuracy for equal statistics (number of sweeps). However, as for $N>2$, the number of degrees of freedom that have to be updated during a sweep is larger for the system that is updated by the internal space sub-worm algorithm, the dual formulation of the $\CPn{N-1}$ model from \cite{Gattringer1} may in general be cheaper to simulate.\\
Nevertheless, the internal space sub-worm algorithm could still be of interest, as it might also be applicable to other systems where no alternative formulation in terms of fewer degrees of freedom exists.\\
Finally it should be mentioned that the two worm algorithms that have been discussed in this paper, unfortunately, do not seem to perform markedly better than the over-heat bath \cite{Campostrini}, the cluster \cite{Jansen} or the loop algorithm \cite{Wolff}. Finally, one advantage of the worm algorithms that remains is, that they do not give rise to any sign-problem when turning on a chemical potential. Furthermore, the worm algorithms do not suffer from topological slowing down, because all topological sectors are mixed in terms of flux variables. This makes reconstructing topological information challenging, as discussed in the Appendix.

\section{Acknowledgements}

We thank Roman Vetter, the author of \cite{Vetter}, for pointing out the ergodicity problem that occurs when applying a naive worm to the flux-variable representation from \cite{Chandrasekharan}, and for providing his results and the corresponding computer program. All numerical simulations have been carried out on the two ETH clusters Brutus and Euler.

\clearpage
\begin{appendices}
\section{The topological term in terms of dual variables}\label{sec:thetaterm}

In the continuum, the topological charge density for the $\of{1+1}$ dimensional $\CPn{N-1}$ model is given by (see \cite{Berg,Rastelli1})
\[
q\of{x}\,=\,\frac{1}{2\,\pi}\epsilon_{\mu\,\nu}\,\partial_{\mu}\,A_{\nu}\ ,\label{eq:ctopochargedens}
\]
where $A_{\mu}$ is the auxiliary $\Un{1}$ gauge field from \eqref{eq:cgaugefield}. One can then define a topological action term,
\[
S_{t}\,=\,-\ii\,\Theta\,\sum\limits_{x}\,q\of{x}\ ,\label{eq:topoaction}
\]
where $\Theta$ is a new free parameter usually called \emph{theta-angle} and \eqref{eq:topoaction} is called \emph{theta-term}. Adding \eqref{eq:topoaction} to the original $\CPn{N-1}$ action does not affect the classical equations of motion so that the classical physics is independent of $\Theta$. In the quantum theory however, physical results will depend on the value of $\Theta$.\\

On the lattice, we write instead
\[
S_{t}\,=\,-\frac{\ii\,\Theta}{2\,\pi}\,\sum\limits_{x,\mu<\nu}\,\tilde{Q}_{x,\mu\,\nu}\ ,\label{eq:ltopoaction}
\]
where $\tilde{Q}_{x,\mu\,\nu}$, in terms of the phases $\theta_{x,\mu}\in\left[-\pi,\pi\right)$ of the auxiliary $\Un{1}$ link-variables $U_{x,\mu}=\e^{\ii\,\theta_{x,\mu}}$, is given by
\[
\tilde{Q}_{x,\mu\,\nu}\,=\,\underbrace{\theta_{x,\mu}+\theta_{x+\hat{\mu},\nu}-\theta_{x+\hat{\nu},\mu}-\theta_{x,\nu}}_{Q_{x,\mu\,\nu}\,\in\,\roint{-4\pi,4\pi}}\,-\,2\,\pi\,\underbrace{n_{x,\mu\,\nu}}_{\mathclap{\hspace{40pt}\in\,\cof{-2,-1,0,1,2}}}\ ,\label{eq:auxu1topocharge}
\]
with $n_{x,\mu\,\nu}\in\cof{-2,-1,0,1,2}$ chosen such that $\tilde{Q}_{x,\mu\,\nu}\in\roint{-\pi,\pi}$.\\

Alternatively, one can express $\tilde{Q}_{x,\mu\,\nu}$ also directly in terms of inner-products between pairs of $z$-variables (i.e. $z^{\dagger}\of{x}\cdot z\of{x+\hat{\mu}}$) around plaquettes (in \cite{Berg,Rastelli1}, the products were taken around triangles instead of around plaquettes):
\begin{multline}
\tilde{Q}_{x,\mu\,\nu}\,=\,-\frac{\ii}{2}\bcof{\log\sfof{\sof{z^{\dagger}\of{x}\cdot z\of{x+\hat{\mu}}}\,\sof{z^{\dagger}\of{x+\hat{\mu}}\cdot z\of{x+\hat{\mu}+\hat{\nu}}}\\
\of{z^{\dagger}\of{x+\hat{\mu}+\hat{\nu}}\cdot z\of{x+\hat{\nu}}}\,\sof{z^{\dagger}\sof{x+\hat{\nu}}\cdot z\of{x}}}\\
-\,\log\sfof{\sof{z\of{x}\cdot z^{\dagger}\of{x+\hat{\mu}}}\,\sof{z\of{x+\hat{\mu}}\cdot z^{\dagger}\of{x+\hat{\mu}+\hat{\nu}}}\\
\sof{z\of{x+\hat{\mu}+\hat{\nu}}\cdot z^{\dagger}\of{x+\hat{\nu}}}\,\sof{z\of{x+\hat{\nu}}\cdot z^{\dagger}\of{x}}}}\ .\label{eq:quartictopocharge}
\end{multline}
Note that \eqref{eq:ltopoaction} is in general only a topological term in two dimensions.\\

In what follows, we will derive the form of topological term for the $\CPn{N-1}$ partition function $\tilde{Z}_{A}$ from \eqref{eq:cpnpartf3}, using the definition \eqref{eq:auxu1topocharge} in \eqref{eq:ltopoaction}.

\subsection{Dualization of partition function including a topological term}
After adding \eqref{eq:ltopoaction} with \eqref{eq:auxu1topocharge} to the lattice partition function \eqref{eq:lpartf3}, the corresponding partition function becomes:
\begin{multline}
\tilde{Z}_{A}\,=\,\int\DD{z^{\dagger},z,U}\,\exp\bof{\beta\,\sum\limits_{x,\nu}\sof{z^{\dagger}\of{x}\e^{\sum\limits_{i}\mu_{i}\lambda_{i}\delta_{\nu,d}}U_{\nu}\of{x}z\of{x+\hat{\nu}}\\
+\,z^{\dagger}\of{x}\e^{-\sum\limits_{i}\mu_{i}\lambda_{i}\delta_{\nu,d}}U^{\dagger}_{\nu}\of{x-\hat{\nu}}z\of{x-\hat{\nu}}}\,+\,\ii\,\bar{\Theta}\sum\limits_{\mu>\nu}\,\tilde{Q}_{x,\mu\,\nu}}\ ,\label{eq:cpnpartf3incltopo1}
\end{multline}
where $U_{\nu}\of{x}$ are the $\Un{1}$ link variables corresponding to the auxiliary gauge field $A_{\nu}$, the $\lambda_{i},i\in\cof{1,N-1}$ are the $\of{N-1}$ diagonal $\SU{N}$ generators, $\bar{\Theta}=\frac{\Theta}{2\,\pi}$ is the \emph{reduced theta angle} and $\tilde{Q}_{x,\nu\,\mu}$ is the topological charge per plaquette as defined above. Proceeding as in Sec.~\ref{ssec:conservedcurrnts}, we use the identity
\[
\e^{x\of{t+1/t}}\,=\,\sum\limits_{k=-\infty}^{\infty} I_{k}\of{2\,x}\,t^{k}\ ,
\]
and defining $\tilde{\mu}_{a}=\sum_{i}\mu_{i}\lambda_{i,a\,a}$, $a\in\cof{1,\ldots,N}$, so that \eqref{eq:cpnpartf3incltopo1} can be written as
\begin{multline}
\tilde{Z}_{A}\,=\,\int\DD{r,\phi,\theta}\prod\limits_{x,\nu}\e^{-2\,\beta}\bof{\prod\limits_{a=1}^{N}\bcof{\sum\limits_{k_{x,\nu}^{a}=-\infty}^{\infty}\e^{\sof{\ii\theta_{x,\nu}+\tilde{\mu}_{a}\delta_{\nu,d}+\ii\ssof{\phi^{a}_{x+\hat{\nu}}-\phi^{a}_{x}}}k_{x,\nu}^{a}}\,I_{k_{x,\nu}^{a}}\of{2\,\beta\,r^{a}_{x}r^{a}_{x+\hat{\nu}}}}}\\
\cdot\prod\limits_{\mu>\nu}\sum\limits_{n_{x,\nu\,\mu}=-2}^{2}\Pi\of{\frac{Q_{x,\nu\,\mu}}{2\,\pi}-n_{x,\nu\,\mu}}\exp\bof{\ii\bar{\Theta}\overbrace{\sof{\underbrace{Q_{x,\nu\,\mu}}_{\mathclap{\in\of{-4\pi,4\pi}}}-2\,\pi\underbrace{n_{x,\nu\,\mu}}_{\mathclap{\hspace{40pt}\in\cof{-2,-1,0,1,2}}}}}^{\tilde{Q}_{x,\nu\,\mu}\in\of{-\pi,\pi}}} ,\label{eq:lpartf3int}
\end{multline}
where $Q_{x,\nu\,\mu}=\theta_{x,\nu}+\theta_{x+\hat{\nu},\mu}-\theta_{x+\hat{\mu},\nu}-\theta_{x,\mu}$ and $\Pi\of{x}=\begin{cases}1,\text{ if }-1/2<x<1/2\\0,\text{ else}\end{cases}$ is the Heaviside-Pi function.
Following a trick introduced in \cite{Wolff}, we write the exponential on the last line of \eqref{eq:lpartf3int} as
\begin{multline}
\exp\sof{\ii\,\bar{\Theta}\of{Q_{x,\nu\,\mu}-2\,\pi\,n_{x,\nu\,\mu}}}\,=\,\sum\limits_{m_{x,\nu\,\mu}=-\infty}^{\infty}\underbrace{\frac{\sin\ssof{\pi\ssof{\bar{\Theta}-m_{x,\nu\,\mu}}}}{\pi\ssof{\bar{\Theta}-m_{x,\nu\,\mu}}}}_{j_{0}\ssof{\pi\ssof{\bar{\Theta}-m_{x,\nu\,\mu}}}}\,\e^{\ii\,m_{x,\nu\,\mu}\of{Q_{x,\nu\,\mu}-2\,\pi\,n_{x,\nu\,\mu}}}\\
\,=\,\sum\limits_{m_{x,\nu\,\mu}=-\infty}^{\infty}j_{0}\ssof{\pi\ssof{\bar{\Theta}-m_{x,\nu\,\mu}}}\,\e^{\ii\,m_{x,\nu\,\mu}\,Q_{x,\nu\,\mu}}\ ,
\end{multline}
with auxiliary, integer plaquette variables $m_{x,\nu\,\mu}$, so that the sum over $n_{x,\nu\,\mu}$ and the Heaviside $\Pi$ function can be dropped. Now we can continue as in Sec.~\ref{ssec:conservedcurrnts}, by using
\[
I_{k}\of{2\,x}\,=\,\sum\limits_{l=0}^{\infty}\frac{x^{k+2\,l}}{\of{k+l}!\,l!}\,=\,\sum\limits_{l=0}^{\infty}\frac{x^{\abs{k}+2\,l}}{\of{\abs{k}+l}!\,l!}\ ,
\]
and integrating out $r^{a}_{x}$, $\phi^{a}_{x}$ and $\theta_{x,\nu}$ to end up with:
\begin{multline}
\tilde{Z}_{A}\,=\,\sum\limits_{\cof{k,\,l,\,m}}\bcof{\prod\limits_{x}\bof{\prod\limits_{\nu}\e^{-2\,\beta}{\delta\sof{\sum\limits_{a}k_{x,\nu}^{a}+\sum\limits_{\mu>\nu}\sof{m_{x,\nu\,\mu}-m_{x-\hat{\mu},\nu\,\mu}}-\sum\limits_{\mu<\nu}\sof{m_{x,\mu\,\nu}-m_{x-\hat{\mu},\mu\,\nu}}}}\\
\cdot\bof{\prod\limits_{\mu>\nu}j_{0}\ssof{\pi\ssof{\bar{\Theta}-m_{x,\nu\,\mu}}}}\bof{\prod\limits_{a}\e^{\tilde{\mu}_{a}\,k_{x,\nu}^{a}\delta_{\nu,d}}\,\frac{\beta^{\abs{k_{x,\nu}^{a}}+2\,l_{x,\nu}^{a}}}{\of{\abs{k_{x,\nu}^{a}}+l_{x,\nu}^{a}}!\,l_{x,\nu}^{a}!}}}\\
\frac{\prod\limits_{a}\delta\sof{\sum\limits_{\nu}\sof{k_{x,\nu}^{a}-k_{x-\hat{\nu}}^{a}}}\sof{\sum\limits_{\nu}\sof{\frac{1}{2}\sof{\sabs{k_{x,\nu}^{a}}+\sabs{k_{x-\hat{\nu},\nu}^{a}}}+l_{x,\nu}^{a}+l_{x-\hat{\nu},\nu}^{a}}}!}{\sof{N-1+\sum\limits_{a}\sum\limits_{\nu}\sof{\frac{1}{2}\sof{\sabs{k_{x,\nu}^{a}}+\sabs{k_{x-\hat{\nu},\nu}^{a}}}+l_{x,\nu}^{a}+l_{x-\hat{\nu},\nu}^{a}}}!}}\ ,\label{eq:cpnpartf3incltopo2}
\end{multline}
where $k_{x,\nu}^{a}\in\mathbb{Z}$, $l_{x,\nu}^{a}\in\mathbb{N}_{0}$ are the flux-variables introduced in Sec.\ref{ssec:conservedcurrnts} and $m_{x,\nu\,\mu}\in\mathbb{Z}$ are new plaquette occupation numbers. Unfortunately, for $\bar{\Theta}\neq 0$, the plaquette weights $j_{0}\of{\pi\of{\bar{\Theta}-m_{x,\nu\,\mu}}}$ can be negative, which introduces a sign problem when trying to obtain the sum over the plaquette variables stochastically using importance sampling. Note that the plaquette variables $m_{x,\nu\,\mu}$ couple to the flux variables $k_{x,\nu}$ through the "on-link" constraints, i.e. the delta function on the first line of \eqref{eq:cpnpartf3incltopo2}. 

\subsection{Topological charge and susceptibility in terms of dual variables}
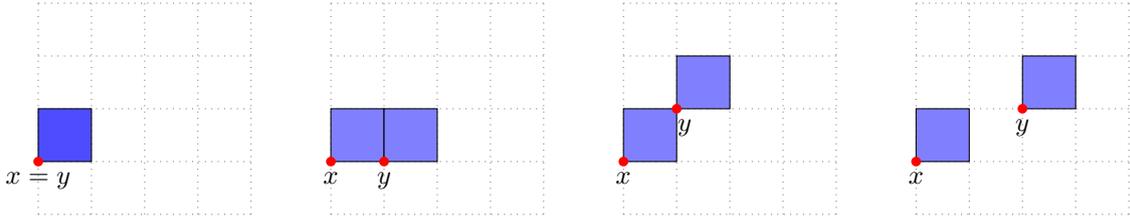
\begin{figure}[h]
\centering
\input{tikzimg/topoplaquettes.tex}
\caption{Due to the delta-function on the first line of \eqref{eq:cpnpartf3incltopo2}, a change in a plaquette variable $m_{x,\nu\,\mu}$ requires that either also its neighboring plaquettes get changed, or that for an unchanged neighboring plaquette, the $k$-variable that lives on the boundary to that plaquette gets updated. As for vanishing $\obar{\Theta}$, the $m_{x,\nu\,\mu}$ can be non-zero only if the corresponding plaquette has been hit by a derivative in \eqref{eq:qsqexp}, one always has to update also $k$-variables. The four cases depicted in the figure give rise to different dependencies of the $k$-variables on changes of the values of the plaquette variables.} 
\label{fig:toposusc}
\end{figure}

The expectation value for the topological charge and its susceptibility are given by the first and second derivatives of $\log\of{Z_{A}}$ with respect to $\ii\,\Theta$, i.e. :
\begin{align}
\avof{Q}\,&=\,-\frac{\ii}{2\,\pi}\partd{\log\of{Z_{A}}}{\bar{\Theta}}\,=\,-\frac{\ii}{2\,\pi}\sum_{x}\partd{\log\of{Z_{A}}}{\bar{\Theta}_{x}}\bigg|_{\bar{\Theta}_{x}=\bar{\Theta}\,\forall x}\ ,\label{eq:qexp}\\
\savof{Q^2}-\savof{Q}^2\,&=\,\frac{-1}{\of{2\,\pi}^2}\sum_{x,y}\spartd{\log\of{Z_{A}}}{\bar{\Theta}_{x}}{\bar{\Theta}_{y}}\bigg|_{\bar{\Theta}_{x}=\bar{\Theta}\,\forall x}\label{eq:qsqexp}\ ,
\end{align}
where, to carry out the derivatives, it can be used that for spherical Bessel function of the first kind, we have:
\[
\totd{j_{l}\of{x}}{x}{}\,=\,\frac{1}{2\,l+1}\sof{l\,j_{l-1}\of{x}-\of{l+1}\,j_{l+1}\of{x}} .
\]
As $j_{0}\ssof{x}$ is just the $\op{sinc}$-function, setting $\bar{\Theta}=0$ requires that
\[
j_{0}\ssof{\pi\ssof{\bar{\Theta}+m}}\bigg|_{\bar{\Theta}=0}\,=\,\begin{cases} 1\,&\text{if }m=0\\
0\,&\text{else} \end{cases}\ ,
\]
while for the derivatives
\begin{align}
\totd{j_{0}\ssof{\pi\ssof{\bar{\Theta}+m}}}{\bar{\Theta}}{}\bigg|_{\bar{\Theta}=0}\,&=\,\begin{cases} \frac{\of{-1}^{m}}{m}\,&\text{  if }m\neq 0\\
0\,&\text{  else}\end{cases}\ ,\\
\totd{j_{0}\ssof{\pi\ssof{\bar{\Theta}+m}}}{\bar{\Theta}}{2}\bigg|_{\bar{\Theta}=0}&=\,\begin{cases} -\frac{2\,\of{-1}^{m}}{m^2}\,&\text{if }m\neq 0\\
\frac{-\pi^2}{3}&\text{else}\end{cases}\ ,\label{eq:sphbesselsd}
\end{align}
$\bar{\Theta}=0$ does not constrain $m$ and we have to carry out the summation over all the $m_{x,\nu\,\mu}$ in \eqref{eq:qexp} and \eqref{eq:qsqexp}. To measure the topological susceptibility \eqref{eq:qsqexp}, one has to consider the four cases depicted in Fig.~\ref{fig:toposusc} when summing over the $m_{x,\mu\,\nu}$, which differ by the number of link variables and site-weights that depend just on one of the plaquette variables $\of{m_{x,\mu\,\nu},\,m_{y,\mu\,\nu}}$ or on both. So far we only tried to do this summation numerically, which is computationally rather expensive.\\

In the strong coupling limit ($\beta=0$) however, we have that $k_{x,\nu}^{a}=0\,\forall x,\nu,a$ in \eqref{eq:cpnpartf3incltopo2} and therefore all $m_{x,\nu\,\mu}$ have to be equal, which for $\bar{\Theta}=0$ in fact means, that $m_{x,\nu\,\mu}=0\,\forall x,\nu,\mu$. From \eqref{eq:qsqexp} and \eqref{eq:sphbesselsd} one can then directly read off that
\[
\savof{Q^2}-\savof{Q}^2\,=\,\frac{V}{\of{2\,\pi}^2}\frac{\pi^2}{3}\ ,
\]
which is the correct strong coupling result.\\

For the dual formulations \eqref{eq:cpnpartf2a} and \eqref{eq:cpnpartf1} it is more involved to incorporate such a topological term: we have seen above that the plaquette variables that comes from the topological term couple to the $k$-variables by modifying the on-link constraints. However, for the dual formulations \eqref{eq:cpnpartf2a} and \eqref{eq:cpnpartf1}, these constraints are automatically satisfied due to the anti-symmetry of the $k_{x,\nu}^{a\,b}$ variables with respect to their internal space indices $\of{a,b}$. This makes it impossible to incorporate the topological term \eqref{eq:topoaction} in a similar way for these partition functions. Instead one could start with the definition \eqref{eq:quartictopocharge} of the topological charge density $\tilde{Q}_{x,\mu\,\nu}$. This will give rise to plaquette variables which also carry internal space indices which are then compatible with the anti-symmetry of the $k$-variables.
\end{appendices}

\end{document}

%% file: tikzimg/constraints.tex
\begin{tikzpicture}[scale=0.7,nodes={inner sep=0}]
  \pgfpointtransformed{\pgfpointxy{1}{1}};
  \pgfgetlastxy{\vx}{\vy}
  \begin{scope}[node distance=\vx and \vy]
    \node[left] at (0,8-4) {$a$};
    \node[above] at (4,8) {$b$};
    \foreach \i in {1,...,7} {
        \draw [thin,dotted,gray] (\i,1) -- (\i,7)  node[solid,black,above] at (\i,7.3) {$\i$};
    }
    \foreach \i in {1,...,7} {
        \draw [thin,dotted,gray] (1,\i) -- (7,\i) node[solid,black,left] at (0.7,8-\i) {$\i$};
    }
    \foreach \i in {1,...,7} {
      \draw [thin,black!100] (7,8-\i) -- (\i,8-\i) -- (\i,8-1);
    }
    \draw [dashed,red!100] (1,8-2) -- (7,8-2);
    \draw [dashed,red!100] (1,8-5) -- (7,8-5);
    \draw [dashed,red!100] (1,8-6) -- (7,8-6);
    \draw [thick,blue!100] (7,8-5) -- (5,8-5) -- (5,8-1);
    \draw [thick,blue!100] (7,8-6) -- (6,8-6) -- (6,8-1);
    \node[draw,circle,inner sep=1.25,fill,color=red] at (6,8-5) (s1) {};
    \node[above right=0.05 and 0.0 of s1] {$k_{x,\nu}^{5\,6}$};
    \node[draw,circle,inner sep=1.25,fill,color=red] at (5,8-6) (sm1) {};
    \node[draw,circle,inner sep=0.75,fill,color=white] at (5,8-6) {};
    \node[below left=0.1 and 0.0 of sm1] {$-k_{x,\nu}^{5\,6}$};
    \draw[<->,thin,dashed,black!100] (sm1) -- (s1); 
    
    \node[draw,circle,inner sep=1.25,fill,color=red] at (5,8-2) (p1) {};
    \node[above right=0.05 and 0.0 of p1] {$k_{x,\nu}^{2\,5}$};
    \node[draw,circle,inner sep=1.25,fill,color=red] at (2,8-5) (pm1) {};
    \node[draw,circle,inner sep=0.75,fill,color=white] at (2,8-5) {};
    \node[below left=0.1 and 0.0 of pm1] {$-k_{x,\nu}^{2\,5}$};
    \draw[<->,thin,dashed,black!100] (pm1) -- (p1);
 
    \node[draw,circle,inner sep=1.25,fill,color=red] at (6,8-2) (p2) {};
    \node[draw,circle,inner sep=0.75,fill,color=white] at (6,8-2) {};
    \node[above right=0.05 and 0.0 of p2] {$k_{x,\nu}^{2\,6}$};
    \node[draw,circle,inner sep=1.25,fill,color=red] at (2,8-6) (pm2) {};
    \node[below left=0.1 and 0.0 of pm2] {$-k_{x,\nu}^{2\,6}$};
    \draw[<->,thin,dashed,black!100] (pm2) -- (p2);
  \end{scope}
\end{tikzpicture}

%% file: tikzimg/ordinaryworm.tex
\begin{tikzpicture}[scale=0.5,nodes={inner sep=0,scale=0.8}]
  \pgfpointtransformed{\pgfpointxy{1}{1}};
  \pgfgetlastxy{\vx}{\vy}
  \begin{scope}[node distance=\vx and \vy]
    \def\dxa{7};
    \def\dya{9};
    
    \def\xa{0};
    \def\ya{\dya};
    \node at (\xa-1,\ya+3) {Start:};
    \node at (\xa+3,\ya+5.5) {\small insert ext. fields?};
    \foreach \i in {1,...,5} {
        \draw [very thin,dotted,gray] (\xa+\i,\ya+1) -- (\xa+\i,\ya+5) node[solid,black,above] at (\xa+\i,\ya+5.3) {};
    }
    \foreach \i in {1,...,5} {
        \draw [very thin,dotted,gray] (\xa+1,\ya+\i) -- (\xa+5,\ya+\i) node[solid,black,left] at (\xa+0.7,\ya+6-\i) {};
    }
  
    \node[draw,circle,inner sep=1.25,fill,color=red,opacity=0.5] at (\xa+3.9,\ya+2) {};
    \node[draw,circle,inner sep=1.25,fill,color=red,opacity=0.5] at (\xa+4,\ya+1.9) {};
    \node[draw,circle,inner sep=0.5,fill,color=white] at (\xa+4,\ya+1.9) {};
    \node[opacity=0.5] at (\xa+3.9,\ya+1.5) {\small $\phi^{a\,b}_{x}$, $\phi^{b\,a}_{x}$};
    \node at (\xa+3.9,\ya+1.) {\small \textbf{?}};
    
    \draw [->,thin,black] (\xa+6,\ya+3) -- (\xa+7,\ya+3);

    \def\xa{\dxa};
    \node at (\xa+3,\ya+5.5) {\small move sink?};
    \foreach \i in {1,...,5} {
        \draw [very thin,dotted,gray] (\xa+\i,\ya+1) -- (\xa+\i,\ya+5) node[solid,black,above] at (\xa+\i,\ya+5.3) {};
    }
    \foreach \i in {1,...,5} {
        \draw [very thin,dotted,gray] (\xa+1,\ya+\i) -- (\xa+5,\ya+\i) node[solid,black,left] at (\xa+0.7,\ya+6-\i) {};
    }
  
  	\node[draw,circle,inner sep=1.25,fill,color=red] at (\xa+3.9,\ya+2) {};
    \draw [ultra thick,red!100,opacity=0.5] (\xa+4.05,\ya+2) -- (\xa+4.95,\ya+2);
    \node[draw,circle,inner sep=1.25,fill,color=red,opacity=0.5] at (\xa+5.1,\ya+2) {};
    \node[draw,circle,inner sep=0.5,fill,color=white,opacity=0.5] at (\xa+5.1,\ya+2) {};
    \node[opacity=0.5] at (\xa+5.2,\ya+1.5) {\small $\phi^{b\,a}_{y}$};
    \node at (\xa+4.6,\ya+2.3) {\small \textbf{?}};
    \node at (\xa+3.9,\ya+1.5) {\small $\phi^{a\,b}_{x}$};
    
    \draw [->,thin,black] (\xa+6,\ya+3) -- (\xa+7,\ya+3);

    \def\xa{2*\dxa};
    \node at (\xa+3,\ya+5.5) {\small move sink?};
    \foreach \i in {1,...,5} {
        \draw [very thin,dotted,gray] (\xa+\i,\ya+1) -- (\xa+\i,\ya+5) node[solid,black,above] at (\xa+\i,\ya+5.3) {};
    }
    \foreach \i in {1,...,5} {
        \draw [very thin,dotted,gray] (\xa+1,\ya+\i) -- (\xa+5,\ya+\i) node[solid,black,left] at (\xa+0.7,\ya+6-\i) {};
    }
  
    \draw [ultra thick,red!100,opacity=1] (\xa+4.05,\ya+2) -- (\xa+5,\ya+2);
    \draw [ultra thick,red!100,opacity=0.5] (\xa+5,\ya+2) -- (\xa+5,\ya+2.95);
    \node[draw,circle,inner sep=1.25,fill,color=red,opacity=0.5] at (\xa+5,\ya+3.1) {};
    \node[draw,circle,inner sep=0.5,fill,color=white,opacity=0.5] at (\xa+5,\ya+3.1) {};
    \node[opacity=0.5] at (\xa+5.2,\ya+3.5) {\small $\phi^{b\,a}_{y}$};
    \node at (\xa+4.7,\ya+2.6) {\small \textbf{?}};
    \node[draw,circle,inner sep=1.25,fill,color=red] at (\xa+3.9,\ya+2) {};
    \node at (\xa+3.9,\ya+1.5) {\small $\phi^{a\,b}_{x}$};
    
    \draw [->,thin,black] (\xa+3,\ya+0) -- (\xa+3,\ya-1);
    \node[thick] at (\xa+3,\ya-1.5) {\small $\cdots$};
    \draw [->,thin,black] (\xa+3,\ya-2) -- (\xa+3,\ya-3);

    \def\xa{2*\dxa};
    \def\ya{0};
    \node at (\xa+3,\ya+5.5) {\small move sink?};
    \foreach \i in {1,...,5} {
        \draw [very thin,dotted,gray] (\xa+\i,\ya+1) -- (\xa+\i,\ya+5) node[solid,black,above] at (\xa+\i,\ya+5.3) {};
    }
    \foreach \i in {1,...,5} {
        \draw [very thin,dotted,gray] (\xa+1,\ya+\i) -- (\xa+5,\ya+\i) node[solid,black,left] at (\xa+0.7,\ya+6-\i) {};
    }
  
    \draw [ultra thick,red!100,opacity=1] (\xa+4.05,\ya+2) -- (\xa+5,\ya+2) -- (\xa+5,\ya+3) -- (\xa+5,\ya+4) -- (\xa+4,\ya+4) -- (\xa+3,\ya+4) -- (\xa+3,\ya+3);
    \draw [ultra thick,red!100,opacity=0.5] (\xa+3,\ya+3) -- (\xa+3.95,\ya+3);
    \node[draw,circle,inner sep=1.25,fill,color=red,opacity=0.5] at (\xa+4.1,\ya+3) {};
    \node[draw,circle,inner sep=0.5,fill,color=white,opacity=0.5] at (\xa+4.1,\ya+3) {};
    \node[opacity=0.5] at (\xa+4.5,\ya+2.5) {\small $\phi^{b\,a}_{y}$};
    \node at (\xa+3.7,\ya+2.6) {\small \textbf{?}};
    \node[draw,circle,inner sep=1.25,fill,color=red] at (\xa+3.9,\ya+2) {};
    \node at (\xa+3.9,\ya+1.5) {\small $\phi^{a\,b}_{x}$};

    \draw [->,thin,black] (\xa,\ya+3) -- (\xa-1,\ya+3);

    \def\xa{\dxa};
    \node at (\xa+3,\ya+5.5) {\small move sink?};
    \foreach \i in {1,...,5} {
        \draw [very thin,dotted,gray] (\xa+\i,\ya+1) -- (\xa+\i,\ya+5) node[solid,black,above] at (\xa+\i,\ya+5.3) {};
    }
    \foreach \i in {1,...,5} {
        \draw [very thin,dotted,gray] (\xa+1,\ya+\i) -- (\xa+5,\ya+\i) node[solid,black,left] at (\xa+0.7,\ya+6-\i) {};
    }
  
    \draw [ultra thick,red!100,opacity=1] (\xa+4.05,\ya+2) -- (\xa+5,\ya+2) -- (\xa+5,\ya+3) -- (\xa+5,\ya+4) -- (\xa+4,\ya+4) -- (\xa+3,\ya+4) -- (\xa+3,\ya+3) -- (\xa+4,\ya+3);
    \draw [ultra thick,red!100,opacity=0.5] (\xa+4,\ya+3) -- (\xa+4,\ya+2.05);
    \node[draw,circle,inner sep=1.25,fill,color=red,opacity=0.5] at (\xa+4,\ya+1.9) {};
    \node[draw,circle,inner sep=0.5,fill,color=white,opacity=0.5] at (\xa+4,\ya+1.9) {};
    \node at (\xa+3.7,\ya+2.3) {\small \textbf{?}};
    \node[draw,circle,inner sep=1.25,fill,color=red] at (\xa+3.9,\ya+2) {};
    \node at (\xa+3.4,\ya+1.5) {\small $\phi^{a\,b}_{x}$, };
    \node[opacity=0.5] at (\xa+4.6,\ya+1.5) {\small $\phi^{b\,a}_{y}$};
    
    \draw [->,thin,black] (\xa,\ya+3) -- (\xa-1,\ya+3);

    \def\xa{0};
    
    \node at (\xa+3,\ya+5.5) {\small remove ext. fields?};
    \foreach \i in {1,...,5} {
        \draw [very thin,dotted,gray] (\xa+\i,\ya+1) -- (\xa+\i,\ya+5) node[solid,black,above] at (\xa+\i,\ya+5.3) {};
    }
    \foreach \i in {1,...,5} {
        \draw [very thin,dotted,gray] (\xa+1,\ya+\i) -- (\xa+5,\ya+\i) node[solid,black,left] at (\xa+0.7,\ya+6-\i) {};
    }
  
    \draw [ultra thick,red!100,opacity=1] (\xa+4.05,\ya+2) -- (\xa+5,\ya+2) -- (\xa+5,\ya+3) -- (\xa+5,\ya+4) -- (\xa+4,\ya+4) -- (\xa+3,\ya+4) -- (\xa+3,\ya+3) -- (\xa+4,\ya+3) -- (\xa+4,\ya+2.05);
    \node[draw,circle,inner sep=1.25,fill,color=red,opacity=0.5] at (\xa+4,\ya+1.9) {};
    \node[draw,circle,inner sep=0.5,fill,color=white,opacity=0.5] at (\xa+4,\ya+1.9) {};
    \node at (\xa+4,\ya+1) {\small \textbf{?}};
    \node[draw,circle,inner sep=1.25,fill,color=red,opacity=0.5] at (\xa+3.9,\ya+2) {};
    \node[opacity=0.5] at (\xa+3.9,\ya+1.5) {\small $\phi^{a\,b}_{x}$, $\phi^{b\,a}_{x}$};

    \draw [->,thin,black] (\xa+3,\ya+6) -- (\xa+3,\ya+6.75);
    \node[text width=5cm,align=center] at (\xa+3,\ya+7.6) {\small take measurements and pick new location $x=x_0$};
    \draw [->,thin,black] (\xa+3,\ya+8.5) -- (\xa+3,\ya+9.25);
  \end{scope}
\end{tikzpicture}

%% file: tikzimg/subworm.tex
\begin{tikzpicture}[scale=0.38,nodes={inner sep=0}]
  \pgfpointtransformed{\pgfpointxy{1}{1}};
  \pgfgetlastxy{\vx}{\vy}
  \begin{scope}[node distance=\vx and \vy]
    \def\dxa{8};
    \def\dya{5};
    \def\xa{0};
    \def\ya{\dya};
    \node[left,scale=0.75] at (\xa-1,\ya+6-3) {$x$};
    \foreach \i in {1,...,5} {
        \draw [thin,dotted,gray] (\xa+\i,\ya+1) -- (\xa+\i,\ya+5) node[solid,black,above] at (\xa+\i,\ya+5.3) {};
    }
    \foreach \i in {1,...,5} {
        \draw [thin,dotted,gray] (\xa+1,\ya+\i) -- (\xa+5,\ya+\i) node[solid,black,left] at (\xa+0.7,\ya+6-\i) {};
    }
    \foreach \i in {1,...,5} {
      \draw [thin,black!100] (\xa+5,\ya+6-\i) -- (\xa+\i,\ya+6-\i) -- (\xa+\i,\ya+6-1);
    }
    \draw [thin,dashed,black] (\xa+1,\ya+5) -- (\xa+5,\ya+1);

    \def\pxa{4};
    \def\pya{3};
    \node[left,scale=0.65] at (\xa+0.75,\ya+6-\pya) {$a_{0}$};
    \draw [thin,black] (\xa+0.85,\ya+6-\pya) -- (\xa+1,\ya+6-\pya);
    \node[left,scale=0.65] at (\xa+0.75,\ya+6-\pxa) {$b_{0}$};
    \draw [thin,black] (\xa+0.85,\ya+6-\pxa) -- (\xa+1,\ya+6-\pxa);
    \draw [thin,dotted,red] (\xa+\pxa+0.1,\ya+6-\pya-0.1) -- (\xa+\pya+0.1,\ya+6-\pxa-0.1);
    \draw [thin,dotted,red] (\xa+\pxa-0.1,\ya+6-\pya+0.1) -- (\xa+\pya-0.1,\ya+6-\pxa+0.1);
	\node[draw,thick,cross,inner sep=1.5,color=red] at (\xa+\pxa+0.1,\ya+6-\pya-0.1) {};
	\node[draw,thick,cross,rotate=45,inner sep=1.5,color=red] at (\xa+\pya+0.1,\ya+6-\pxa-0.1) {};
	\node[draw,thick,cross,rotate=45,inner sep=1.5,color=red] at (\xa+\pxa-0.1,\ya+6-\pya+0.1) {};
	\node[draw,thick,cross,inner sep=1.5,color=red] at (\xa+\pya-0.1,\ya+6-\pxa+0.1) {}; 
    
\draw[->,thin,black] (\xa+\dxa-2,\ya+0.5) -- (\xa+\dxa-1,\ya+0.5);

    \def\ya{0};
    \node[left,scale=0.75] at (\xa-1,\ya+6-3) {$x+\hat{\nu}$};
    \foreach \i in {1,...,5} {
        \draw [thin,dotted,gray] (\xa+\i,\ya+1) -- (\xa+\i,\ya+5)  node[solid,black,above] at (\xa+\i,\ya+5.3) {};
    }
    \foreach \i in {1,...,5} {
        \draw [thin,dotted,gray] (\xa+1,\ya+\i) -- (\xa+5,\ya+\i) node[solid,black,left] at (\xa+0.7,\ya+6-\i) {};
    }
    \foreach \i in {1,...,5} {
      \draw [thin,black!100] (\xa+5,\ya+6-\i) -- (\xa+\i,\ya+6-\i) -- (\xa+\i,\ya+6-1);
    }
    \draw [thin,dashed,black] (\xa+1,\ya+5) -- (\xa+5,\ya+1);

    \def\pxa{4};
    \def\pya{3};
    \node[left,scale=0.65] at (\xa+0.75,\ya+6-\pya) {$a_{0}$};
    \draw [thin,black] (\xa+0.85,\ya+6-\pya) -- (\xa+1,\ya+6-\pya);
    \node[left,scale=0.65] at (\xa+0.75,\ya+6-\pxa) {$b_{0}$};
    \draw [thin,black] (\xa+0.85,\ya+6-\pxa) -- (\xa+1,\ya+6-\pxa);      

    \def\xa{\dxa};
    \def\ya{\dya};
    \foreach \i in {1,...,5} {
        \draw [thin,dotted,gray] (\xa+\i,\ya+1) -- (\xa+\i,\ya+5) node[solid,black,above] at (\xa+\i,\ya+5.3) {};
    }
    \foreach \i in {1,...,5} {
        \draw [thin,dotted,gray] (\xa+1,\ya+\i) -- (\xa+5,\ya+\i) node[solid,black,left] at (\xa+0.7,\ya+6-\i) {};
    }
    \foreach \i in {1,...,5} {
      \draw [thin,black!100] (\xa+5,\ya+6-\i) -- (\xa+\i,\ya+6-\i) -- (\xa+\i,\ya+6-1);
    }
    \draw [thin,dashed,black] (\xa+1,\ya+5) -- (\xa+5,\ya+1);

    \def\pxa{4};
    \def\pya{3};
    \def\pxb{4};
    \def\pyb{2};
    \def\pxc{3};
    \def\pyc{2};
    \node[left,scale=0.65] at (\xa+0.75,\ya+6-\pya) {$a=a_{0}$};
    \draw [thin,black] (\xa+0.85,\ya+6-\pya) -- (\xa+1,\ya+6-\pya);
    \node[left,scale=0.65] at (\xa+0.75,\ya+6-\pxa) {$b_{0}$};
    \draw [thin,black] (\xa+0.85,\ya+6-\pxa) -- (\xa+1,\ya+6-\pxa);
    \node[left,scale=0.65] at (\xa+0.75,\ya+6-\pyb) {$b$};
    \draw [thin,black] (\xa+0.85,\ya+6-\pyb) -- (\xa+1,\ya+6-\pyb);
    \draw [thin,dotted,red] (\xa+\pxa,\ya+6-\pya) -- (\xa+\pya,\ya+6-\pxa);
    \draw [thin,dotted,red] (\xa+\pxb,\ya+6-\pyb) -- (\xa+\pyb,\ya+6-\pxb);
    \draw [thin,dotted,red] (\xa+\pxc,\ya+6-\pyc) -- (\xa+\pyc,\ya+6-\pxc);
	\node[draw,thick,cross,rotate=45,inner sep=1.5,color=red] at (\xa+\pxa,\ya+6-\pya) {};
	\node[draw,thick,cross,inner sep=1.5,color=red] at (\xa+\pya,\ya+6-\pxa) {}; 
	\node[draw,thick,cross,inner sep=1.5,color=red] at (\xa+\pxb,\ya+6-\pyb) {};
	\node[draw,thick,cross,rotate=45,inner sep=1.5,color=red] at (\xa+\pyb,\ya+6-\pxb) {};
	\node[draw,thick,circle,inner sep=1.,fill,color=red] at (\xa+\pxc,\ya+6-\pyc) {};
	\node[draw,thick,circle,inner sep=1.,color=red] at (\xa+\pyc,\ya+6-\pxc) {};

\draw[->,thin,black] (\xa+\dxa-2,\ya+0.5) -- (\xa+\dxa-1,\ya+0.5);

    \def\ya{0};
    \foreach \i in {1,...,5} {
        \draw [thin,dotted,gray] (\xa+\i,\ya+1) -- (\xa+\i,\ya+5)  node[solid,black,above] at (\xa+\i,\ya+5.3) {};
    }
    \foreach \i in {1,...,5} {
        \draw [thin,dotted,gray] (\xa+1,\ya+\i) -- (\xa+5,\ya+\i) node[solid,black,left] at (\xa+0.7,\ya+6-\i) {};
    }
    \foreach \i in {1,...,5} {
      \draw [thin,black!100] (\xa+5,\ya+6-\i) -- (\xa+\i,\ya+6-\i) -- (\xa+\i,\ya+6-1);
    }
    \draw [thin,dashed,black] (\xa+1,\ya+5) -- (\xa+5,\ya+1);
    
    \def\pxa{4};
    \def\pya{3};
    \def\pxb{2};
    \def\pyb{3};
    \def\pxc{2};
    \def\pyc{3};
    \node[left,scale=0.65] at (\xa+0.75,\ya+6-\pya) {$a=a_{0}$};
    \draw [thin,black] (\xa+0.85,\ya+6-\pya) -- (\xa+1,\ya+6-\pya);
    \node[left,scale=0.65] at (\xa+0.75,\ya+6-\pxa) {$b_{0}$};
    \draw [thin,black] (\xa+0.85,\ya+6-\pxa) -- (\xa+1,\ya+6-\pxa);
    \node[left,scale=0.65] at (\xa+0.75,\ya+6-\pxb) {$b$};
    \draw [thin,black] (\xa+0.85,\ya+6-\pxb) -- (\xa+1,\ya+6-\pxb);
    \draw [thin,dotted,red] (\xa+\pxb+0.1,\ya+6-\pyb-0.1) -- (\xa+\pyb+0.1,\ya+6-\pxb-0.1);
    \draw [thin,dotted,red] (\xa+\pxc-0.1,\ya+6-\pyc+0.1) -- (\xa+\pyc-0.1,\ya+6-\pxc+0.1);
	\node[draw,thick,cross,inner sep=1.5,color=red] at (\xa+\pxb+0.1,\ya+6-\pyb-0.1) {};
	\node[draw,thick,cross,rotate=45,inner sep=1.5,color=red] at (\xa+\pyb+0.1,\ya+6-\pxb-0.1) {};
	\node[draw,thick,circle,inner sep=1.,fill,color=red] at (\xa+\pxc-0.1,\ya+6-\pyc+0.1) {};
	\node[draw,thick,circle,inner sep=1.,color=red] at (\xa+\pyc-0.1,\ya+6-\pxc+0.1) {};

    \def\xa{2*\dxa};
    \def\ya{\dya};
    \foreach \i in {1,...,5} {
        \draw [thin,dotted,gray] (\xa+\i,\ya+1) -- (\xa+\i,\ya+5) node[solid,black,above] at (\xa+\i,\ya+5.3) {};
    }
    \foreach \i in {1,...,5} {
        \draw [thin,dotted,gray] (\xa+1,\ya+\i) -- (\xa+5,\ya+\i) node[solid,black,left] at (\xa+0.7,\ya+6-\i) {};
    }
    \foreach \i in {1,...,5} {
      \draw [thin,black!100] (\xa+5,\ya+6-\i) -- (\xa+\i,\ya+6-\i) -- (\xa+\i,\ya+6-1);
    }
    \draw [thin,dashed,black] (\xa+1,\ya+5) -- (\xa+5,\ya+1);
    \def\pxa{4};
    \def\pya{3};
    \def\pxb{4};
    \def\pyb{1};
    \def\pxc{3};
    \def\pyc{2};
    \def\pxd{2};
    \def\pyd{1};
    \node[left,scale=0.65] at (\xa+0.75,\ya+6-\pya) {$a_{0}$};
    \draw [thin,black] (\xa+0.85,\ya+6-\pya) -- (\xa+1,\ya+6-\pya);
    \node[left,scale=0.65] at (\xa+0.75,\ya+6-\pxa) {$b_{0}$};
    \draw [thin,black] (\xa+0.85,\ya+6-\pxa) -- (\xa+1,\ya+6-\pxa);
    \node[left,scale=0.65] at (\xa+0.75,\ya+6-\pyc) {$a$};
    \draw [thin,black] (\xa+0.85,\ya+6-\pyc) -- (\xa+1,\ya+6-\pyc);
    \node[left,scale=0.65] at (\xa+0.75,\ya+6-\pyb) {$b$};
    \draw [thin,black] (\xa+0.85,\ya+6-\pyb) -- (\xa+1,\ya+6-\pyb);
    \draw [thin,dotted,red] (\xa+\pxa,\ya+6-\pya) -- (\xa+\pya,\ya+6-\pxa);
    \draw [thin,dotted,red] (\xa+\pxb,\ya+6-\pyb) -- (\xa+\pyb,\ya+6-\pxb);
    \draw [thin,dotted,red] (\xa+\pxc,\ya+6-\pyc) -- (\xa+\pyc,\ya+6-\pxc);
    \draw [thin,dotted,red] (\xa+\pxd,\ya+6-\pyd) -- (\xa+\pyd,\ya+6-\pxd);
	\node[draw,thick,cross,rotate=45,inner sep=1.5,color=red] at (\xa+\pxa,\ya+6-\pya) {};
	\node[draw,thick,cross,inner sep=1.5,color=red] at (\xa+\pya,\ya+6-\pxa) {}; 
	\node[draw,thick,cross,inner sep=1.5,color=red] at (\xa+\pxb,\ya+6-\pyb) {};
	\node[draw,thick,cross,rotate=45,inner sep=1.5,color=red] at (\xa+\pyb,\ya+6-\pxb) {};
	\node[draw,thick,circle,inner sep=1.,fill,color=red] at (\xa+\pxc,\ya+6-\pyc) {};
	\node[draw,thick,circle,inner sep=1.,color=red] at (\xa+\pyc,\ya+6-\pxc) {};
	\node[draw,thick,circle,inner sep=1.,fill,color=red] at (\xa+\pxd,\ya+6-\pyd) {};
	\node[draw,thick,circle,inner sep=1.,color=red] at (\xa+\pyd,\ya+6-\pxd) {};

\draw[->,thin,black] (\xa+\dxa-2,\ya+0.5) -- (\xa+\dxa-1,\ya+0.5);
\draw[->,thin,black] (\xa+\dxa-2,\ya+0.5-0.2) -- (\xa+\dxa-2+0.5,\ya+0.5-0.2) --(\xa+\dxa-2+0.5,\ya+0.5-2*\dya) -- (\xa+\dxa-1,\ya+0.5-2*\dya);

    \def\ya{0};
    \foreach \i in {1,...,5} {
        \draw [thin,dotted,gray] (\xa+\i,\ya+1) -- (\xa+\i,\ya+5)  node[solid,black,above] at (\xa+\i,\ya+5.3) {};
    }
    \foreach \i in {1,...,5} {
        \draw [thin,dotted,gray] (\xa+1,\ya+\i) -- (\xa+5,\ya+\i) node[solid,black,left] at (\xa+0.7,\ya+6-\i) {};
    }
    \foreach \i in {1,...,5} {
      \draw [thin,black!100] (\xa+5,\ya+6-\i) -- (\xa+\i,\ya+6-\i) -- (\xa+\i,\ya+6-1);
    }
    \draw [thin,dashed,black] (\xa+1,\ya+5) -- (\xa+5,\ya+1);
    \def\pxa{4};
    \def\pya{3};
    \def\pxb{1};
    \def\pyb{3};
    \def\pxc{2};
    \def\pyc{3};
    \def\pxd{1};
    \def\pyd{2};
    \node[left,scale=0.65] at (\xa+0.75,\ya+6-\pya) {$a_{0}$};
    \draw [thin,black] (\xa+0.85,\ya+6-\pya) -- (\xa+1,\ya+6-\pya);
    \node[left,scale=0.65] at (\xa+0.75,\ya+6-\pxa) {$b_{0}$};
    \draw [thin,black] (\xa+0.85,\ya+6-\pxa) -- (\xa+1,\ya+6-\pxa);
    \node[left,scale=0.65] at (\xa+0.75,\ya+6-\pxc) {$a$};
    \draw [thin,black] (\xa+0.85,\ya+6-\pxc) -- (\xa+1,\ya+6-\pxc);
    \node[left,scale=0.65] at (\xa+0.75,\ya+6-\pxb) {$b$};
    \draw [thin,black] (\xa+0.85,\ya+6-\pxb) -- (\xa+1,\ya+6-\pxb);
    \draw [thin,dotted,red] (\xa+\pxb,\ya+6-\pyb) -- (\xa+\pyb,\ya+6-\pxb);
    \draw [thin,dotted,red] (\xa+\pxc,\ya+6-\pyc) -- (\xa+\pyc,\ya+6-\pxc);
    \draw [thin,dotted,red] (\xa+\pxd,\ya+6-\pyd) -- (\xa+\pyd,\ya+6-\pxd);
	\node[draw,thick,cross,inner sep=1.5,color=red] at (\xa+\pxb,\ya+6-\pyb) {};
	\node[draw,thick,cross,rotate=45,inner sep=1.5,color=red] at (\xa+\pyb,\ya+6-\pxb) {};
	\node[draw,thick,circle,inner sep=1.,fill,color=red] at (\xa+\pxc,\ya+6-\pyc) {};
	\node[draw,thick,circle,inner sep=1.,color=red] at (\xa+\pyc,\ya+6-\pxc) {};
	\node[draw,thick,circle,inner sep=1.,fill,color=red] at (\xa+\pxd,\ya+6-\pyd) {};
	\node[draw,thick,circle,inner sep=1.,color=red] at (\xa+\pyd,\ya+6-\pxd) {};

    \def\xa{3*\dxa};
    \def\ya{\dya};
    \foreach \i in {1,...,5} {
        \draw [thin,dotted,gray] (\xa+\i,\ya+1) -- (\xa+\i,\ya+5) node[solid,black,above] at (\xa+\i,\ya+5.3) {};
    }
    \foreach \i in {1,...,5} {
        \draw [thin,dotted,gray] (\xa+1,\ya+\i) -- (\xa+5,\ya+\i) node[solid,black,left] at (\xa+0.7,\ya+6-\i) {};
    }
    \foreach \i in {1,...,5} {
      \draw [thin,black!100] (\xa+5,\ya+6-\i) -- (\xa+\i,\ya+6-\i) -- (\xa+\i,\ya+6-1);
    }
    \draw [thin,dashed,black] (\xa+1,\ya+5) -- (\xa+5,\ya+1);
    \def\pxa{4};
    \def\pya{3};
    \def\pxb{4};
    \def\pyb{1};
    \def\pxc{3};
    \def\pyc{2};
    \def\pxd{2};
    \def\pyd{1};
    \node[left,scale=0.65] at (\xa+0.75,\ya+6-\pya) {$a_{0}$};
    \draw [thin,black] (\xa+0.85,\ya+6-\pya) -- (\xa+1,\ya+6-\pya);
    \node[left,scale=0.65] at (\xa+0.75,\ya+6-\pxa) {$b=b_{0}$};
    \draw [thin,black] (\xa+0.85,\ya+6-\pxa) -- (\xa+1,\ya+6-\pxa);
    \node[left,scale=0.65] at (\xa+0.75,\ya+6-\pyb) {$a$};
    \draw [thin,black] (\xa+0.85,\ya+6-\pyb) -- (\xa+1,\ya+6-\pyb);
    \draw [thin,dotted,red] (\xa+\pxa,\ya+6-\pya) -- (\xa+\pya,\ya+6-\pxa);
    \draw [thin,dotted,red] (\xa+\pxb,\ya+6-\pyb) -- (\xa+\pyb,\ya+6-\pxb);
    \draw [thin,dotted,red] (\xa+\pxc,\ya+6-\pyc) -- (\xa+\pyc,\ya+6-\pxc);
    \draw [thin,dotted,red] (\xa+\pxd,\ya+6-\pyd) -- (\xa+\pyd,\ya+6-\pxd);
	\node[draw,thick,cross,rotate=45,inner sep=1.5,color=red] at (\xa+\pxa,\ya+6-\pya) {};
	\node[draw,thick,cross,inner sep=1.5,color=red] at (\xa+\pya,\ya+6-\pxa) {}; 
	\node[draw,thick,circle,inner sep=1.,color=red] at (\xa+\pxb,\ya+6-\pyb) {};
	\node[draw,thick,circle,inner sep=1.,fill,color=red] at (\xa+\pyb,\ya+6-\pxb) {};
	\node[draw,thick,circle,inner sep=1.,fill,color=red] at (\xa+\pxc,\ya+6-\pyc) {};
	\node[draw,thick,circle,inner sep=1.,color=red] at (\xa+\pyc,\ya+6-\pxc) {};
	\node[draw,thick,circle,inner sep=1.,fill,color=red] at (\xa+\pxd,\ya+6-\pyd) {};
	\node[draw,thick,circle,inner sep=1.,color=red] at (\xa+\pyd,\ya+6-\pxd) {};
	
	\draw[->,thin,black] (\xa+\dxa-2,\ya+0.5) -- (\xa+\dxa-1,\ya+0.5);
	\node[black,right,scale=0.75] at (\xa+\dxa,\ya+1) {$x\leftarrow x+\hat{\nu}$};
	\node[black,right,scale=0.75] at (\xa+\dxa,\ya) {$\nu\leftarrow \op{rand\_dir}\of{}$};
	
    \def\ya{0};
    \foreach \i in {1,...,5} {
        \draw [thin,dotted,gray] (\xa+\i,\ya+1) -- (\xa+\i,\ya+5)  node[solid,black,above] at (\xa+\i,\ya+5.3) {};
    }
    \foreach \i in {1,...,5} {
        \draw [thin,dotted,gray] (\xa+1,\ya+\i) -- (\xa+5,\ya+\i) node[solid,black,left] at (\xa+0.7,\ya+6-\i) {};
    }
    \foreach \i in {1,...,5} {
      \draw [thin,black!100] (\xa+5,\ya+6-\i) -- (\xa+\i,\ya+6-\i) -- (\xa+\i,\ya+6-1);
    }
    \draw [thin,dashed,black] (\xa+1,\ya+5) -- (\xa+5,\ya+1);
    \def\pxa{4};
    \def\pya{3};
    \def\pxb{1};
    \def\pyb{4};
    \def\pxc{2};
    \def\pyc{3};
    \def\pxd{1};
    \def\pyd{2};
    \node[left,scale=0.65] at (\xa+0.75,\ya+6-\pya) {$a_{0}$};
    \draw [thin,black] (\xa+0.85,\ya+6-\pya) -- (\xa+1,\ya+6-\pya);
    \node[left,scale=0.65] at (\xa+0.75,\ya+6-\pxa) {$b=b_{0}$};
    \draw [thin,black] (\xa+0.85,\ya+6-\pxa) -- (\xa+1,\ya+6-\pxa);
    \node[left,scale=0.65] at (\xa+0.75,\ya+6-\pxb) {$a$};
    \draw [thin,black] (\xa+0.85,\ya+6-\pxb) -- (\xa+1,\ya+6-\pxb);
    \draw [thin,dotted,red] (\xa+\pxa,\ya+6-\pya) -- (\xa+\pya,\ya+6-\pxa);
    \draw [thin,dotted,red] (\xa+\pxb,\ya+6-\pyb) -- (\xa+\pyb,\ya+6-\pxb);
    \draw [thin,dotted,red] (\xa+\pxc,\ya+6-\pyc) -- (\xa+\pyc,\ya+6-\pxc);
    \draw [thin,dotted,red] (\xa+\pxd,\ya+6-\pyd) -- (\xa+\pyd,\ya+6-\pxd);
	\node[draw,thick,cross,inner sep=1.5,color=red] at (\xa+\pxa,\ya+6-\pya) {};
	\node[draw,thick,cross,rotate=45,inner sep=1.5,color=red] at (\xa+\pya,\ya+6-\pxa) {}; 
	\node[draw,thick,circle,inner sep=1.,color=red] at (\xa+\pxb,\ya+6-\pyb) {};
	\node[draw,thick,circle,inner sep=1.,fill,color=red] at (\xa+\pyb,\ya+6-\pxb) {};
	\node[draw,thick,circle,inner sep=1.,fill,color=red] at (\xa+\pxc,\ya+6-\pyc) {};
	\node[draw,thick,circle,inner sep=1.,color=red] at (\xa+\pyc,\ya+6-\pxc) {};
	\node[draw,thick,circle,inner sep=1.,fill,color=red] at (\xa+\pxd,\ya+6-\pyd) {};
	\node[draw,thick,circle,inner sep=1.,color=red] at (\xa+\pyd,\ya+6-\pxd) {};

    \def\xa{3*\dxa};
    \def\ya{-\dya};

    \node[left,scale=0.75] at (\xa-3,\ya+6-3) {$x$};    
    
    \foreach \i in {1,...,5} {
        \draw [thin,dotted,gray] (\xa+\i,\ya+1) -- (\xa+\i,\ya+5) node[solid,black,above] at (\xa+\i,\ya+5.3) {};
    }
    \foreach \i in {1,...,5} {
        \draw [thin,dotted,gray] (\xa+1,\ya+\i) -- (\xa+5,\ya+\i) node[solid,black,left] at (\xa+0.7,\ya+6-\i) {};
    }
    \foreach \i in {1,...,5} {
      \draw [thin,black!100] (\xa+5,\ya+6-\i) -- (\xa+\i,\ya+6-\i) -- (\xa+\i,\ya+6-1);
    }
    \draw [thin,dashed,black] (\xa+1,\ya+5) -- (\xa+5,\ya+1);
    \def\pxa{4};
    \def\pya{3};
    \def\pxb{3};
    \def\pyb{1};
    \def\pxc{3};
    \def\pyc{2};
    \def\pxd{2};
    \def\pyd{1};
    
    \node[left,scale=0.65] at (\xa+0.75,\ya+6-\pya) {$b=a_{0}$};
    \draw [thin,black] (\xa+0.85,\ya+6-\pya) -- (\xa+1,\ya+6-\pya);
    \node[left,scale=0.65] at (\xa+0.75,\ya+6-\pxa) {$b_{0}$};
    \draw [thin,black] (\xa+0.85,\ya+6-\pxa) -- (\xa+1,\ya+6-\pxa);
    \node[left,scale=0.65] at (\xa+0.75,\ya+6-\pyb) {$a$};
    \draw [thin,black] (\xa+0.85,\ya+6-\pyb) -- (\xa+1,\ya+6-\pyb);
    \draw [thin,dotted,red] (\xa+\pxa+0.1,\ya+6-\pya-0.1) -- (\xa+\pya+0.1,\ya+6-\pxa-0.1);
    \draw [thin,dotted,red] (\xa+\pxa-0.1,\ya+6-\pya+0.1) -- (\xa+\pya-0.1,\ya+6-\pxa+0.1);    
    
    \draw [thin,dotted,red] (\xa+\pxb,\ya+6-\pyb) -- (\xa+\pyb,\ya+6-\pxb);
    \draw [thin,dotted,red] (\xa+\pxc,\ya+6-\pyc) -- (\xa+\pyc,\ya+6-\pxc);
    \draw [thin,dotted,red] (\xa+\pxd,\ya+6-\pyd) -- (\xa+\pyd,\ya+6-\pxd);
	\node[draw,thick,cross,rotate=45,inner sep=1.5,color=red] at (\xa+\pxa+0.1,\ya+6-\pya-0.1) {};
	\node[draw,thick,cross,inner sep=1.5,color=red] at (\xa+\pya+0.1,\ya+6-\pxa-0.1) {}; 
	
	\node[draw,thick,cross,inner sep=1.5,color=red] at (\xa+\pxa-0.1,\ya+6-\pya+0.1) {};
	\node[draw,thick,cross,rotate=45,inner sep=1.5,color=red] at (\xa+\pya-0.1,\ya+6-\pxa+0.1) {};
	
	\node[draw,thick,circle,inner sep=1.,color=red] at (\xa+\pxb,\ya+6-\pyb) {};
	\node[draw,thick,circle,inner sep=1.,fill,color=red] at (\xa+\pyb,\ya+6-\pxb) {};
	\node[draw,thick,circle,inner sep=1.,fill,color=red] at (\xa+\pxc,\ya+6-\pyc) {};
	\node[draw,thick,circle,inner sep=1.,color=red] at (\xa+\pyc,\ya+6-\pxc) {};
	\node[draw,thick,circle,inner sep=1.,fill,color=red] at (\xa+\pxd,\ya+6-\pyd) {};
	\node[draw,thick,circle,inner sep=1.,color=red] at (\xa+\pyd,\ya+6-\pxd) {};
	
	\draw[->,thin,black] (\xa+\dxa-2,\ya+0.5) -- (\xa+\dxa-1,\ya+0.5);
	\node[black,right,scale=0.75] at (\xa+\dxa,\ya+1) {$x\leftarrow x$};
	\node[black,right,scale=0.75] at (\xa+\dxa,\ya) {$\nu\leftarrow \op{rand\_dir}\of{}$};
			
    \def\ya{-2*\dya};
     \node[left,scale=0.75] at (\xa-3,\ya+6-3) {$x+\hat{\nu}$};
    \foreach \i in {1,...,5} {
        \draw [thin,dotted,gray] (\xa+\i,\ya+1) -- (\xa+\i,\ya+5)  node[solid,black,above] at (\xa+\i,\ya+5.3) {};
    }
    \foreach \i in {1,...,5} {
        \draw [thin,dotted,gray] (\xa+1,\ya+\i) -- (\xa+5,\ya+\i) node[solid,black,left] at (\xa+0.7,\ya+6-\i) {};
    }
    \foreach \i in {1,...,5} {
      \draw [thin,black!100] (\xa+5,\ya+6-\i) -- (\xa+\i,\ya+6-\i) -- (\xa+\i,\ya+6-1);
    }
    \draw [thin,dashed,black] (\xa+1,\ya+5) -- (\xa+5,\ya+1);
    \def\pxa{4};
    \def\pya{3};
    \def\pxb{1};
    \def\pyb{3};
    \def\pxc{2};
    \def\pyc{3};
    \def\pxd{1};
    \def\pyd{2};
    \node[left,scale=0.65] at (\xa+0.75,\ya+6-\pya) {$b=a_{0}$};
    \draw [thin,black] (\xa+0.85,\ya+6-\pya) -- (\xa+1,\ya+6-\pya);
    \node[left,scale=0.65] at (\xa+0.75,\ya+6-\pxa) {$b_{0}$};
    \draw [thin,black] (\xa+0.85,\ya+6-\pxa) -- (\xa+1,\ya+6-\pxa);
    \node[left,scale=0.65] at (\xa+0.75,\ya+6-\pxb) {$a$};
    \draw [thin,black] (\xa+0.85,\ya+6-\pxb) -- (\xa+1,\ya+6-\pxb);

    \draw [thin,dotted,red] (\xa+\pxb,\ya+6-\pyb) -- (\xa+\pyb,\ya+6-\pxb);
    \draw [thin,dotted,red] (\xa+\pxc,\ya+6-\pyc) -- (\xa+\pyc,\ya+6-\pxc);
    \draw [thin,dotted,red] (\xa+\pxd,\ya+6-\pyd) -- (\xa+\pyd,\ya+6-\pxd);
	\node[draw,thick,circle,inner sep=1.,color=red] at (\xa+\pxb,\ya+6-\pyb) {};
	\node[draw,thick,circle,inner sep=1.,fill,color=red] at (\xa+\pyb,\ya+6-\pxb) {};
	\node[draw,thick,circle,inner sep=1.,fill,color=red] at (\xa+\pxc,\ya+6-\pyc) {};
	\node[draw,thick,circle,inner sep=1.,color=red] at (\xa+\pyc,\ya+6-\pxc) {};
	\node[draw,thick,circle,inner sep=1.,fill,color=red] at (\xa+\pxd,\ya+6-\pyd) {};
	\node[draw,thick,circle,inner sep=1.,color=red] at (\xa+\pyd,\ya+6-\pxd) {};   
    
  \end{scope}
\end{tikzpicture}

%% file: tikzimg/topoplaquettes.tex
\begin{tikzpicture}[scale=0.7,nodes={inner sep=0}]
  \pgfpointtransformed{\pgfpointxy{1}{1}};
  \pgfgetlastxy{\vx}{\vy}
  \begin{scope}[node distance=\vx and \vy]
    \def\dxa{5.5};
    \def\dya{5};
    \def\xa{0};
    \def\ya{\dya};
    \foreach \i in {1,...,5} {
        \draw [thin,dotted,gray] (\xa+\i,\ya+1) -- (\xa+\i,\ya+5)  node[solid,black,above] at (\xa+\i,\ya+5.3) {};
    }
    \foreach \i in {1,...,5} {
        \draw [thin,dotted,gray] (\xa+1,\ya+\i) -- (\xa+5,\ya+\i) node[solid,black,left] at (\xa+0.7,\ya+6-\i) {};
    }

    \node[black,below] at (\xa+1,\ya+2-0.2) {$x=y$};
	\draw[black,fill=blue!70] (\xa+1,\ya+2) rectangle (\xa+2,\ya+3);
	\node[draw,thick,circle,inner sep=1,fill,color=red] at (\xa+1,\ya+2) {};

    \def\xa{\dxa};
    \def\ya{\dya};
    \foreach \i in {1,...,5} {
        \draw [thin,dotted,gray] (\xa+\i,\ya+1) -- (\xa+\i,\ya+5)  node[solid,black,above] at (\xa+\i,\ya+5.3) {};
    }
    \foreach \i in {1,...,5} {
        \draw [thin,dotted,gray] (\xa+1,\ya+\i) -- (\xa+5,\ya+\i) node[solid,black,left] at (\xa+0.7,\ya+6-\i) {};
    }
    \node[black,below] at (\xa+1,\ya+2-0.2) {$x$};
	\draw[black,fill=blue!50] (\xa+1,\ya+2) rectangle (\xa+2,\ya+3);
	\node[draw,thick,circle,inner sep=1,fill,color=red] at (\xa+1,\ya+2) {};
    \node[black,below] at (\xa+2,\ya+2-0.2) {$y$};
    \draw[black,fill=blue!50] (\xa+2,\ya+2) rectangle (\xa+3,\ya+3);
    \node[draw,thick,circle,inner sep=1,fill,color=red] at (\xa+2,\ya+2) {};
    
    \def\xa{2*\dxa};
    \def\ya{\dya};
    \foreach \i in {1,...,5} {
        \draw [thin,dotted,gray] (\xa+\i,\ya+1) -- (\xa+\i,\ya+5)  node[solid,black,above] at (\xa+\i,\ya+5.3) {};
    }
    \foreach \i in {1,...,5} {
        \draw [thin,dotted,gray] (\xa+1,\ya+\i) -- (\xa+5,\ya+\i) node[solid,black,left] at (\xa+0.7,\ya+6-\i) {};
    }
    
    \node[black,below] at (\xa+1,\ya+2-0.2) {$x$};
	\draw[black,fill=blue!50] (\xa+1,\ya+2) rectangle (\xa+2,\ya+3);
	\node[draw,thick,circle,inner sep=1,fill,color=red] at (\xa+1,\ya+2) {};
    \node[black,below] at (\xa+2+0.15,\ya+3-0.2) {$y$};
    \draw[black,fill=blue!50] (\xa+2,\ya+3) rectangle (\xa+3,\ya+4);
    \node[draw,thick,circle,inner sep=1,fill,color=red] at (\xa+2,\ya+3) {};
    
    \def\xa{3*\dxa};
    \def\ya{\dya};
    \foreach \i in {1,...,5} {
        \draw [thin,dotted,gray] (\xa+\i,\ya+1) -- (\xa+\i,\ya+5)  node[solid,black,above] at (\xa+\i,\ya+5.3) {};
    }
    \foreach \i in {1,...,5} {
        \draw [thin,dotted,gray] (\xa+1,\ya+\i) -- (\xa+5,\ya+\i) node[solid,black,left] at (\xa+0.7,\ya+6-\i) {};
    }
    \node[black,below] at (\xa+1,\ya+2-0.2) {$x$};
	\draw[black,fill=blue!50] (\xa+1,\ya+2) rectangle (\xa+2,\ya+3);
	\node[draw,thick,circle,inner sep=1,fill,color=red] at (\xa+1,\ya+2) {};
    \node[black,below] at (\xa+3,\ya+3-0.2) {$y$};
    \draw[black,fill=blue!50] (\xa+3,\ya+3) rectangle (\xa+4,\ya+4);
    \node[draw,thick,circle,inner sep=1,fill,color=red] at (\xa+3,\ya+3) {};
  \end{scope}
\end{tikzpicture}